\documentclass[vecphys]{svmultm}

\usepackage[breaklinks,colorlinks,bookmarks=false,citecolor=blue,linkcolor=red,urlcolor=blue]{hyperref}    
\usepackage{makeidx}     
\usepackage{graphicx}    
\usepackage{multicol}    
\usepackage{amsfonts}    
\usepackage{amssymb}     
\usepackage{cite}    

\makeindex             

\newcommand{\myframe}[1]{#1}
\newcommand{\nn}{\nonumber}

\newcommand{\be}{\begin{equation}}
\newcommand{\ee}{\end{equation}}
\newcommand{\bit}{\begin{itemize}}
\newcommand{\eit}{\end{itemize}}
\newcommand{\bea}{\begin{eqnarray}}
\newcommand{\eea}{\end{eqnarray}}
\newcommand{\stateX}{\(\sqrt{3}\)\(\times\)\(\sqrt{3}\)}
\newcommand{\stateO}{\(q\)\(=\)\(0\)}
\newcommand{\Neel}{N\'{e}el }
\newcommand{\Kagome}{kagom\'e}
\newcommand{\kagome}{{\Kagome} }
\newcommand{\UC}{unit cell}
\newcommand{\SWT}{spin-wave theory}
\def\ChapExp{P.~Lemmens and P.~Millet}
\def\ChapOneD{H.-J.~Mikeska and A.K.~Kolezhuk}
\def\ChapQPT{S.~Sachdev}
\def\ChapED{N.~Laflorencie and D.~Poilblanc}
\def\ChapSWT{N.B.~Ivanov and D.~Sen}
\def\ChapFT{D.C.~Cabra and P.~Pujol}
\def\ChapCCM{D.J.J.~Farnell and R.F.~Bishop}

\begin{document}

\title*{Quantum magnetism in two dimensions: From semi-classical \Neel order
to magnetic disorder 
}
\titlerunning{Quantum magnetism in two dimensions}
\author{J.~Richter\inst{1}, J.~Schulenburg\inst{2}, \and
A.~Honecker\inst{3}}
\authorrunning{Richter, Schulenburg, and Honecker}
\institute{Institut f\"ur Theoretische Physik, Otto-von-Guericke-Universit\"at
Magdeburg, P.O.Box 4120, D-39016 Magdeburg, Germany
\texttt{johannes.richter@physik.uni.magdeburg.de}
\and Universit\"atsrechenzentrum, Otto-von-Guericke-Universit\"at
Magdeburg, P.O.Box 4120, D-39016 Magdeburg, Germany
\texttt{joerg.schulenburg@ovgu.de}
\and Institut f\"ur Theoretische Physik, TU Braunschweig,
Mendelsohnstr. 3, D-38106 Braunschweig, Germany
\& Universit\"at Hannover, Institut f\"ur Theoretische
Physik, Appelstrasse 2, D-30167 Hannover, Germany
 \\
\texttt{andreas.honecker@u-cergy.fr}}
%
%
\maketitle

\vspace*{4mm}
Published version: Chapter 2 in
\href{http://dx.doi.org/10.1007/b96825}%
{\it Quantum Magnetism},
ed by U.~Schollw\"ock, J.~Richter, D.J.J.~Farnell, R.F.~Bishop,
\href{http://dx.doi.org/10.1007/BFb0119592}%
{Lecture Notes in Physics {\bf 645}, 85-153}
(Springer-Verlag, Berlin Heidelberg 2004)
\vspace*{6mm} \\
{\bf Abstract:}  

In this article we focus on the ground state  and the low-lying excitations
of the $s=1/2$ Heisenberg antiferromagnet (HAFM) on the 11 two-dimensional (2D) 
uniform Archimedean lattices.

Although we know from the Mermin-Wagner theorem that 
thermal fluctuations are strong enough to
destroy magnetic long-range order (LRO) for Heisenberg spin systems at 
any finite temperature 
in one  and two dimensions, 
 the role of quantum fluctuations is less understood.
 While  the ground state of the
 one-dimensional (1D) quantum HAFM is not long-range ordered,  
the quantum HAFM e.g.\ on the 2D square  and  triangular lattices exhibits
semi-classical N\'eel like LRO.
However, in two dimensions there are many other lattices with different coordination
numbers and topologies, and there is no general statement concerning
zero-temperature N\'eel-like LRO. Recent experimental results on
CaV$_4$O$_9$ and SrCu$_2$(BO$_3$)$_2$
demonstrate the possibility of non-N\'{e}el
 ordered ground states and signal that the $s=1/2$ HAFM  on 
2D lattices with appropriate 
topology may have a ground state without semi-classical LRO.

Based on extensive large-scale exact diagonalization studies of the ground
state and the low-lying excitations for the spin-1/2 HAFM on the Archimedean
lattices we compare and discuss the ground-state features of all 11
lattices. In this manner we obtain some insight in the influence of
lattice topology on magnetic ordering of quantum antiferromagnets in
two dimensions. From our results we conclude that
the ground state of the spin-1/2 HAFM on most of the Archimedean lattices 
(in particular the four bipartite ones) turns out to be semi-classically
N\'eel-like ordered. However, we find that the interplay of competition
of bonds (geometric frustration and non-equivalent nearest neighbor bonds)
and quantum fluctuations gives rise to a quantum paramagnetic ground state 
without semi-classical LRO 
for two lattices. The first one is the famous kagom\'e lattice, for
which this statement is  well-known by numerous studies during the last
decade. Remarkably, we find one additional lattice
among the 11 uniform Archimedean lattices, the so-called star lattice, with
a quantum paramagnetic ground state.
For both these  Archimedean lattices the ground state is
highly degenerate in the classical limit $s \to \infty$, although notably their
quantum ground states are fundamentally different.

Furthermore, we present numerical results for the
magnetization curve of the HAFM on all
11 Archimedean lattices. The magnetization process is discussed
in some detail for the square, triangular and \kagome lattices.
One focus are plateaus appearing in the magnetization curve
due to quantum fluctuations and geometric frustration.
In particular, the kagom\'e lattice
exhibits a rich spectrum of magnetization plateaus.
Another focus are magnetization jumps arising on the \kagome
and the star lattice just below the saturation field.
These magnetization jumps may be understood analytically
by using independent local magnon excitations.

Some related $s=1/2$ models are also discussed briefly.
Particular attention is given to the 2D Shastry-Sutherland model.
For this model, we discuss quantum phase transitions and discuss
the magnetization curve in comparison with experiments
on SrCu$_2$(BO$_3$)$_2$.

\vspace*{2.8cm}
\noindent
This preprint version differs in the following respects from the
book chapter:
\begin{itemize}
\item Value of the spin-wave result \cite{miyake92} of the sublattice magnetization
on the triangular lattice  on page \pageref{mSlSWT} corrected:
\emph{linear}  SWT: $m^{sl}=0.2387=0.4774\;m^{sl}_{\mathrm{class}}$.
For information: the \emph{second-order} SWT theory result of Ref.~\cite{miyake92}
reads
$m^{sl}=s-0.2613+0.0055/s=0.2497=0.4994\;m^{sl}_{\mathrm{class}}$ for $s=1/2$.
\item Value of the
ground-state energy per bond for the maple-leaf lattice
on page \pageref{p:maple-leaf} corrected: $E_0/\mathrm{bond}=-0.2137$.
\item Value of the square of the order parameter for the
$N=36$ \kagome lattice on page \pageref{kag36orderParam} corrected: $(m^+)^2= 0.059128$.
\item Some references updated.
\end{itemize}

\newpage
\setcounter{tocdepth}{2}
\tableofcontents

\newpage

\section{Introduction} \label{intro}

The subject of quantum spin-half
antiferromagnetism in two-dimensional (2D) systems
has attracted a great deal of interest in recent times in
connection with the magnetic properties of layered cuprate high-temperature
superconductors \cite{barnes91dd,manousakis91dd,dagotto94dd} 
and with the recent progress in synthesizing novel 
quasi-2D magnetic materials exhibiting a spin-gap behavior
like CaV$_4$O$_9$ \cite{taniguchi95dd}
or SrCu$_2$(BO$_3$)$_2$ \cite{kageyama99dd}.
Another striking feature is the plateau structure in the magnetization
process of frustrated quasi-two-dimensional magnetic materials like
SrCu$_2$(BO$_3$)$_2$ \cite{kageyama99dd}
or Cs$_2$CuBr$_4$ \cite{TOKMIG02} 
(for more details concerning the experiments see chapter by \ChapExp\
in this book).
However, low-dimensional quantum spin systems are
 of interest in their own right  as examples of strongly interacting
 quantum  many-body systems.
 Although we know from the Mermin-Wagner theorem \cite{mermin66dd} that 
 thermal fluctuations are strong enough to
 destroy magnetic long-range order (LRO) for Heisenberg spin systems in one
and two
dimensions at any finite temperature, 
 the role of quantum fluctuations is less understood.
For the  magnetic ordering in the ground state (GS)
the transition from one to two dimensions  seems
to be crucial. It is well known that the GS of the one-dimensional
 Heisenberg quantum antiferromagnet does not possess \Neel LRO
(see chapter by \ChapOneD\ in this book).  
On the other hand  
as a result of intensive work in the late eighties it is now
 well-established that the GS of the 
Heisenberg antiferromagnet  on the square lattice
exhibits semi-classical \Neel LRO 
(see for example the reviews \cite{barnes91dd,manousakis91dd}).
 However, Anderson's and Fazekas'
 investigations \cite{anderson73dd,fazekas74} of the
 triangular lattice led to the conjecture that quantum fluctuations
 plus frustration may be sufficient to destroy the N\'eel-like
 LRO in
 two dimensions.

 Besides frustration,
 there is another mechanism favoring the ``melting'' of 
N\'eel ordering in the
 ground states of unfrustrated Heisenberg antiferromagnets,
 namely the competition of non-equivalent nearest-neighbor (NN) 
bonds leading to the
 formation of local singlets of two (or even four) 
coupled spins.
By contrast to frustration, which yields
competition in quantum as well as in classical systems, this type of
competition is present only in quantum systems. 

Several notations for the quantum phases without semi-classical \Neel order
are used in the literature, where one often finds the terms `quantum disorder'
or `quantum spin liquid'. However, these quantum phases may exhibit quite
different complex properties.
We shall prefer the notation `quantum paramagnet' (see, e.g.\ \cite{sachdev})
to stress their common feature, namely the absence of magnetic order at $T=0$.

A more specific classification of GS phases of 2D
quantum magnets has been proposed recently by 
 Lhuillier, Sindzingre, Fouet and  Misguich
\cite{lhuillier00dec,lhuillier01sep,lhuillier01oct,lhuillier02dec,lhuillier03}.
Besides the semi-classical \Neel like LRO, these authors also
characterize three quantum GS phases, namely the so-called valence bond crystal,
the type I spin liquid and the type II spin liquid (for more details see 
\cite{lhuillier00dec,lhuillier01sep,lhuillier01oct,lhuillier02dec,lhuillier03} 
and also section \ref{summary_1}). 

We note that quantum paramagnetic phases may be observed also in
three-dimensional strongly frustrated quantum magnets like the
Heisenberg antiferromagnet on the pyrochlore lattice\cite{canals98}
although the tendency to order is more pronounced in three than in two
dimensions.

In this review we focus on the GS of the 2D isotropic  
Heisenberg antiferromagnet (HAFM)
\be \label{ham}
 H= \sum_{<i,j>}J_{ij}{\bf S}_i{\bf S}_j =
\sum_{<i,j>} J_{ij}\left( S_i^xS_j^x+S_i^yS_j^y+S_i^zS_j^z 
   \right)
\ee
and consider the 
extreme quantum case of spin quantum number $1/2$.
Of course, there is a long history of investigations of this model.
Nevertheless, much interesting new physics has been discovered in recent
years. The 2D systems are of particular interest because the
competition between quantum fluctuations and interactions seems to be well
balanced, and fine tuning of this competition 
may lead to zero-temperature transitions between
semi-classical and quantum phases (see chapter by \ChapQPT\ in this book
and also section \ref{qpt}).

The calculation of the GS of the spin half HAFM is challenging. Besides 
the conventional methods like spin-wave theory and general quantum-many body
techniques like the coupled cluster method also 
new numerical methods like quantum
Monte Carlo and exact diagonalization are  powerful instruments. However,
only a few of them (e.g.\ exact diagonalization or the coupled cluster method)
are universally applicable, whereas some  methods  suffer from the sign 
problem in frustrated systems. More details regarding analytical and numerical
methods can be found in chapters by \ChapSWT; \ChapFT; \ChapED; \ChapCCM.
The majority of the results presented in this chapter were obtained by exact
diagonalization using the program package {\tt spinpack} \cite{spinpack}.

Quantum magnetism in 2D systems is a very broad field. To be specific and
different from other existing reviews 
we focus our discussion on the ground
state properties of the spin half HAFM on the 
11 uniform Archimedean lattices (tilings). These
lattices are the prototypes of 2D arrangements of spins and vary in their
geometrical and topological properties. Hence they present an ideal
possibility for a systematic study of the interplay of   
lattice geometry and magnetic interactions in 2D quantum spin systems.
Many of the lattices considered find their realization in nature either in
a pure or in a modified form. Furthermore, almost all lattices can be
transformed into each other by bond or site depletion/addition.
One now has the opportunity to study GS transitions caused by modifying
the strength of some bonds \cite{phd}.  

With regard to other aspects of 2D quantum magnetism 
like e.g.\ finite temperature properties we recommend among others the Refs.\ 
\cite{barnes91dd,manousakis91dd,lhuillier01sep,lhuillier01oct,chakravarty89,chubukov94,qpt_ri01,mila02}.

The plan of this review is as follows. In section \ref{secArchGitt} we describe
the main geometrical features of the 11 uniform Archimedean lattices and
discuss their mutual relationships. In section \ref{criteria} we discuss
several criteria for semi-classical N\'{e}el like order in quantum
antiferromagnets with a particular focus on the information that
can be extracted from exact diagonalization of finite lattices.
The subsequent section \ref{mag1} is devoted to the analysis of the
magnetic ground-state ordering of the spin-half HAFM on the
Archimedean lattices,
where we consider separately bipartite  (section \ref{bipart}) and frustrated
(sections \ref{canted} and \ref{seckagome})  lattices. The findings for all
these lattices are compared and summarized in section \ref{summary_1}.
Readers uninterested in the detailed discussion of the particular lattices
are referred to this section \ref{summary_1}.
In section \ref{qpt} we consider briefly quantum phase transitions occurring
in the 2D HAFM due to the interplay of competition in the interactions and
strong quantum fluctuations.
In the final section \ref{m_h} we discuss the magnetization
process of the spin-half HAFM on the Archimedean lattices using
the square (section \ref{secSquare}), triangular (section \ref{secTriag})
and \kagome lattice (section \ref{secKag}) as main examples. We further
discuss exact eigenstates that appear for the \kagome and star lattices
in section \ref{secImag} and the relation between the Shastry-Sutherland
model and SrCu$_2$(BO$_3$)$_2$ in section \ref{ExpShaSu}.

\section{Archimedean  Lattices} \label{secArchGitt}
\subsection{Characteristics and geometry} \label{geometry}

In 2D magnetism we are faced with a large number of different lattices
with differing coordination numbers and topologies and therefore we
cannot expect a 
general statement concerning zero-temperature semi-classical N\'eel-like 
LRO in 2D quantum spin systems. 
Nevertheless, we can try to find some systematics concerning the main
geometric features relevant for the magnetic ordering in antiferromagnets.

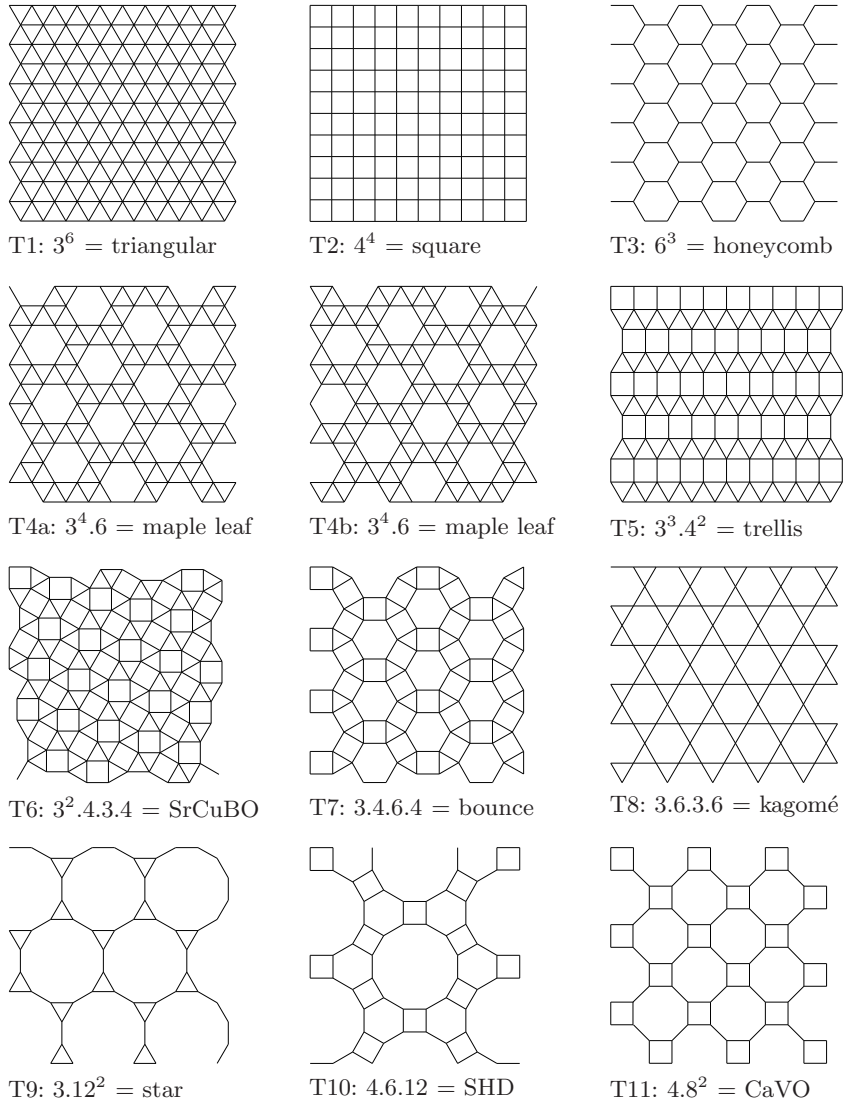
\begin{figure}
\begin{center}
 \setlength{\unitlength}{1mm}
 \begin{picture}(120,38)
 \put( 1,8){{\includegraphics[height=2.9cm]{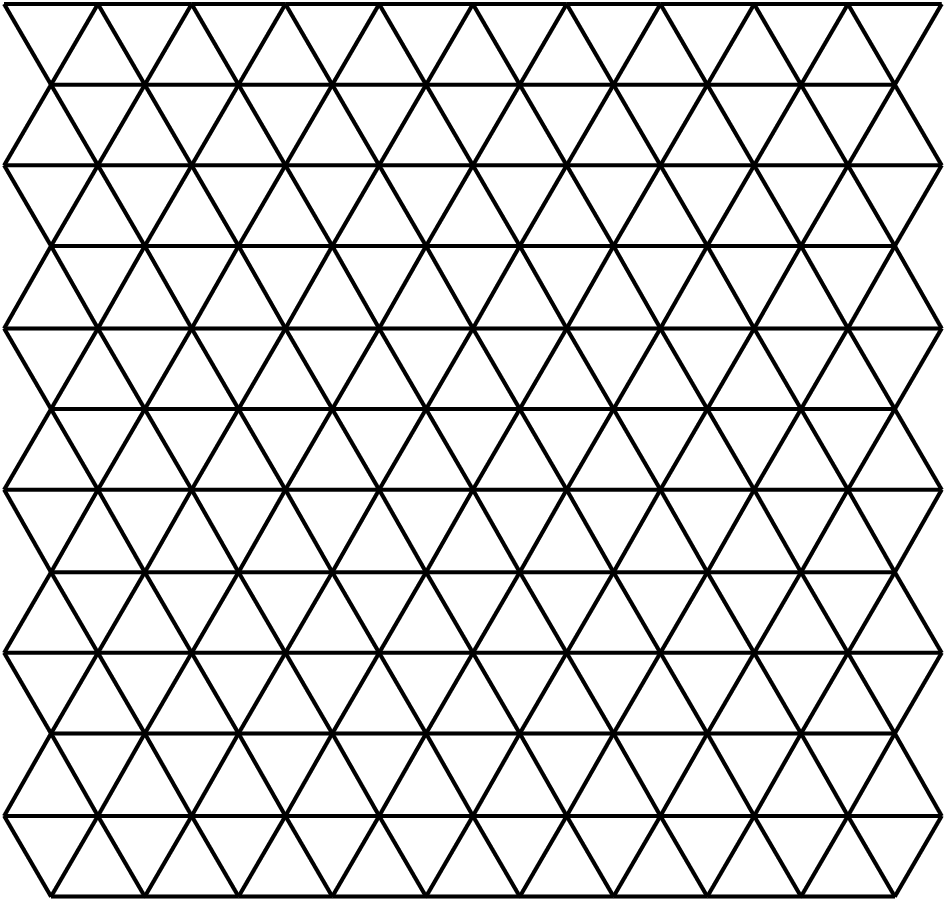}} }
 \put( 1,5){\makebox(0,0)[lc]{T1: 3$^6$ = triangular}}
 \put(41,8){{\includegraphics[height=2.9cm]{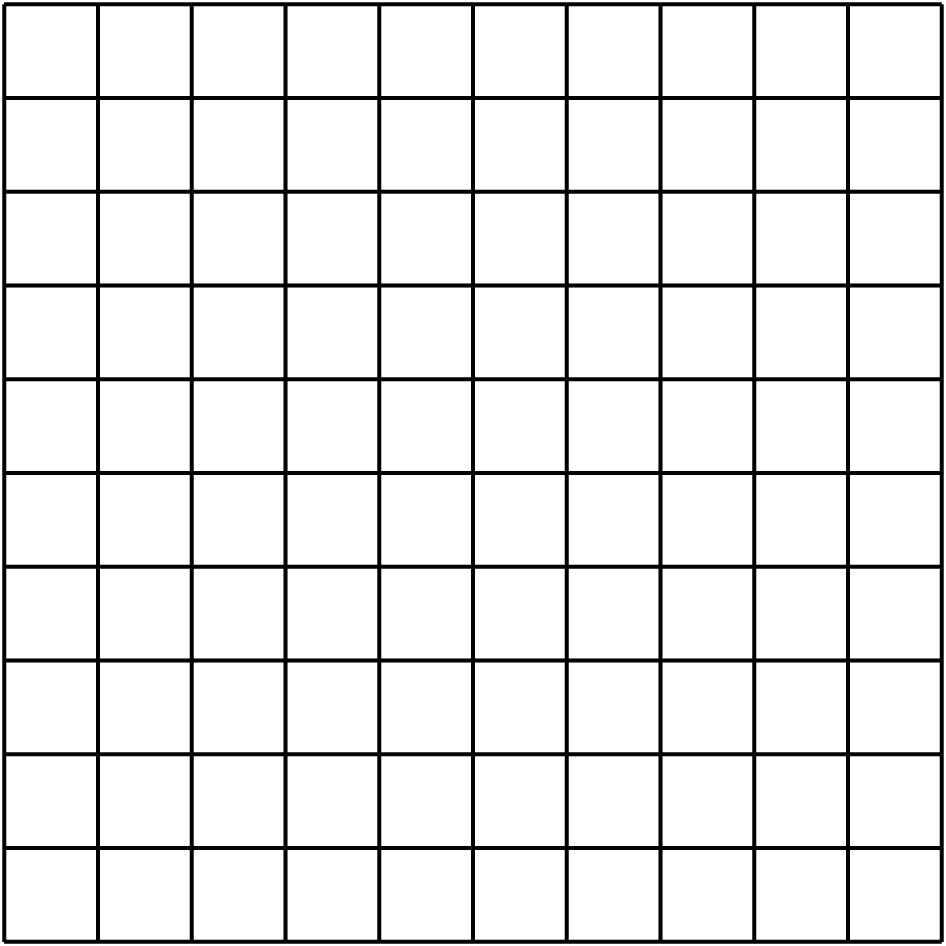}} }
 \put(41,5){\makebox(0,0)[lc]{T2: 4$^4$ = square}}
 \put(81,8){{\includegraphics[height=2.9cm]{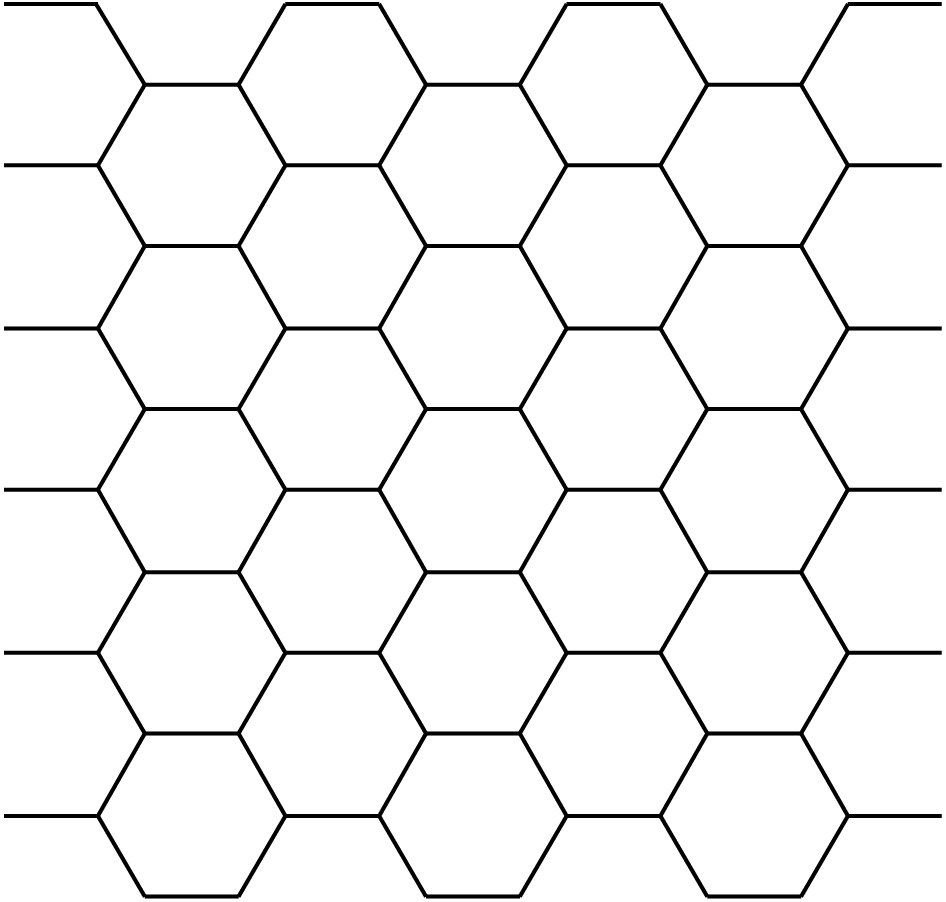}} }
 \put(81,5){\makebox(0,0)[lc]{T3: 6$^3$ = honeycomb}}
 \end{picture}
 \\
 \setlength{\unitlength}{1mm}
 \begin{picture}(120,37)
 \put( 1,8){{\includegraphics[height=2.9cm]{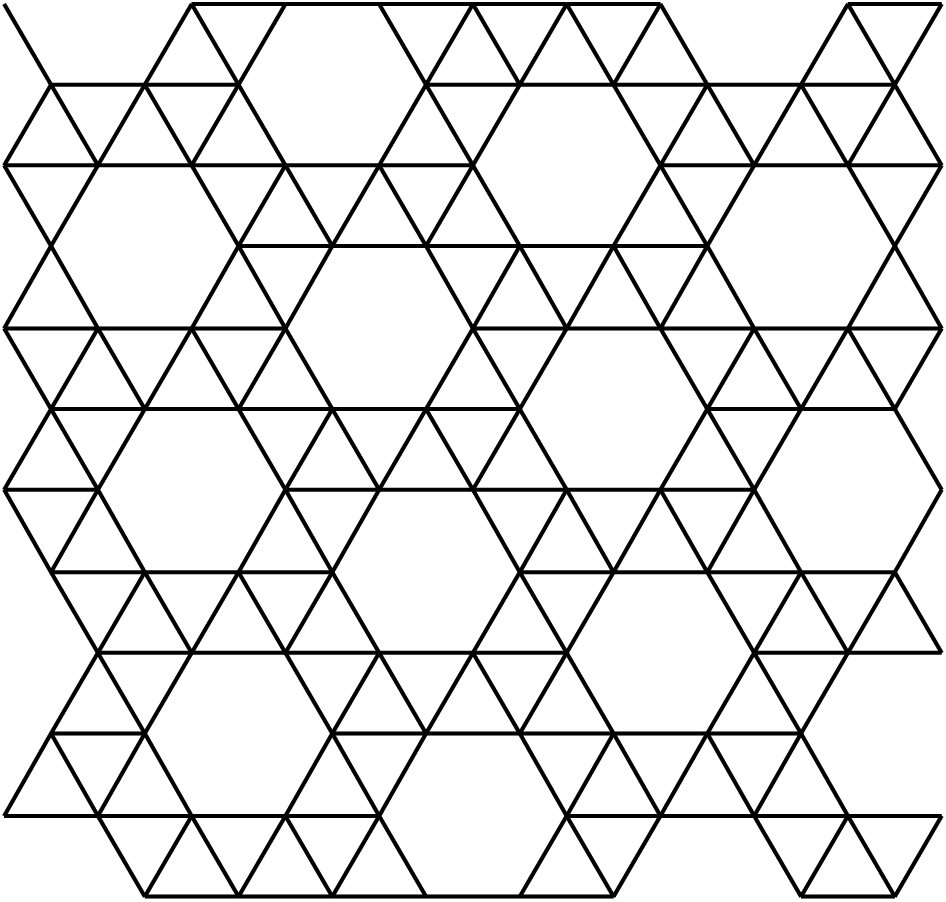}} }
 \put( 1,5){\makebox(0,0)[lc]{T4a: 3$^4$.6 = maple leaf}}
 \put(41,8){{\includegraphics[height=2.9cm]{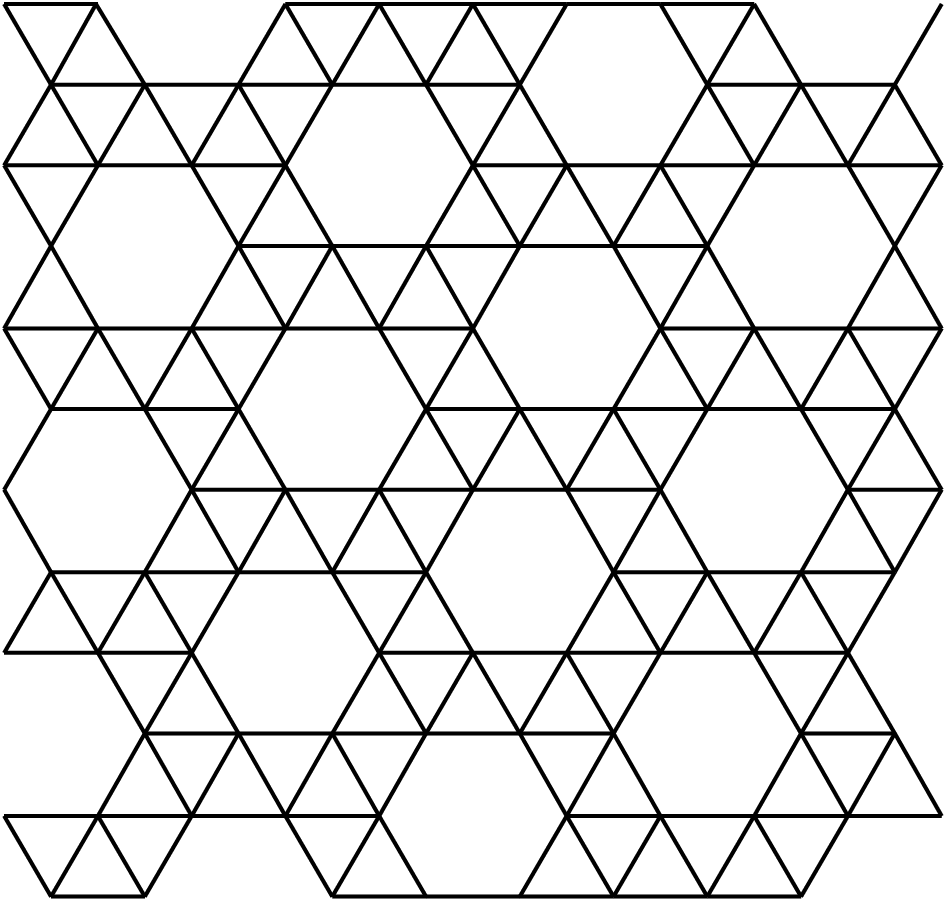}} }
 \put(41,5){\makebox(0,0)[lc]{T4b: 3$^4.6$ = maple leaf}}
 \put(81,8){{\includegraphics[height=2.9cm]{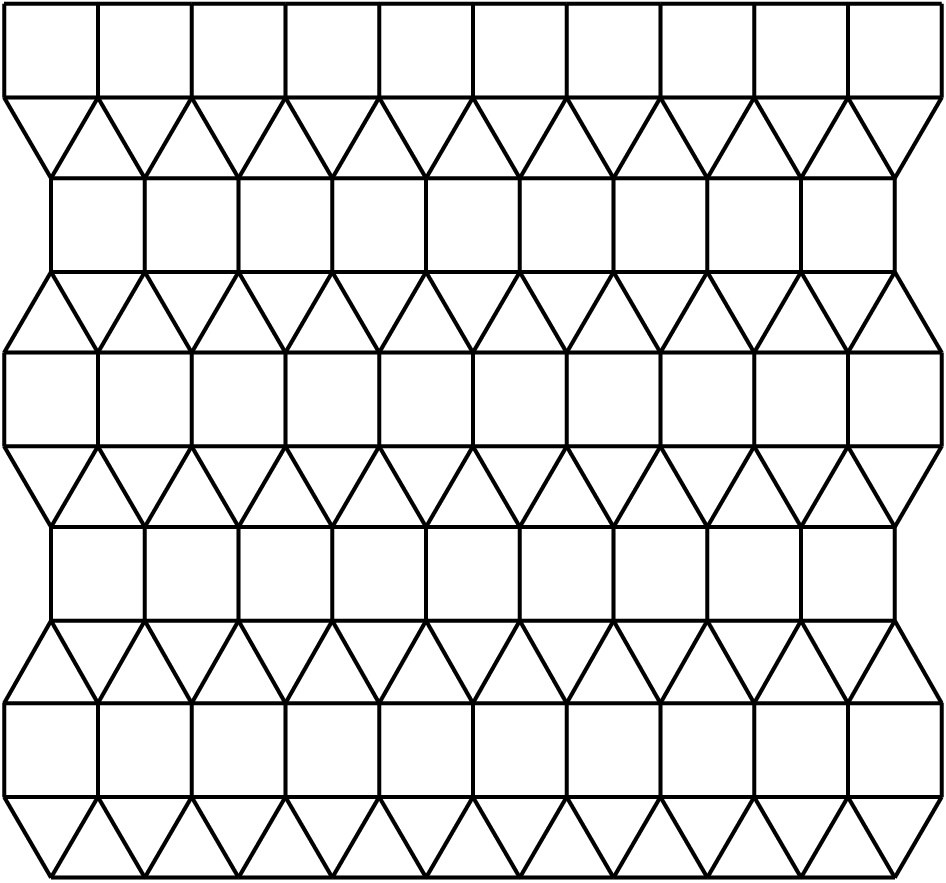}} }
 \put(81,5){\makebox(0,0)[lc]{T5: 3$^3$.4$^2$ = trellis}}
 \end{picture}
 \\
 \setlength{\unitlength}{1mm}
 \begin{picture}(120,37)
 \put( 1,8){{\includegraphics[height=2.9cm]{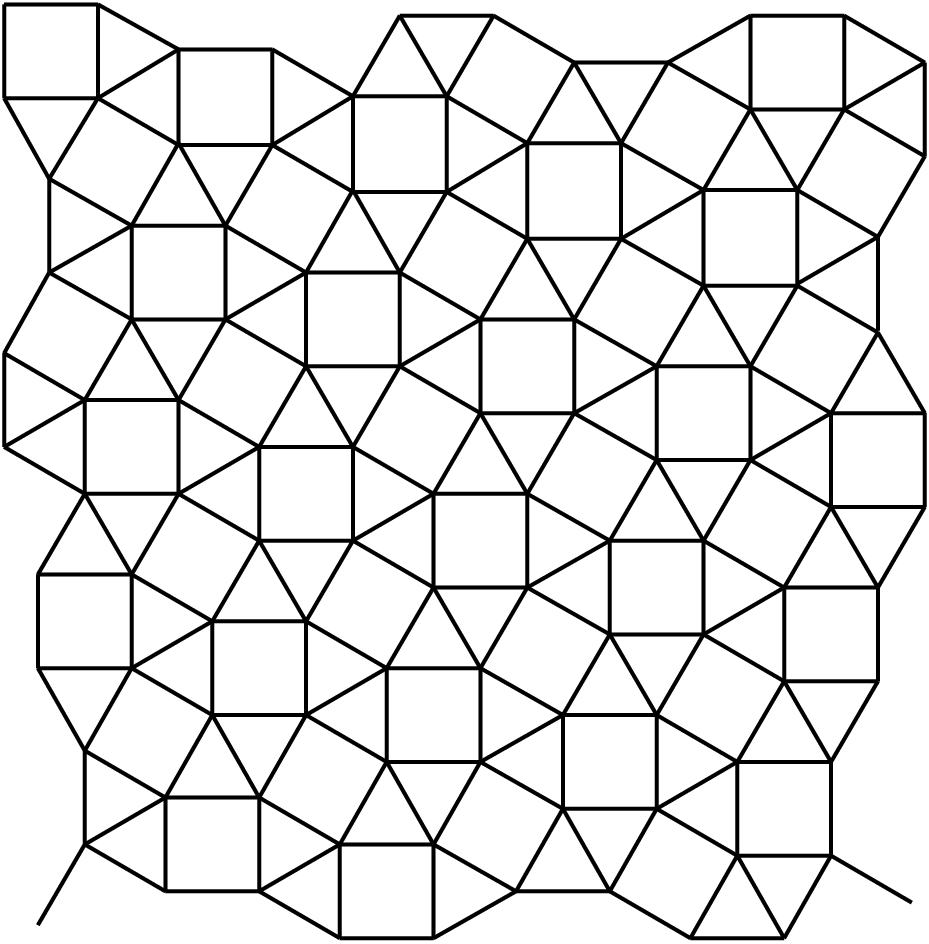}} }
 \put( 1,5){\makebox(0,0)[lc]{T6: 3$^2$.4.3.4 = SrCuBO}}
 \put(41,8){{\includegraphics[height=2.9cm]{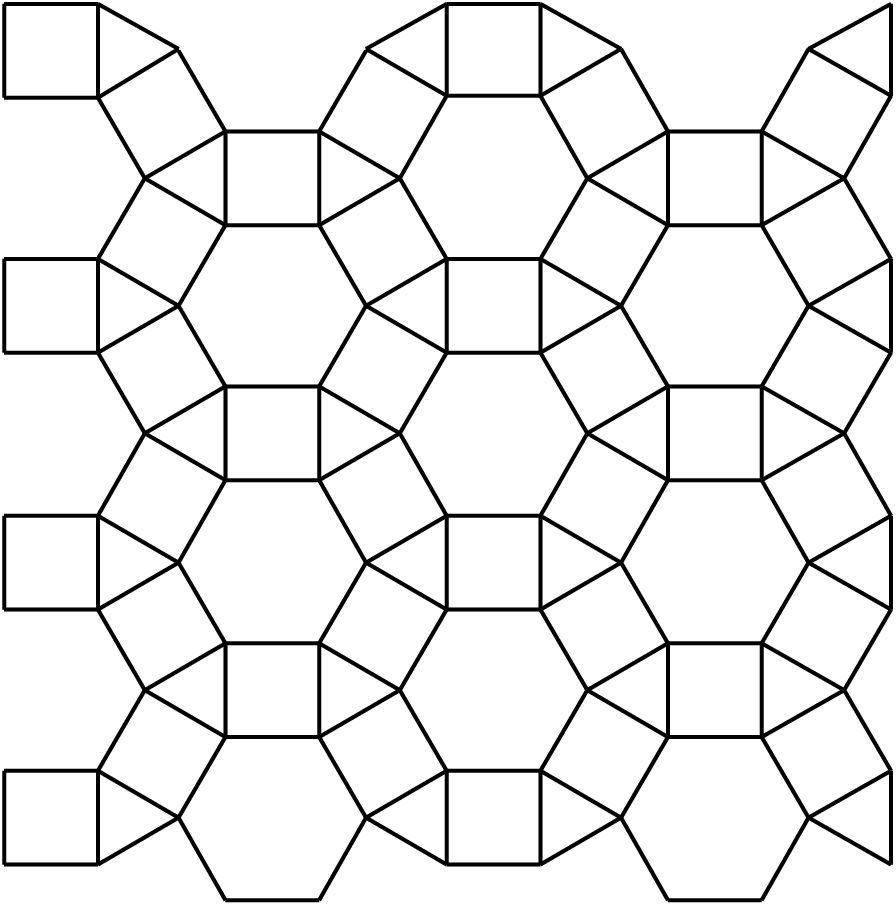}} }
 \put(41,5){\makebox(0,0)[lc]{T7: 3.4.6.4 = bounce}}
 \put(81,8){{\includegraphics[height=2.9cm]{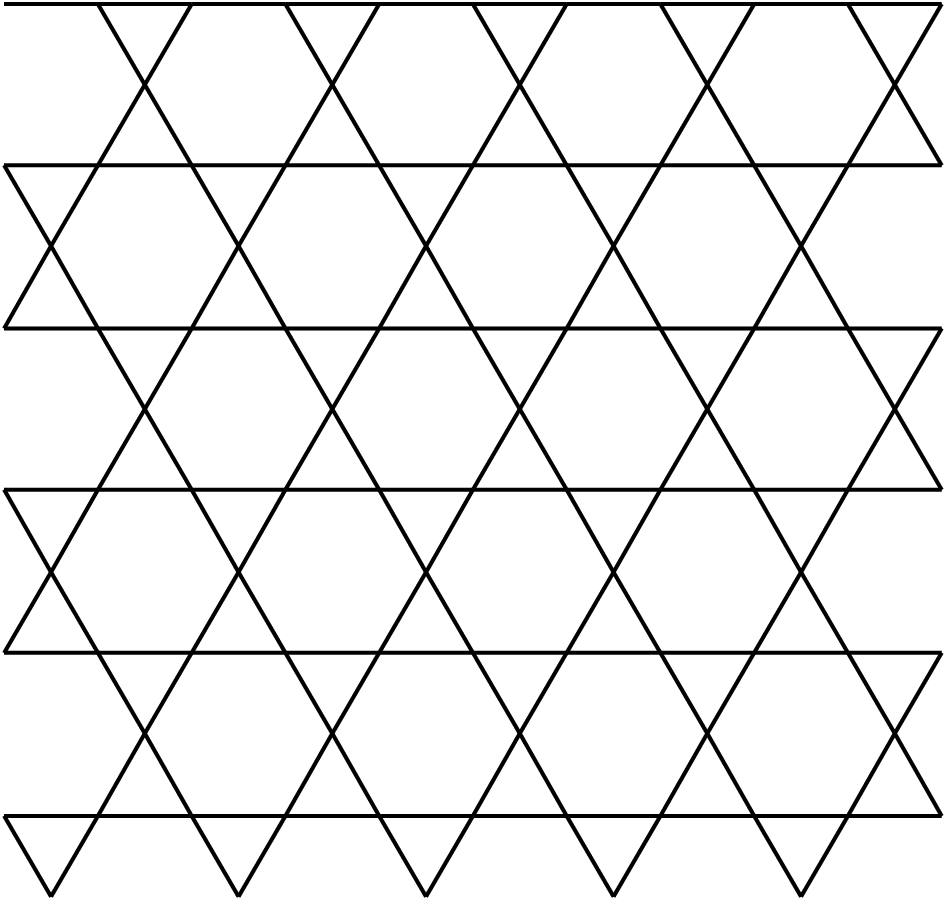}} }
 \put(81,5){\makebox(0,0)[lc]{T8: 3.6.3.6 = \kagome}}
 \end{picture}
 \\
 \setlength{\unitlength}{1mm}
 \begin{picture}(120,37)
 \put( 1, 8){{\includegraphics[height=2.9cm]{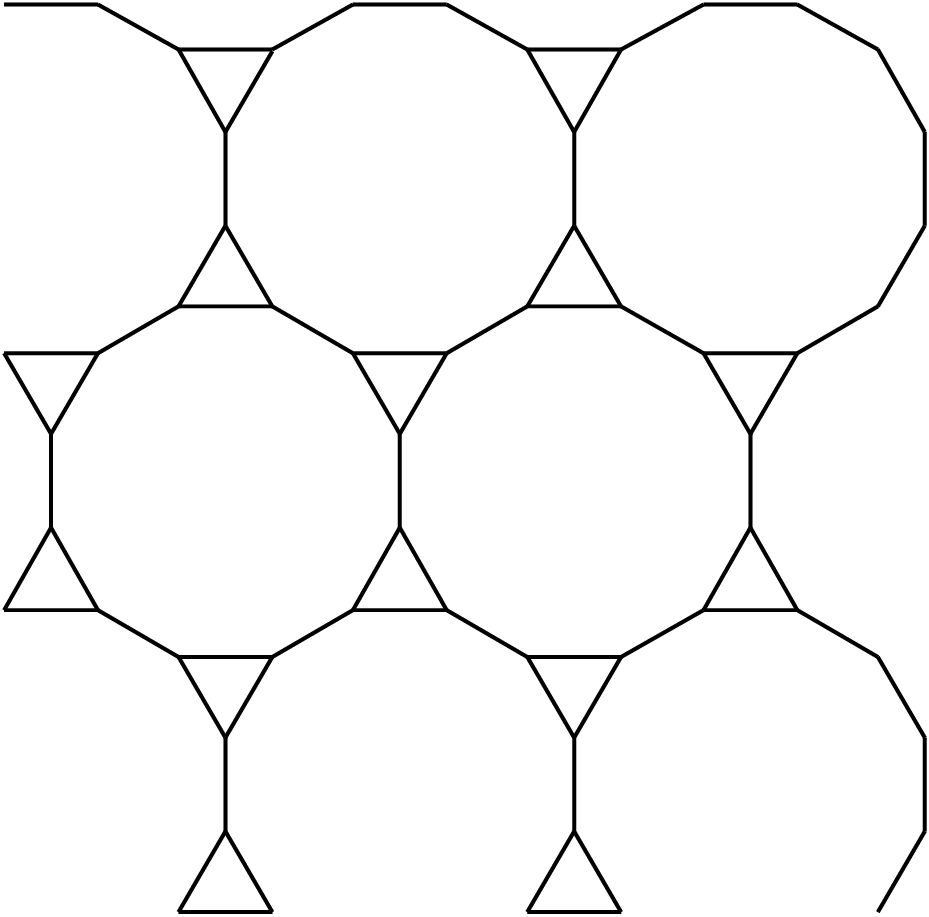}} }
 \put( 1,5){\makebox(0,0)[lc]{T9: 3.12$^2$ = star}}
 \put(41, 8){{\includegraphics[height=2.9cm]{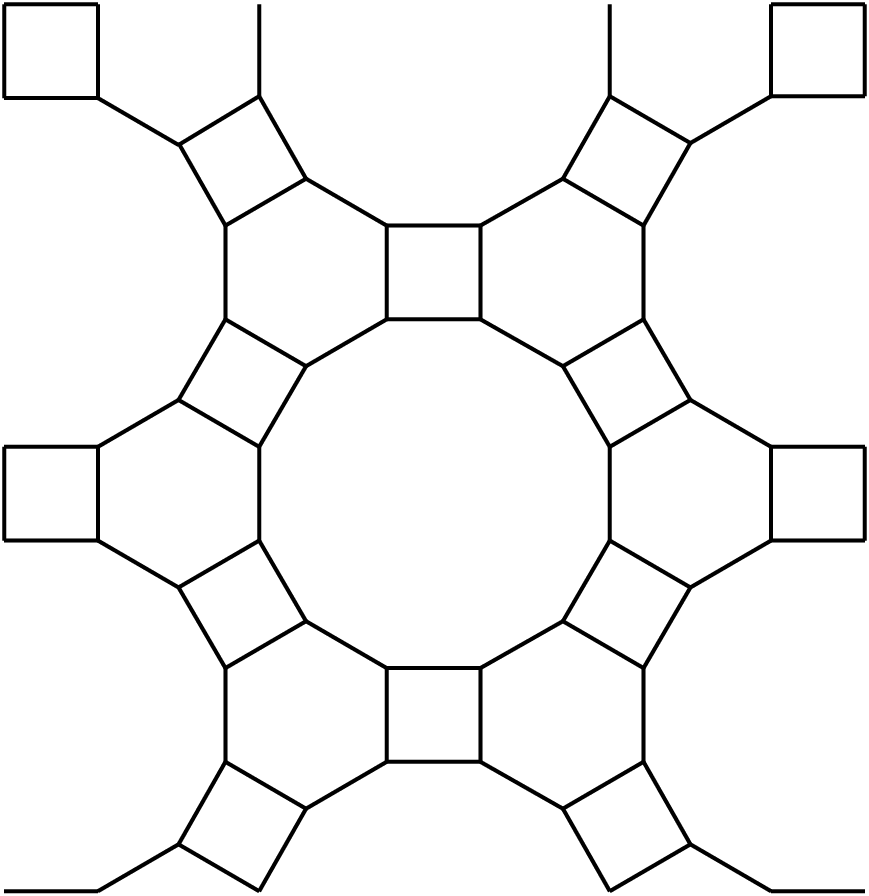}} }
 \put(41,5){\makebox(0,0)[lc]{T10: 4.6.12 = SHD}}
 \put(81, 8){{\includegraphics[height=2.9cm]{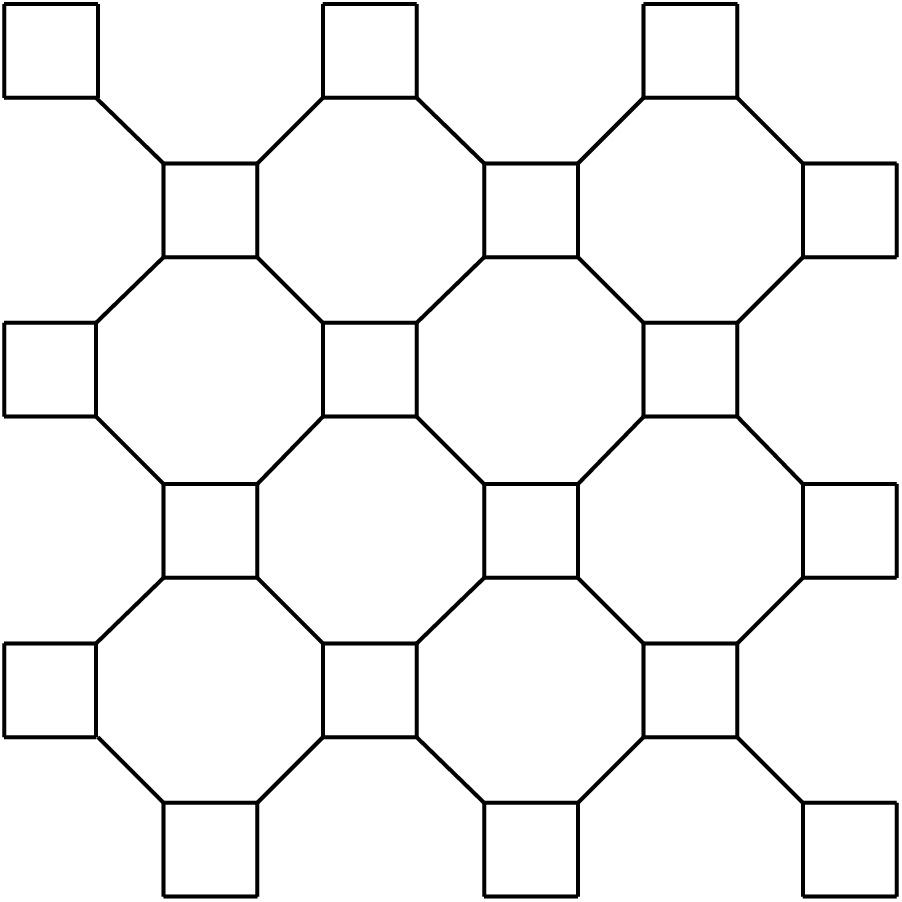}} }
 \put(81,5){\makebox(0,0)[lc]{T11: 4.8$^2$ = CaVO}}
 \end{picture}

\vspace{-0.5cm}
\end{center}
\caption[Archimedische Tilings]{ \label{archTilings}
 The 11 Archimedean tilings T1$\ldots$T11. 
 The  mathematical description $n_1.n_2.n_3\ldots n_r$
 by numbers $n_i$ separated by dots 
 corresponds to the number of vertices of the polygons arranged 
around a vertex.
The tilings T1, T2, T3, T8 are well-known as triangular (T1),
square (T2), honeycomb (T3) and \kagome (T8) lattices. 
For the other lattices
no standardized names are available. For T4, T5, T6, T10 and T11 
we employ the names maple-leaf (T4), trellis (T5), 
SrCuBO or Shastry-Sutherland (T6),
SHD (i.e.\ square-hexagonal-dodecagonal, T10) and CaVO (T11) lattice
previously used in
papers dealing with magnetic properties of these lattices (see also section
\ref{mag1}). We shall denote the tilings T7 and T9 by the names bounce (T7) and
star (T9) lattice, proposed in \cite{suding99}.
}
\end{figure}

The 11 uniform Archimedean tilings (lattices) shown in Fig.\ \ref{archTilings} 
represent the prototypes of 2D arrangements of regular polygons. The first 
investigations of 2D regular tilings go back to Johannes Kepler 
({\it Harmonice Mundi}, 1619).
2D (spin) lattices are obtained from the tilings by putting sites (spins) 
on each vertex connecting neighboring polygons.
The HAFM for these lattices is obtained by assuming antiferromagnetic exchange
bonds $J=1$ on each edge of the polygons.

The Archimedean lattices vary in coordination number $z$ (from 3 to 6)
and in topology (frustrated and nonfrustrated;
equivalent nearest-neighbor (NN) bonds and non-equivalent NN bonds).
Therefore a systematic study of the influence of lattice geometry on
magnetic ordering may be made.

Among them we have three 2D lattices
built by a periodic arrangement of {\bf identical} regular polygons, 
namely the square lattice (T2), the triangular lattice (T1) and the
honeycomb lattice (T3).  
Other uniform  tilings are obtained by combining different regular polygons
such as hexagons and triangles or hexagons, squares and triangles with the
restriction that all lattice sites are equivalent and all polygons have
identical edge length. Under these geometric restrictions precisely 11
uniform Archimedean tilings are possible, where one tiling exists 
in two  enantiomorphic forms  (left and right handed). Only two of them,
namely the square lattice (T2), and the triangular lattice (T1)
are primitive lattices having only one site per geometric unit cell; 
all other ones have at least two sites per \UC.
More information can be found, for example, in Ref.\ \cite{gruenbaum}.

In this section we will illustrate the Archimedean tilings and discuss their
main geometric properties.
As mentioned above, they represent the prototypes of 2D tilings, from
which a large variety of 2D lattices can be derived.
As a result we obtain bipartite, i.e.\ non-frustrated 
(only even polygons, tilings
T2, T3, T10, T11) as well as non-bipartite, i.e. frustrated spin lattices 
(tilings with odd polygons (triangles), i.e.\ T1, T4, T5, T6, T7,
T8, T9). Furthermore, we can differentiate between lattices with
only equivalent NN bonds (T1, T2, T3, T8) and lattices with 
non-equivalent NN bonds (T4,
T5, T6, T7, T9, T10, T11).

\begin{figure}
\begin{center}
\myframe{{\includegraphics[height=5.5cm]{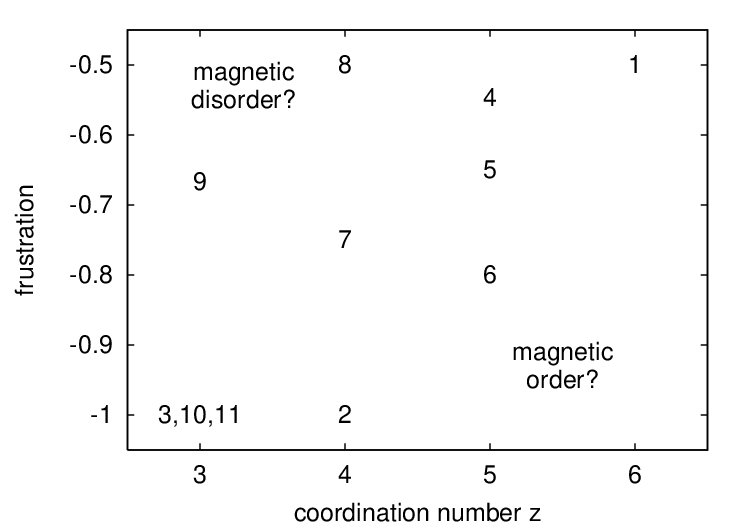}}}
\end{center}
\caption[Parameterraum (Frustration,OZ)]{ \label{fig_ov1}
 Location of the  Archimedean tilings in parameter space spanned by 
 frustration (classical GS energy per bond divided by $s^2$, see text) and 
coordination number $z$.
}
\end{figure}

The degree of geometric frustration and the coordination number are
important quantities that strongly influence the magnetic properties.
In order to give a more precise characteristics of the frustration, we use
an idea proposed by Kobe and coworkers
\cite{kobe95} and consider the GS energy of the classical HAFM
(i.e.\ the spins $\bf S$ are ordinary classical vectors of length $s=1/2$).
Non-frustrated lattices (T2, T3, T10, T11) have minimal energy per bond 
$E_0^{\mathrm{class}}/\mathrm{bond}=-1s^2$. 
Geometric frustration leads to unsatisfied bonds yielding an increase of
classical GS energy.
This increase of energy can be used as a measure of frustration.
The tilings with maximal frustration are 
the triangular lattice (T1) and the \kagome lattice (T8)
having $E_0^{\mathrm{class}}/\mathrm{bond}=-s^2/2$. The combination of strong
frustration and low coordination number $z$ favors strong quantum fluctuations. 
In Fig.\ \ref{fig_ov1} we show the location of the lattices in a parameter space
spanned by the coordination number $z$ and the frustration.
The suppression of classical N\'{e}el-like 
LRO is most likely in the upper left corner in Fig.\ \ref{fig_ov1},
whereas in the opposite region N\'{e}el ordered systems are expected.

\subsection{Relationships between the lattices}
\label{relation}

\begin{figure}
\begin{center}
 \myframe{{\includegraphics[height=3.5cm]{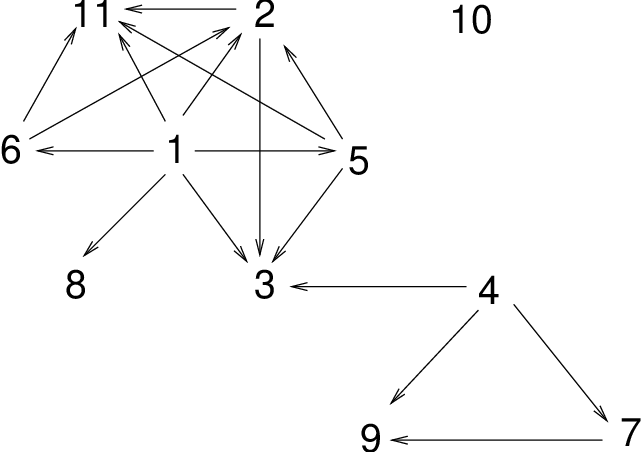}}}
\end{center}
\caption[Tilings-Verwandtschaften]{ \label{fig_relat}
 Relationships (arrows) between the tilings (represented by numbers).
 A related tiling is obtained from an initial tiling by 
 removing  certain edges (bonds) and  a subsequent appropriate
 distortion. 
}
\end{figure}

\begin{figure}
\begin{center}
 \setlength{\unitlength}{1mm}
 \begin{picture}(55,55) 
 \put(0,8){ \myframe{\hspace{-3.3cm}\includegraphics[height=3.5cm]{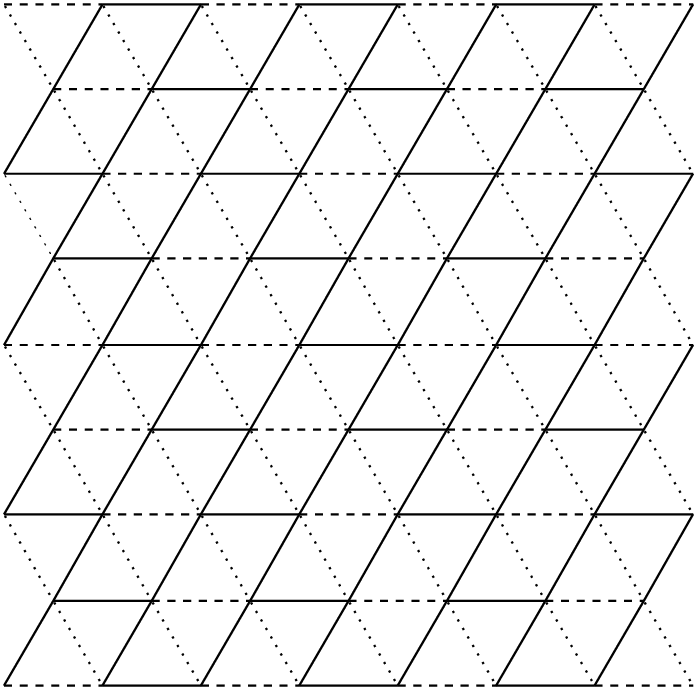}} }
 \put(5,3){\hspace{-3.3cm}\makebox(0,0)[lc]{T1: 3$^6$}}
 \put(0,8){ \myframe{\hspace{+0.75cm}\includegraphics[height=3.5cm]{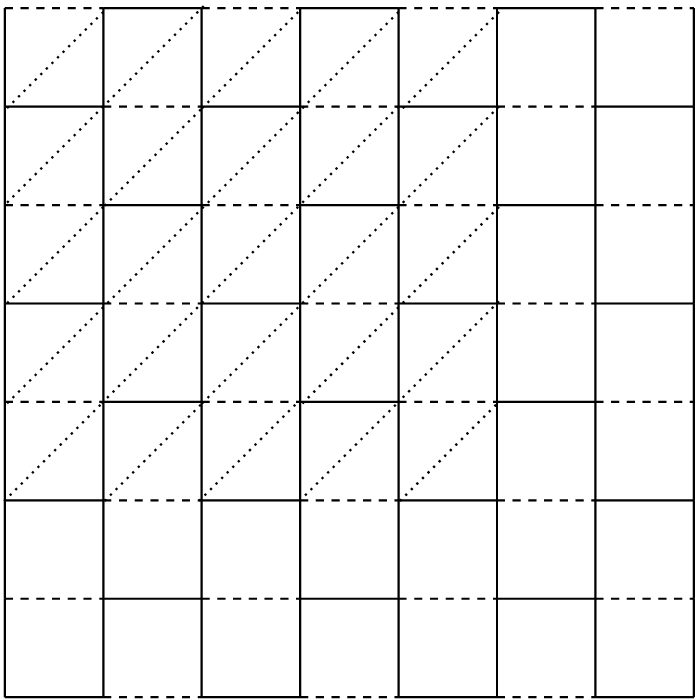}} }
 \put(5,3){\hspace{+0.75cm}\makebox(0,0)[lc]{T2: 4$^4$}}
 \put(0,8){ \myframe{\hspace{+4.8cm}\includegraphics[height=3.5cm]{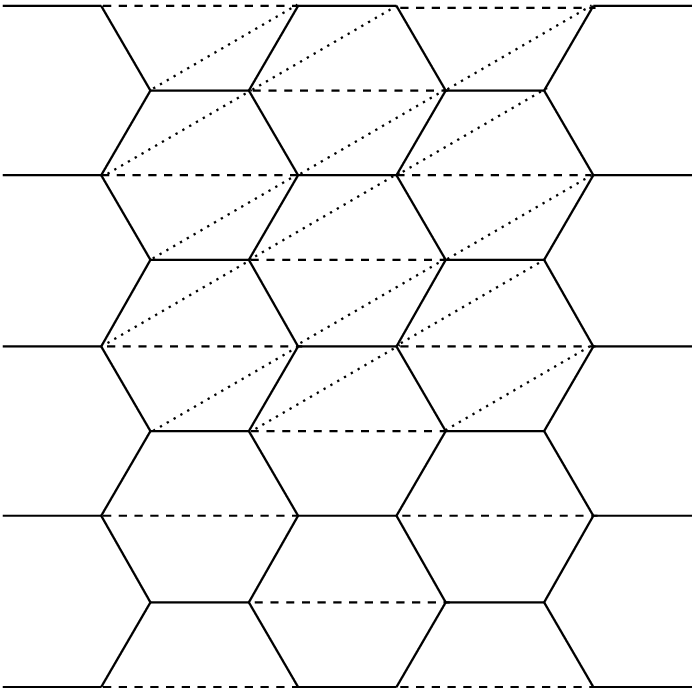}} }
 \put(5,3){\hspace{+4.8cm}\makebox(0,0)[lc]{T3: 6$^3$}}
 \end{picture}
\end{center}
\caption[T1-2-3 relationships]{ \label{t123}
Relationships between triangular (T1, left),
square (T2, middle) and honeycomb lattice (T3, right), see text for details.
}
\end{figure}
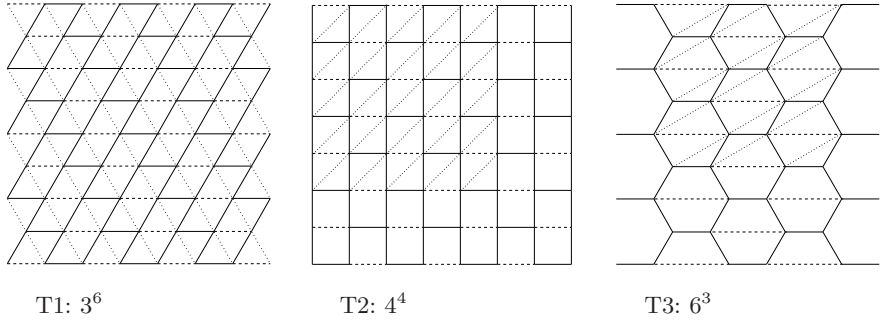

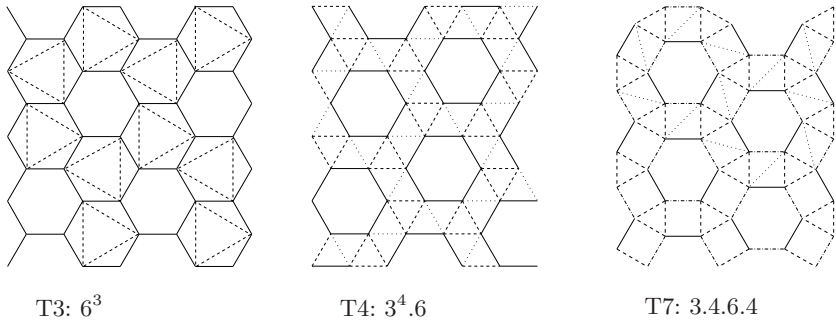
\begin{figure}
\begin{center}
 \setlength{\unitlength}{1mm}
 \begin{picture}(55,50)
 \put(0,8){ \myframe{\hspace{-3.3cm}\includegraphics[height=3.5cm]{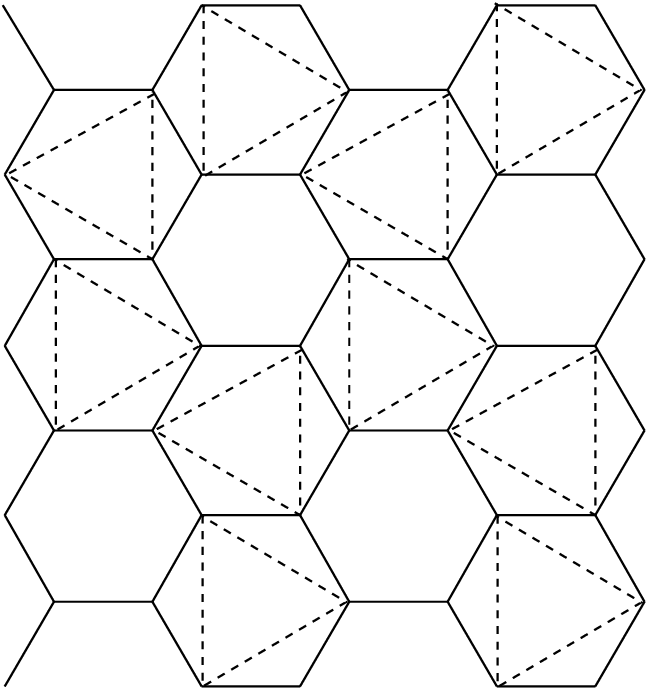}} }
 \put(5,3){\hspace{-3.3cm}\makebox(0,0)[lc]{T3: 6$^3$}}
 \put(0,8){ \myframe{\hspace{+0.75cm}\includegraphics[height=3.5cm]{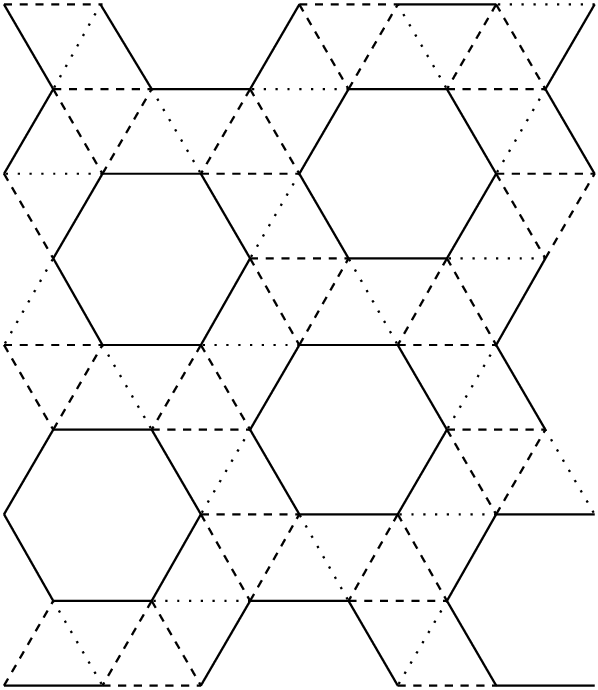}} }
 \put(5,3){\hspace{+0.75cm}\makebox(0,0)[lc]{T4: 3$^4$.6}}
 \put(0,8){ \myframe{\hspace{+4.8cm}\includegraphics[height=3.5cm]{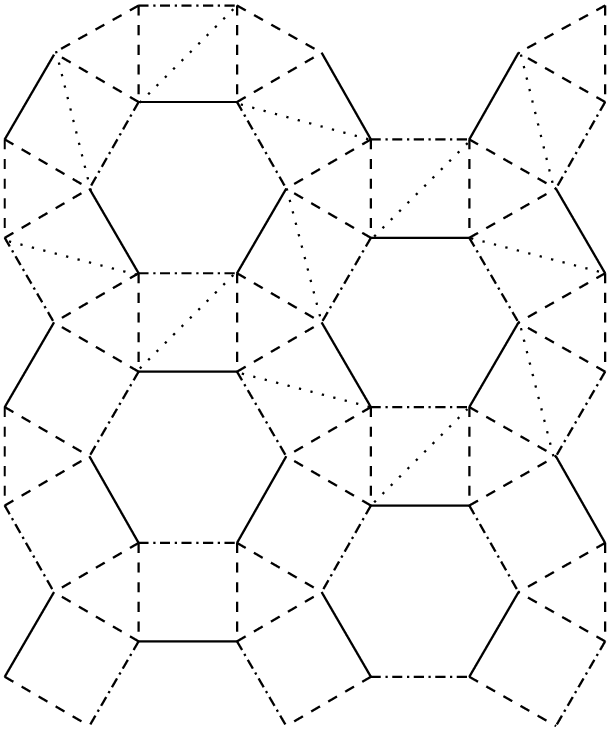}} }
 \put(5,3){\hspace{+4.8cm}\makebox(0,0)[lc]{T7: 3.4.6.4}}
 \end{picture}
\end{center}
\caption[T3-4-7 relationships]{ \label{t347}
Relationships between the honeycomb (left), maple-leaf (middle) and bounce
lattice (right), see text for details.
}
\end{figure}

As mentioned above, we interpret the edges of the polygons as exchange 
bonds which connect the spins sitting on the vertices. In real magnetic
systems often we are faced with the situation
that bonds may vary in strength for instance due to lattice distortions.
Hence it is interesting to consider also bonds varying in strength.
In particular, a given lattice may interpolate into another different
lattice as selected bonds are forced to reach the limit $J'=0$.
The relationships between the Archimedean 
lattices based on removing bonds are summarized in 
Fig.\ \ref{fig_relat}.\footnote{
It is also possible to find transformations between lattices by removing 
sites (site depletion). That is not considered here.}
A continuous change of those bonds from $J'=1$ to $J'=0$ is therefore
accompanied by a transition or a crossover between the ground states
of the related lattices.
We illustrate some of these relationships between 
lattices in Figs.\ \ref{t123} and \ref{t347}.

Fig.\ \ref{t123} shows the relationships between triangular,
 square and honeycomb lattices.
The square lattice is obtained from the triangular lattice by omitting 
the dotted bond $J''$. The geometric distortion of the square lattice
obtained in this manner is irrelevant for the HAFM because the interaction
matrix $J_{ij}$ of the distorted lattice is identical to the regular
lattice.\footnote{Of course, the distorted lattices obtained by removing
bonds may also be transformed to the
regular (non-distorted) lattice by an appropriate shift of the sites.}
The honeycomb lattice is then obtained from the square lattice by omitting
the dashed bonds $J'$ (the model with variable $J'$ is known as the
$J-J'$ model on the square lattice and shows interesting quantum phase
transitions \cite{sing88,krueger00,qpt_ri01}). 

In Fig.\ \ref{t347} the relationships between the honeycomb,
the maple-leaf and the bounce lattice are shown. Starting from the 
maple-leaf lattice one obtains the bounce lattice by 
omitting the dotted bonds
$J_D$. Further removing the dashed-dotted bond in the bounce lattice one 
obtains the star lattice (T9, not shown in Fig.\ \ref{t347}). 
On the other hand the honeycomb lattice is obtained from the  
maple-leaf lattice by removing the dashed bonds $J_T$. Again the distortion
of the lattices is irrelevant for the HAFM.

\section{Criteria for N\'{e}el like order}
\label{criteria}
\subsection{Order parameter}
\label{chapOrder}
The definition of the magnetic order parameter is usually related to the  
classical ground state (GS). Thus supposing that in the classical GS
a spin at site $i$ is directed along the unit vector ${\bf e}_i$, 
we choose the spin orientation ${\bf e}_i$
as local $z$-direction ${\bf e}^{z'}_i={\bf e}_i$, 
which may in general vary from site to site.
In order to break the rotational symmetry we add a field to the Hamiltonian
(\ref{ham})
\be \label{sym_break}
H' = H - h \sum_i {\bf e}^{z'}_i {\bf S}_i.
\ee
We define the order-parameter operator as  
\be \label{op_m_z}
\hat m^z = \frac{1}{N} \sum_i S^{z'}_i = 
\frac{1}{N}\sum_i {\bf e}^{z'}_i{\bf S}_i  .
\ee
Then the order parameter for a GS spontaneously breaking 
the rotational symmetry of $H$ is defined as 
\be \label{ord_para}
m^z= \lim_{h \to 0}\lim_{N \to \infty}
\langle \hat m^z \rangle,
\ee 
where  $\langle \hat O  \rangle $ means the expectation value of the
operator $\hat O$ in the GS.  
This definition of the order parameter corresponds, e.g., to the order
parameter used in spin-wave theory (SWT).
However, symmetry breaking is introduced in this case by the
Holstein-Primakoff transformation, which  starts from a symmetry broken 
classical GS.  

In order to be more specific let us consider a classical spin system having a 
{\bf planar} magnetic GS ordering. We choose 
the $z$-$x$ plane of a fixed global  
coordinate system to describe the order.
Then the relation between the spin 
${\bf S}'_i$ in the local coordinate system 
and the spin ${\bf S}_i$ 
in the global coordinate system 
is given by 
\begin{equation} \label{spintrafo}
 {\bf S}'_i={\hat U}(\phi_i) {\bf S}_i 
    = (\cos(\phi_i)S^x_i -\sin(\phi_i)S^z_i  \; , \;
        S^y_i \; , \;
       \sin(\phi_i)S^x_i +\cos(\phi_i)S^z_i) ,
\end{equation}
where $\phi_i$ is the angle between the local ${\bf e}^{z'}_i$ and 
the global $z$ axis ${\bf e}^{z}$.  
The last component in (\ref{spintrafo}) enters the order-parameter operator
$\hat m^z$ in (\ref{op_m_z}).

The definition of the order-parameter operator (\ref{op_m_z}) 
yields the well-known order parameter  of the ferromagnet
($\phi_i=0$) $ m^z= \lim_{h \to 0; \; N \to \infty}
\frac{1}{N}\langle\sum_i S^{z'}_i \rangle = \lim_{h \to 0; \; N \to \infty}  
\frac{1}{N}\langle\sum_i
S^z_i \rangle$ (magnetization)
as well as
the order parameter  for the conventional 
two-sublattice N\'{e}el antiferromagnet 
($\phi_{i \in A}=0$, $\phi_{i \in B}=\pi$)   $m^z_s=  
\lim_{h \to 0; \; N \to \infty}\frac{1}{N}\langle\sum S^{z'}_i
\rangle = \lim_{h \to 0; \; N \to \infty}
\frac{1}{N}\langle \sum \epsilon_i S^z_i 
\rangle$ (staggered 
magnetization), where 
the staggered factor $\epsilon_i$ is $\epsilon_i=+1$ ($\epsilon_i=-1$) for 
sites belonging to sublattice $A$($B$). The staggered magnetization can 
be expressed by the sublattice magnetizations
$S^z_A=\sum_{i \in A} S^z_i$ and $S^z_B=\sum_{i \in B} S^z_i$, we have
$m^z_s = \frac{1}{N}\langle S^z_A - S^z_B
\rangle_{h \to 0, N \to \infty}$.    
The general definition (\ref{ord_para})
is also applicable for non-collinear (canted) spin structures appearing on
frustrated lattices.
For example, the classical GS of the 
HAFM on the triangular lattice consists of three sublattices $A$, $B$, $C$
 with an angle of 120$^\circ$ between the sublattice spins, i.e. we have 
$\phi_{i \in A}=0$, $\phi_{i \in B}=2\pi/3$ and $\phi_{i \in C}=4\pi/3$. 
Consequently we find 
$m^z =  \frac{1}{N}\langle 
S^z_A + \sqrt{3}S^x_B/2 - S^z_B/2 - \sqrt{3}S^x_C/2 - S^z_C/2
\rangle_{h \to 0, N \to \infty}$.
The extension to arbitrary non-collinear spin structures is 
straightforward.

The situation is changed 
for the HAFM on finite lattices considered in numerical studies because the
GS of a finite system cannot possess the spontaneous 
symmetry breaking used for the infinite lattice (eqs. (\ref{sym_break}) - 
(\ref{ord_para})).  Therefore the square of the order-parameter operator 
$(\hat m^z)^2$ has to be used.
Furthermore, we have to take into account the fact that the 
GS of finite antiferromagnetic systems with even number of sites
$N$ is a rotationally invariant singlet state.\footnote{ 
Note that although there is a rigorous proof for the singlet character of
the GS of finite systems only 
for the HAFM on bipartite lattices \cite{marsh55,lieb62}, much 
numerical evidence suggests that the same statement is true for nonbipartite
frustrated antiferromagnetic Heisenberg systems (see e.g.\ \cite{ri_jmmm95}).}
Then the magnetic correlations are equally distributed over all three
components 
$\langle S^x_i S_j^x\rangle = \langle S^y_i S_j^y \rangle=
\langle S^z_i S_j^z\rangle$. Thus,
taking into account this symmetry, one defines the relevant order
parameter for finite systems as 
\be \label{ord_square}
\bar m = \sqrt{\langle \big ( \frac{1}{N} \sum_i {\bf S}_i'\big)^2\rangle}=
\sqrt{3\langle ({\hat m}^z)^2 \rangle } \, . 
\ee
One may write this order parameter as
\bea \label{m_s}
\bar m &=&
\sqrt{\langle\big(\frac{1}{N}\sum_i \epsilon_i{\bf S}_i\big)^2\rangle} =
\sqrt{\frac{1}{N^2}\sum_{i,j} \epsilon_i\epsilon_j\langle {\bf S}_i
{\bf S}_j\rangle}\\ 
&=& \sqrt{\frac{3}{N^2}\langle (S^z_A)^2+(S^z_B)^2-2S^z_AS^z_B \rangle} \nn
\eea
for bipartite antiferromagnets and 
\begin{equation}
\bar m =\sqrt{\frac{3}{N^2}\langle (S^z_A)^2+(S^z_B)^2+(S^z_C)^2
             -S^z_AS^z_B-S^z_AS^z_C-S^z_BS^z_C \rangle}
\end{equation}
for the triangular lattice. Note that
$\langle S^x_iS^z_j\rangle=0$ was used
in the last equation. Obviously the analysis  of 
magnetic order is then based on the 
spin pair correlation function 
$\langle {\bf S}_i{\bf S}_j \rangle$. We notice
that an alternative definition to (\ref{ord_para})  of the order parameter
for infinite systems uses the asymptotic large-distance behavior of the 
spin pair correlation function.  

The order parameter $\bar m$ 
is widely used for finite lattices such as square or triangular
lattices. Finite-size extrapolations of $\bar m$  yield good agreement
with $m^z$ defined in (\ref{ord_para}) calculated e.g.\ by \SWT\ or
the coupled cluster method (see e.g.\
\cite{bernu92,betts98,betts99a}).

However, the definition of the order parameter given above
is to some extent problematic for the following reasons:
(i) The definition is biased because it supposes the same type of ordering
in the quantum system as in the classical system. Investigations of spin
systems with spiral order demonstrate,  
that the characteristic angles $\phi_i$ entering  eq. (\ref{spintrafo}) may
be different in the classical and quantum case \cite{krueger00}.
(ii) There are systems with a huge non-trivial degeneracy of the classical
GS (e.g.\ the HAFM on T8 (\Kagome) and on T9 (star), see section
\ref{seckagome}). The question now arises: which of the large number of
degenerate classical ground states should be used?
(iii) A significant problem is also posed if the classical GS is not known.

We therefore use a universal definition of the order parameter, given by
\begin{equation} \label{mdef}
  m^+=\sqrt{\frac{(M^+)^2}{N^2}}
     =\left(\frac{1}{N^2}
     \sum_{i,j}^{N}  {}|\langle{\bf S}_{i}{\bf S}_{j}\rangle|\right)^{1/2} \, ,
\end{equation}
which avoids the problems listed above.
For bipartite systems this definition is identical to the
staggered magnetization $\bar m$ defined in (\ref{m_s}).
%
%
For spin systems with noncollinear GSs both definitions $\bar m$
and $m^+$ are not identical, although there is a relation between
them. For instance, we have in the classical limit
$(m^+_{\mathrm{class}})^2=\frac{4}{3} ({\bar m}_{\mathrm{class}})^2$
for the HAFM on the triangular lattice. Note
that this relation remains valid also for the
singlet GS of the quantum spin-1/2 HAFM on the triangular lattice.

Finally, we mention that the universality of the definition (\ref{mdef})
of the order parameter may lead to a certain loss of distinction between 
different types of ordering and for the detection of the type of order an
additional inspection of the spin-spin correlation function is necessary.   
 
\subsection{Mechanism of symmetry breaking - the Pisa tower of
quasi-degenerate joint states (QDJS)} 
\label{pisa}

As pointed out already by P.W.~Anderson \cite{anderson52} the spontaneous
symmetry breaking in semi-classically \Neel ordered antiferromagnets at the
thermodynamic limit is revealed in the spectrum of a finite system. This
idea has been picked-up in several papers 
\cite{bernu92,neuberger89,bernu93,bernu94,lecheminant97,waldtmann98dd,ri_acta94,sierp01,fouet01,tomczak01,hasenfratz90,hasenfratz93}
dealing with two-dimensional quantum antiferromagnetism.

In the limit $N \to \infty$ a whole set of non-rotationally invariant 
excited states collapses onto the true GS (e.g.\ the 
semi-classical two-sublattice \Neel state for the HAFM on a bipartite lattice).     
Therefore a large amount of information on possible  N\'{e}el like LRO
is contained in the spectrum of  HAFM on finite lattices.
There are extensive systematic studies for HAFM on the square, triangular
and \kagome lattice 
\cite{bernu92,bernu93,bernu94,lecheminant97,waldtmann98dd,fouet01}
and some recent reviews 
by Lhuillier, Sindzingre, Fouet and  Misguich
\cite{lhuillier00dec,lhuillier01sep,lhuillier01oct,lhuillier02dec,lhuillier03}.
We follow the lines of their studies and illustrate some main features using
the HAFM on the square lattice as an example.

\begin{figure}
\vspace{1.cm}
\begin{center}
 \setlength{\unitlength}{1mm}
 \begin{picture}(55,70)
 \put(0,8){
\myframe{\hspace{-3.9cm}\includegraphics[height=7.5cm]{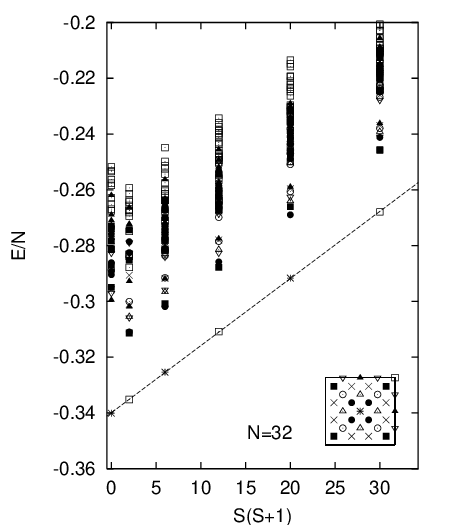}} }
 \put(5,3){\hspace{-2.cm}\makebox(0,0)[lc]{a}}
 \put(-20,70){ \myframe{\includegraphics[height=0.8cm]{pict1.eps}}}
 \put(2,8){ \myframe{\hspace{+2.2cm}\includegraphics[height=7.5cm]{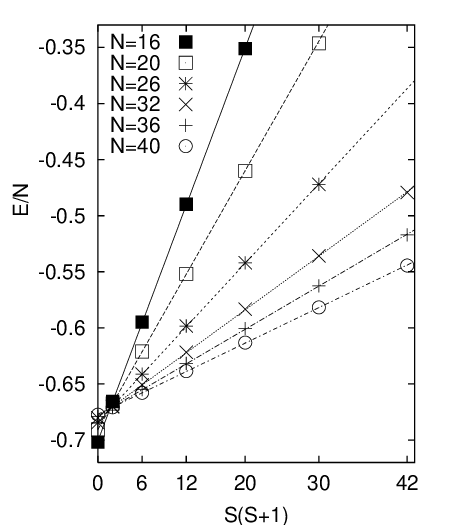}} }
 \put(5,3){\hspace{+7.2cm}\makebox(0,0)[lc]{b}}
 \end{picture}
\end{center}
\caption[sq32 Pisa]{ \label{sqpisa}
HAFM on the square lattice\\
a: Low-energy spectrum for  $N=32$ 
(the inset shows the {\bf k} points in the Brillouin zone).
\hspace{0.3cm}
b: Energy of the QDJS versus $S(S+1)$ for various system sizes $N$.
}
\end{figure}

\begin{figure}

\vspace{1cm}
\begin{center}
 \setlength{\unitlength}{1mm}
 \begin{picture}(55,70)
 \put(0,8){ \myframe{\hspace{-3.9cm}\includegraphics[height=7.5cm]{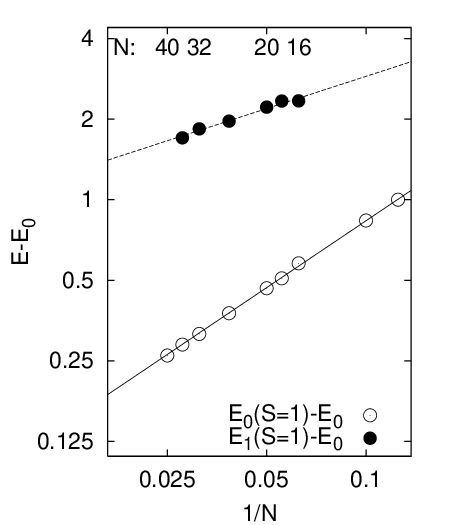}} }
 \put(5,3){\hspace{-2.cm}\makebox(0,0)[lc]{a}}
 \put(2,8){ \myframe{\hspace{+2.2cm}\includegraphics[height=7.5cm]{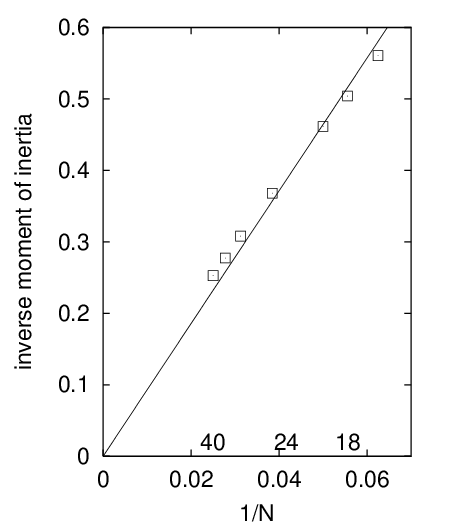}} }
 \put(5,3){\hspace{+7.2cm}\makebox(0,0)[lc]{b}}
 \end{picture}
\end{center}
\caption[sq scal]{ \label{sqgap2}
HAFM on the square lattice\\
a: Finite-size scaling of lowest excitations (double logarithmic scale). 
\hspace{0.3cm}
b: Finite-size scaling of inverse moment of inertia.
}
\end{figure}

For a system with two-sublattice \Neel LRO in the GS
the low-energy part of the spectrum up to $S \approx \sqrt{N}$
is roughly described by the dynamics of
a quantum top, i.e.\ the effective low-energy Hamiltonian reads
\begin{equation}\label{eqtos}
H_{\mathrm{eff}} \simeq E_0 + \frac{{\bf S}^2}{2N\chi_0} + H_{\mathrm{magnons}}
\; \to \; 
 E_{\min}(S) \simeq E_0 + \frac{S(S+1)}{2N\chi_0} + E_{\mathrm{magnons}}
\end{equation}
with $E_0$ as the GS energy, $\chi_0$ as
uniform susceptibility and ${\bf S}^2$ as the square of total spin, 
cf.\ Fig.\ \ref{sqpisa}.
The inverse `moment of inertia' $\frac{1}{N\chi_0}$ 
vanishes in the thermodynamic limit (see Fig.\ \ref{sqgap2}b) 
and therefore the so-called
quasi-degenerate joint states (QDJS) described by (\ref{eqtos}) 
collapse to the symmetry
broken \Neel state in the thermodynamic limit. 
In case of more complex \Neel order e.g. with three sublattices as for the
HAFM on the triangular lattice the basic features of the 
low-energy Hamiltonian as given in eq.\ (\ref{eqtos}) are maintained but the
moment of inertia then contains both in-plane and out-of-plane
susceptibilities.
Also the number of the QDJS for a given total spin $S$  
depends on the number of sublattices in the N\'{e}el state. 
There is only one
QDJS in each sector of $S$ for the two-sublattice HAFM, but e.g. 
$N_S=\min(2S+1,N/2-S+1)$ QDJS for a three-sublattice N\'{e}el
state such as in the triangular lattice \cite{bernu94}.
Furthermore, the translational symmetry of the QDJS depends on the relation
between the translational symmetry of the \Neel state and of 
the lattice.      
For instance,  the size of the magnetic unit cell 
for the N\'{e}el ordered  square-lattice (triangular-lattice) HAFM 
is twice (three times) as large as the geometric \UC.
Consequently, the QDJS belong to
${\bf k}$-vectors ${\bf Q}_1=(0,0)$ and ${\bf Q}_2=(\pi,\pi)$ 
(${\bf Q}_1=(0,0)$ and  ${\bf Q}^+_2=(+4\pi/3,0)$, 
${\bf Q}^-_2=(-4\pi/3,0)$) of lattice translational symmetry
with ${\bf Q}_2$ (${\bf Q}^+_2$ and ${\bf Q}^-_2$)
mapping on the center of the magnetic Brillouin zone. 
However, only ${\bf Q}=(0,0)$ appears for the QDJS
for the honeycomb lattice which has two atoms in the 
geometric \UC\ as well as in the magnetic \UC. 

Indeed, a linear relation between the lowest eigenvalues $E_{\min}(S)$
and $S(S+1)$ and a similar relation for the family of magnon
excitations has been observed for the HAFM on the 2D
square lattice \cite{neuberger89}, the honeycomb lattice \cite{fouet01} 
as well as for the
triangular-lattice HAFM \cite{bernu94}. We show in Fig.~\ref{sqpisa}a the
low-energy spectrum for the   HAFM on the square lattice with $N=32$ sites
in more detail. 

According to (\ref{eqtos}) the collapsing QDJS follow in 
the energy-$S(S+1)$ diagram for
small $S \lesssim \sqrt{N}$ in good approximation  a straight line with 
increasing  inclination (with decreasing slope), see Fig.\ \ref{sqpisa}b,
and are sometimes called the Pisa tower of states. 
The strong deviation from this
linear relation has been used as one argument for the absence of
semi-classical LRO for the HAFM on the 
\kagome lattice \cite{lecheminant97}. A
similar argumentation has been used in \cite{ri_acta94} for the
 $J_1-J_2$ square-lattice HAFM and  in \cite{sierp01} for the 
HAFM on the fractal Sierpi{\'n}ski gasket.

Well separated above the family of QDJS a
second family of levels exists describing the magnon excitations typical
for a HAFM with \Neel ordering. This family represents the  
`softest magnons', i.e.\ magnons of
energy $E^M_{\min}=c|{\bf k}_{\min}|$ with $c$ as lowest spin-wave velocity 
and $|{\bf k}_{\min}|\sim\frac{1}{L}$ ($L=\sqrt{N}$)
as smallest finite wave vector (related to the wave vector ${\bf Q}$
of the corresponding QDJS) allowed by the periodic boundary conditions of the
finite lattice. The energy of these magnons also collapses, however, with
\begin{equation} \label{e_sw}  
  E^M_{\min} \simeq \frac{c}{\sqrt{N}}
\end{equation}
much slower than 
 $E_{\min}(S)\propto \frac{1}{N}$ from eq. (\ref{eqtos}).
This scaling behavior of the QDJS and of the softest magnons 
is shown in  Fig.\ \ref{sqgap2}a, where
the logarithmic scale in  this figure makes it obvious that the slope
of the $E_0(S=1)-E_0$-curve belonging to the QDJS is about twice as large as  
the slope of the $E_1(S=1)-E_0$-curve belonging to the softest magnons.

Finally we want to emphasize two special aspects of the spectrum of
semi-classically \Neel ordered HAFMs. The first one is the so-called 
spin gap, i.e.\ the gap $\Delta= E_{0}(S=1)-E_0(S=0)$ between the first 
triplet excitation
and the singlet GS. According to eqs.\ (\ref{eqtos}) and (\ref{e_sw})
this gap vanishes in the thermodynamic
limit for a \Neel ordered GS. Note that eq.\ (\ref{e_sw}) is
related to the existence of gapless Goldstone modes.
However, a non-vanishing spin gap for $N \to \infty$ is an 
indication for a quantum paramagnetic GS.  
The second aspect is that the singlet GS is unique and  
the lowest singlet excitation above the GS is well separated from it
in a finite system (see Fig.\ \ref{sqpisa}a).
The first triplet excitation above the rotationally invariant singlet GS
generally is the lowest excitation at all.
Therefore the existence of low-lying singlets deep within the spin
gap also can be understood as an indication for a non-\Neel 
ordered GS.

\subsection{Finite-size scaling}
\label{scaling}

Effective continuum field-theory studies as well as \SWT\
and quantum Monte Carlo (QMC) calculations have led to detailed
predictions for the low-energy physics and the finite-size scaling 
of \Neel ordered quantum antiferromagnets
in two dimensions
\cite{neuberger89,hasenfratz90,hasenfratz93,oitmaa94,sandvik97}, which we
will use below in eqs.\ (\ref{gs_scale}), (\ref{m_scale}) and  
(\ref{gap_scale}).
As already discussed in the last section the inverse `moment of inertia'
is obtained from the QDJS and so the spin gap scales in leading order
with $1/N$, see eq.\ (\ref{eqtos}). However, for 
finite-lattice sizes accessible in exact diagonalization the asymptotic
behavior is often not reached and boundary effects are important. Hence,  
the extrapolation to $N \to \infty$ possesses
some uncertainty. According to Refs.\ 
\cite{neuberger89,hasenfratz90,hasenfratz93,oitmaa94,sandvik97}
the GS energy per site $e_0=\frac{E_0}{N}$  
for a semi-classical \Neel state  scales as
\be \label{gs_scale}  
e_0(L)=  A_0 + \frac{A_3 }{L^{3}} + {\cal O}(L^{-4}) 
\ee
where $L=N^{1/2}$ is the linear size of the lattice, 
$A_0=e_0(\infty)$ and $A_3$ is proportional to the spin-wave
velocity $c$.
For the order parameter $m^+$
we use
\be \label{m_scale}  
m^+(L) = B_0 + \frac{B_1 }{L} +  {\cal O}(L^{-2})
\ee
where $B_0=m^+(\infty)$.
For the spin gap $\Delta$ we apply
\be \label{gap_scale}  
\Delta(L) = G_0 +  \frac{G_2 }{L^2}  + {\cal O}(L^{-3}) 
\ee
where $G_0=\Delta(\infty)$.
In case that there are many appropriate  finite lattices with $N\le 36$ 
the large number of data points leads to reliable extrapolation 
to the thermodynamic limit 
\cite{betts98,betts99a}. On the other hand, for systems 
with only a few appropriate finite lattices the extrapolation is much
stronger influenced by the boundary effects and the extrapolated results
exhibit a larger uncertainty. Furthermore, only the leading terms in
(\ref{gs_scale}), (\ref{m_scale}), (\ref{gap_scale}) can be used
in case of a small number of data points. Particular problems may arise
for the spin gap:
(i) Boundary effects are present in both $E_0$ and $E_1$ leading to a
larger error in $E_1-E_0$. (ii)
As discussed in the last section the first triplet excitation belongs to the
QDJS with a definite symmetry. However, it may appear that this symmetry 
is not present in a certain  finite lattice, i.e. the calculated first
excitation belongs to another symmetry and 
consequently it has higher energy leading to an overestimation of the gap.
Therefore the extrapolation of the gap will not be  a main focus of our
discussion of the ordering of the HAFM on the Archimedean tilings in
section \ref{mag1}.  

We use only the leading terms even in case that the number of data points
would allow a scaling including next-to-leading term
in order to have the same systematics for all the 11 Archimedean tilings in
the comparative discussion given below.
By way of illustration we compare the results obtained by both variants of
extrapolation for the square 
lattice (for a comparison with data available in literature, see section
\ref{bipart}):
\bit
\item GS energy per site: $E_0/N=-0.6701$ (leading term only); 
$E_0/N=-0.6685$ (next-to-leading term included);
\item singlet-triplet gap: $\Delta=0.0605$ (leading term only); 
$\Delta=0.0247$ (next-to-leading term included); 
\item order parameter: $m^+=0.3173$ (leading term only); 
$m^+=0.3235$ (next-to-leading term included).
\eit
For the extrapolation altogether 
12 finite square lattices from $N=18$ to $N=40$ are used. 
The consideration of the 
next-to-leading terms changes the
energy by less than 0.2\% 
and the order parameter by less than 2\%. 
The values for the extrapolated 
gap can be understood as a measure of the accuracy of the extrapolation, 
since we know that the excitations about a \Neel ordered GS become gapless
for $N \to \infty$.

Finally, we mention that 
the finite-size scaling for systems with a critical GS or with a GS 
having only short-range 
spin pair correlations $\langle {\bf S}_i{\bf S}_j \rangle$ can be 
different from the above given relations. The concrete relations may
depend on details of GS correlations. Nevertheless
one aspect shall be noted: due to the absence of long-range correlations in
$\langle {\bf S}_i{\bf S}_j \rangle$  the finite-size effects should be
weaker for the GS energy. As a simple example we can consider a
HAFM with a valence-bond GS as realized for the 
Shastry-Sutherland model for stronger frustration (see section \ref{qpt}).  
The GS energy per site is completely independent of $N$ in this case.

%

\section{Magnetic ground-state ordering for the spin half HAFM on the 
Archimedean  lattices} 
\label{mag1}
In this section we present and discuss 
results obtained by exact diagonalization 
for the 11 Archimedean lattices. For some of these lattices such studies 
have not been performed so far or the presented results go beyond the 
system sizes published until now.
If available we also discuss results obtained by other methods to get 
a reliable picture on the magnetic ordering. From our results we conclude that  
three categories of ground state ordering appear: Collinear
two-sublattice \Neel long-range order (LRO), non-collinear (multi-sublattice
\Neel or spiral) LRO and quantum paramagnetic ground states without LRO in
the spin pair correlation $\langle {\bf S}_i{\bf S}_j \rangle$.   
  
At first we consider in sections \ref{bipart}, \ref{canted} and \ref{seckagome} 
each lattice separately and present results for
the GS energy, the spin gap and the order parameter as well as the spectrum. 
In a second step we summarize and compare  in section \ref{summary_1} 
the magnetic ordering on the
various Archimedean lattices. We refer the reader who is not interested in the
detailed discussion of the individual lattices to section \ref{summary_1}.   
\subsection{Semi-classical \Neel ordering on bipartite lattices} 
\label{bipart}
The classical GS for bipartite lattices is the 
perfect \Neel state having a GS energy per bond
$E_0^{\mathrm{class}}/\mathrm{bond}=-s^2=-0.25$
and an order parameter $m^+_{\mathrm{class}}=s=0.5$.
However, this classical order is very sensitive to fluctuations. 
Indeed the 1D HAFM does not exhibit \Neel LRO. For the 2D HAFM we
know  from the Mermin-Wagner theorem\cite{mermin66dd} that 
at arbitrarily small finite temperatures $T$ 
the thermal fluctuations are strong
enough to destroy the \Neel  LRO. 
However, it was for a long time an open question whether also 
quantum fluctuations 
are able to destroy \Neel LRO in 2D at absolute zero.
Each 2D lattice needs its individual consideration
because the strength of quantum fluctuations can vary from lattice to
lattice. Stronger quantum fluctuations 
appear in lattices with low coordination number $z$ and in lattices with
non-equivalent NN bonds. Although this non-equivalency of NN bonds is irrelevant
for classical bipartite HAFM it leads to a competition between the
bonds in quantum models. This quantum competition favors local 
singlet formation weakening that way \Neel ordering
(see, e.g.\ \cite{krueger00,troyer96} and section \ref{qpt}).

\subsubsection{The square lattice (T2)}

Starting from P.W.~Anderson's pioneering work \cite{anderson52}, 
the spin half HAFM on the square lattice has been studied over many decades.
There are some excellent reviews \cite{barnes91dd,manousakis91dd,dagotto94dd}
which can be used to get more detailed information on this work.
Although till now there is no rigorous proof for the 
existence of \Neel LRO\footnote{We
mention that the existence of \Neel LRO was proven for the HAFM with $S \ge
1$ \cite{neves88,affleck88} and for the spin half anisotropic $XXZ$
antiferromagnet \cite{kubo88,wischmann91} on the square lattice.} 
after intensive studies over many decades it became clear in the late
eighties that there is no doubt of semi-classical \Neel LRO at absolute zero. 
The quantum fluctuations lead to a substantial renormalization of the
order parameter (sublattice magnetization), which amounts to about 
60\% of the classical value.  
Experimentally there are some layered
antiferromagnetic inorganic materials
like the parent  compound La$_2$CuO$_4$
for high-$T_c$ superconductors or Sr$_2$CuO$_2$Cl$_2$  
\cite{manousakis91dd,chakravarty89,greven94} but also organic compounds
\cite{WAWLT02} which are well
described by the (quasi-)2D HAFM on the square lattice.

The  spin half HAFM on the square lattice can serve as 
the canonical example for a quantum HAFM
on a 2D bipartite lattice. 
Already about ten years ago Schulz and coworkers \cite{schulz92dd} 
published large-scale exact
diagonalization studies for the GS of systems up to $N=36$. 
Recently Betts and coworkers \cite{betts99a} have presented a 
systematic study of
a complete set of all finite square lattices up to $N=32$. 
In particular, one finds in \cite{betts99a} a
guideline how to find systematically the so-called
defining edge vector in  finite lattices for arbitrary
dimension and type of lattice. 
We use this scheme to generate the finite lattices discussed below. 
We have recalculated and extended Schulz' and Betts' results for systems up 
to $N=40$ sites including the results for the low-lying excitations.
We have presented some of our results  already in
sections \ref{pisa} and \ref{scaling}.

\begin{figure}
\begin{center} {\includegraphics[height=5.5cm]{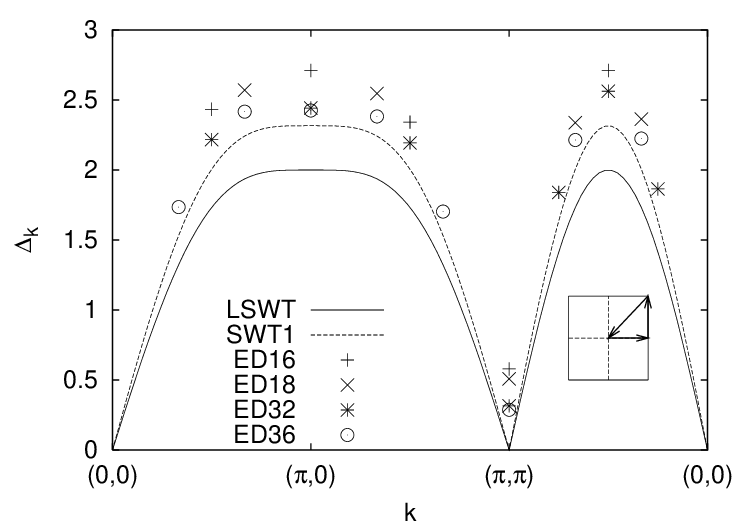}} \end{center}
\caption[Vergleich der Magnonenergie mit Spinwellentheorie]{ \label{sqdisp}
 Energy of the one-magnon excitations $\Delta_k=E_0({\bf k},S=1)-E_0(S=0)$
 versus ${\bf k}$ (dispersion relation).
 The solid and dashed lines show results of the linear (LSWT) and
 higher-order \SWT\ (SWT1) \cite{weihong93}, the points are exact data for
 finite lattices of size $N=$ 16, 18, 32 und 36.
 The inset in the right part of the figure shows
 the path through the Brillouin zone.
}
\end{figure}

The classical GS breaks the translational symmetry of the
lattice.
The magnetic \UC\ is twice as large as the geometric one.
On a finite bipartite lattice the quantum GS is a rotationally 
invariant singlet
state 
(Lieb-Mattis theorem \cite{lieb62}).
As can be seen from Fig.\ \ref{sqpisa}a 
there is one  QDJS  in each sector $S$ (cf.\ section \ref{pisa})
and the translational symmetry of the QDJS alternates between
${\bf Q}_0=(0,0)$ and ${\bf Q}_1=(\pi,\pi)$. Note that ${\bf Q}_0$ and 
${\bf Q}_1$ are different in the geometric but they coincide in the magnetic
Brillouin zone.
The energies of the QDJS are well described by
$E_0+S(S+1)/2N\chi_0$, see eq.\ (\ref{eqtos}). The
family of one-magnon states is well separated from the QDJS. 
Their energies follow the dispersion
relation obtained by \SWT, see Fig.\ \ref{sqdisp}.
The lowest singlet excitation is significantly
above the first triplet excitation.

The GS energy, the first excitation and the square of the order
parameter for the largest lattices with $N=38$ and $N=40$ and for comparison
for $N=36$
 are given in table \ref{t2max}. 
Furthermore, we present for the first time all 
spin-spin correlations for $N=40$ in table
\ref{cor40}. It is obvious that the decay of the spin-spin correlations  is
weak.
\begin{table}
\caption[Der HAFM auf dem Quadratgitter, $N=38$ and $N=40$40]{\label{t2max}
 Ground-state energy $E_0$ (singlet), 
singlet-triplet gap (spin gap) $\Delta=E_0(S=1)-E_0$  and square of the 
order parameter $m^2 \equiv (m^+)^2$
of the  
HAFM on finite square lattices with
 $N=38$, $N=40$ and $N=36$ sites 
(the results for $N=36$ are in agreement with
those of \cite{schulz92dd}). 
$(l_{11},l_{12})$ and $(l_{21},l_{22})$ are the components of the two 
edge vectors
defining the  finite 2D lattice. 
}
\begin{center}
\begin{tabular}{r|rrrr|l|l|l}
 $N$ &
 $l_{11}$ & $l_{12}$ & $l_{21}$ & $l_{22}$ &
  $E_0$ & $\Delta$ & $m^2$
\\ \hline  

36 & 6 & 0 & $0$  & 6 & $-24.4393974$  & 0.287538  & 0.20983715 \\ 
38 & 1 & 7 & $-5$ & 3 & $-25.7607925$  & 0.272791  & 0.20751801 \\ 
40 & 2 & 6 & $-6$ & 2 & $-27.0948503$  & 0.261623  & 0.20361937    
\end{tabular}
\end{center}
\end{table}

\begin{table}
\caption[square $N=40$]{\label{cor40}
Ground-state spin-spin correlations 
$\langle S^z_0 S_{R}^z\rangle= \frac{1}{3}
\langle {\bf S}_0{\bf S}_{R} \rangle$  for all lattice vectors ${\bf R}$  
of the HAFM on a square lattice with
$N=40$  sites. Note that due to the reduced symmetry of the finite lattice 
we have slightly different values  for lattice vectors ${\bf R}=(3,1)$
and $(3,-1)$ as well as for ${\bf R}=(1,2)$
and $(1,-2)$.   
}
\begin{center}
\begin{tabular}{r|r|r|r|r|r|r}
$ \bf{R} \; \; \quad$                     & $(0,0)$  $\quad$ & $(0,1)$  $\quad$   & $(1,1)$ $ \quad $ & $(0,2)$ $ \quad $ & $(1,-2)$ $ \quad$  & $(1,2)$  $ \quad $ \\ \hline 
$\langle S^z_0 S_{R}^z\rangle \quad \; $  & 0.250000 $\; $ & $-0.112895$ $\; $  & 0.069066 $\; $  & 0.061711 $\;$   & $-0.059679$ $\; $  & $-0.059055$  $\; $ \\ \hline 
$ \bf{R} \; \; \quad $                    & $(2,2)$ $ \quad$ & $(0,3)$ $ \quad $  & $(3,1)$ $ \quad $ & $(3,-1)$ $\quad $ & $(3,-2)$ $ \quad$  & $(4,-2)$ $ \quad $ \\ \hline 
$\langle S^z_0 S_{R}^z\rangle \quad \; $  & 0.053826 $\;$  & $-0.055663$ $\; $  & 0.054700 $\; $  & 0.052344 $\; $  & $-0.052074$  $\;$  & 0.050275    $\;$ \\ \hline
\end{tabular}
\end{center}
\end{table}
Altogether 12 finite square lattices from $N=18$ to $N=40$ are used
for the extrapolation of the GS energy and the order
parameter to infinite $N$
(see Figs.\ \ref{vgl_ex_e12}a and \ref{vgl_ex_m12}a).
We compare our results with some corresponding data obtained by other means:
\bit
\item 
GS energy per bond:  our result: $E_0/\mathrm{bond}=-0.3350$; 
high-order SWT \cite{weihong93}:
$E_0/\mathrm{bond}=-s^2-0.157948s-0.006237+0.0000108/s=-0.335233$;
QMC \cite{sandvik97}: $E_0/\mathrm{bond}=-0.334719$; coupled cluster
method (CCM) \cite{bishop00}: $E_0/\mathrm{bond}=-0.3349$;
series expansion \cite{oitmaa91}:
$E_0/\mathrm{bond}=-0.3347$; previous exact
diagonalization up to $N=32$ \cite{betts99a}:
$E_0/\mathrm{bond}=-0.33404$;
%
\item
order parameter (sublattice magnetization): 
our result: $m^+=0.3173 \sim  0.635 \; m^+_{\mathrm{class}}$; 
high-order SWT  \cite{weihong93}: 
$m^+=s-0.1966019+0.0000866(25)/s^2=0.3037$;  QMC \cite{sandvik97}:
$m^+=0.3070$; CCM \cite{bishop00}:
$m^+=0.31$;    series expansion \cite{oitmaa91}:
$m^+=0.307$; previous exact
diagonalization up to $N=32$ \cite{betts99a}: $m^+=0.30676$
\eit
(for the extrapolation of the gap, see section \ref{scaling}).
A more detailed collection of results for the sublattice 
magnetization and the GS energy 
obtained by different methods can be found in \cite{betts99a,bishop00}.

The existence of \Neel LRO for the square lattice 
does not automatically imply the conclusion, that all other bipartite 
lattices are also \Neel long-range ordered. Stronger quantum fluctuations can
appear in lattices with coordination number  $z < 4$ and in lattices with
non-equivalent NN bonds. 

\subsubsection{The honeycomb lattice (T3)} 
For this lattice the geometric and the magnetic (\Neel state) \UC\
are identical and include two sites.
All NN bonds are equivalent but  
the coordination number $z=3$ is less than in the square lattice giving rise
to stronger quantum fluctuations.
Nevertheless there is a lot of evidence obtained by several methods
\cite{reger89,weihong91,oitmaa92,mattsson94,krueger00,fouet01}
and
also from the data presented below, that the GS is a
semi-classical \Neel state.

\begin{figure}

\vspace{1cm}
\begin{center}
 \setlength{\unitlength}{1mm}
 \begin{picture}(55,70)
 \put(0,8){ \myframe{\hspace{-3.9cm}\includegraphics[height=7.5cm]{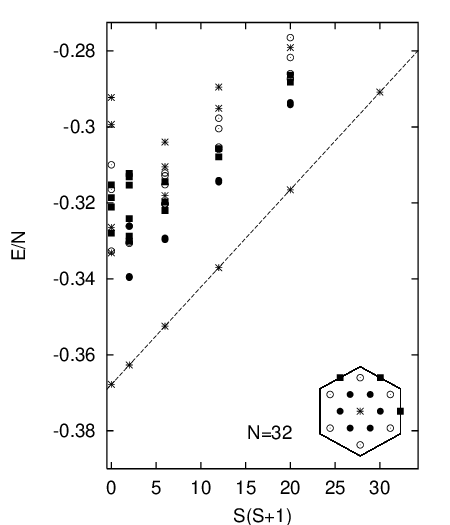}} }
 \put(-20,70){ \myframe{\includegraphics[height=0.8cm]{pict2.eps}}}
 \end{picture}
\end{center}
\caption[sq32 Pisa]{ \label{til3erg}
Low-energy spectrum for  the HAFM on the honeycomb lattice with $N=32$
sites 
(the inset shows the {\bf k} points 
in the Brillouin
zone).
}
\end{figure}

The low-energy spectrum is shown in Fig.\ \ref{til3erg}.
The QDJS are well separated from the other states 
and follow eq.\ (\ref{eqtos}). 
Due to the coincidence of
geometric and magnetic \UC\ they all have translational symmetry vector
${\bf Q}=(0,0)$.
There are no low-lying singlets within the spin gap.
The similarity between the spectra of the square and the honeycomb lattice
is obvious.

The largest lattice considered has $N=38$ sites and is defined by the edge 
vectors
$(3,2);(-2,5)$ 
and has 
GS energy $E_0/\mathrm{bond}=-0.366768$, spin gap $\Delta=0.213953$ and 
square of the order parameter 
$(m^+)^2=0.184396$.

 For the finite-size extrapolation of the GS energy (Fig.\ \ref{vgl_ex_e12}a),
the spin gap  and the order parameter 
(Fig.\ \ref{vgl_ex_m12}a) we have used 14 finite lattices from $N=6$ 
up to $N=38$.
The extrapolation according to formulae (\ref{gs_scale}), 
(\ref{m_scale}), (\ref{gap_scale}) leads 
to the following results: 
\bit
\item
GS energy per bond: $E_0/\mathrm{bond}=-0.3632$\\  
(for comparison: 
QMC \cite{reger89}:
$E_0/\mathrm{bond}=-0.3630$; 2nd order SWT \cite{weihong91}:
$E_0/\mathrm{bond}=-s^2-0.209842s-0.0110084=-0.365929$; series expansion  
\cite{oitmaa92}:
$E_0/\mathrm{bond}=-0.3629$; CCM \cite{krueger00}:
$E_0/\mathrm{bond}=-0.3631$);
\item spin gap:
$\Delta=0.0504$ \\
(for comparison: 
 CCM \cite{krueger00}:
$\Delta=0.02$);
\item order parameter:
$m^+=0.2788 \sim  0.558 \; m^+_{\mathrm{class}}$ \\ 
(for comparison:
QMC \cite{reger89}: 
$m^+=0.235$; 2nd order SWT \cite{weihong91}:
$m^+=0.2418$; series expansion  
\cite{oitmaa92}:
$m^+=0.266$; CCM \cite{krueger00}: $m^+=0.28$). 
\eit
Obviously due to the lower coordination number the magnetization 
is approximately 10\% smaller than for the
square lattice, but the existence of \Neel LRO is not in question. 
%
\subsubsection{The CaVO (T11) and the SHD (T10) lattices}
\label{sectil11}

\begin{figure}

\begin{center}
 \setlength{\unitlength}{1mm}
 \begin{picture}(55,55)
 \put(0,8){ \myframe{\hspace{-2.7cm}\includegraphics[height=4.1cm]{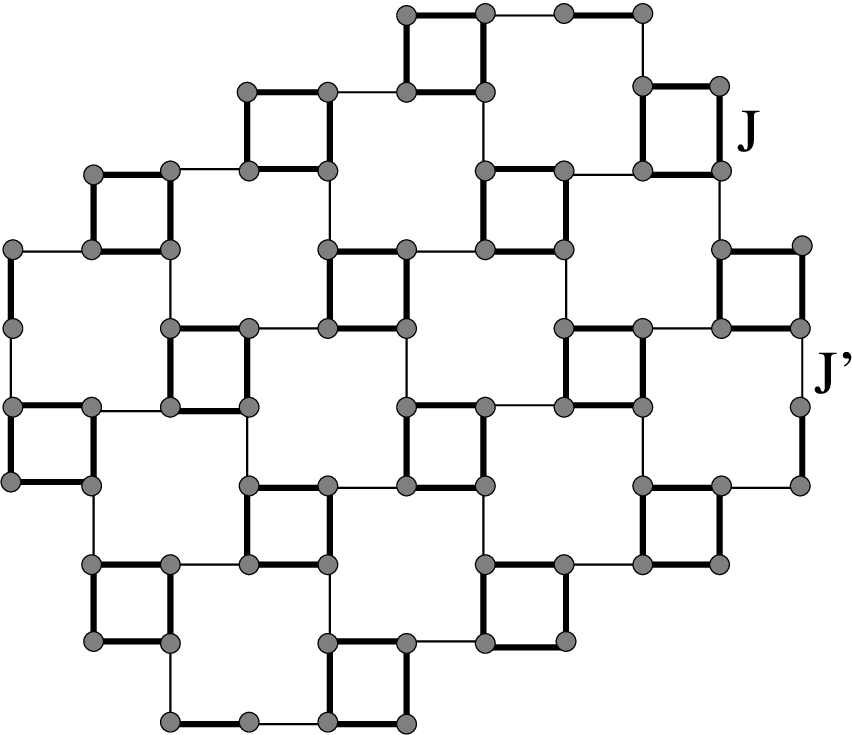}} }
 \put(5,3){\hspace{-2.8cm}\makebox(0,0)[lc]{a}}
 \put(20,9){ \myframe{\hspace{1.1cm}\includegraphics[height=4.0cm]{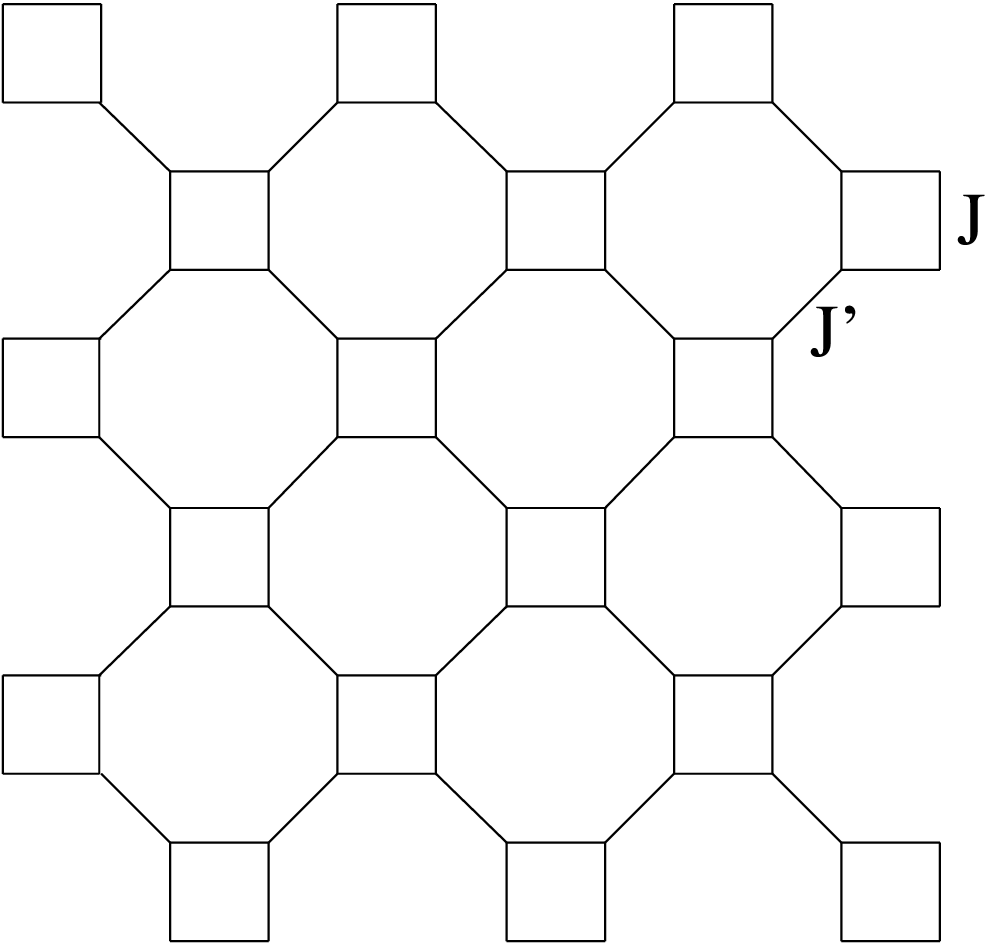}}}
 \put(5,3){\hspace{+3.0cm}\makebox(0,0)[lc]{b}}
 \end{picture}
\end{center}
\caption[Struktur des CaVO-Gitters]{ \label{cavo1}
Arrangement of the V$^{4+}$ atoms (points) in the V-O layers of CaV$_4$O$_9$
(left) and the corresponding Archimedean tiling T11 (right).
}
\end{figure}

\begin{figure}

\vspace{1.5cm}
\begin{center}
 \setlength{\unitlength}{1mm}
 \begin{picture}(55,70)
 \put(0,8){ \myframe{\hspace{-3.9cm}\includegraphics[height=7.5cm]{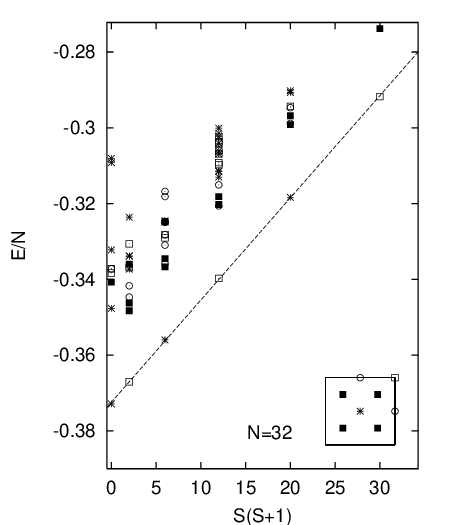}} }
 \put(5,3){\hspace{-2.cm}\makebox(0,0)[lc]{a}}
 \put(-20,68){ \myframe{\includegraphics[height=0.8cm]{pict11.eps}}}
 \put(2,8){
\myframe{\hspace{+2.2cm}\includegraphics[height=7.5cm]{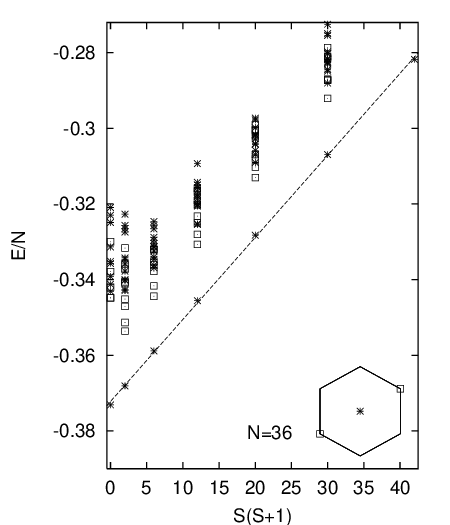}} }
 \put(5,3){\hspace{+7.2cm}\makebox(0,0)[lc]{b}}
 \put(+44,68){ \myframe{\includegraphics[height=0.8cm]{pict10.eps}}}
 \end{picture}
\end{center}
\caption[sq32 Pisa]{ \label{t11_10_erg}
Low-energy spectrum for HAFM on the CaVO (T11) and on the SHD (T10) lattice 
(the insets show the {\bf k} points 
in the  Brillouin
zone).\\
a: CaVO with  $N=32$. \hspace{0.3cm}
b: SHD with $N=36$. 
}
\end{figure}

Both these lattices have non-equivalent NN bonds and low
coordination number $z=3$ leading 
to strong quantum fluctuations. 
The lattice T11 has attracted much attention since 1995 when in
susceptibility measurements  on  CaV$_4$O$_9$
\cite{taniguchi95dd} for the first
time a rotationally invariant quantum paramagnetic GS with a finite 
spin gap of $\Delta\approx 110\mbox{K}$ was  discovered experimentally 
for a quasi-2D antiferromagnetic spin half system.
The underlying lattice of CaV$_4$O$_9$
is a $1/5$ site-depleted square lattice which can be
transformed by an appropriate distortion 
to the Archimedean lattice T11 (see Fig.\ \ref{cavo1}). We use therefore
the name `CaVO' to denote this lattice.
The experimental findings stimulated a series of 
theoretical studies for the spin half HAFM on the CaVO lattice
\cite{troyer96,katoh95,ueda96dd,starykh96,gelfand96,albrecht96a,albrecht96b,weihong97,troyer97,bose97,manuel98,weihong98,richter98dp,korotin99dd,hellberg99,fukumoto99}.
The geometric \UC\ of the CaVO lattice contains four sites. However, the
translational symmetry of the lattice and of the classical \Neel 
GS  do not fit to each other and consequently the magnetic \UC\ must be
chosen as twice as large as the geometric one.  This makes the symmetry of
the QDJS similar to that of the square lattice (see Figs. \ref{t11_10_erg}a 
and \ref{sqpisa}a).

The Archimedean lattice T10 is built by regular squares, hexagons and
dodecagons (SHD) and is therefore more complex than the CaVO lattice.
As far as we know till now no antiferromagnetic material was 
synthesized having the lattice structure of tiling 10. 
The low coordination number, the quantum competition of non-equivalent NN
bonds and the complex structure of the lattice have stimulated the
search for a possible non-\Neel ordered GS for this lattice
\cite{tomczak99,tomczak01}. 
The geometric \UC\ of the SHD lattice contains twelve sites.
The translational symmetry of the lattice and of the classical \Neel 
GS  fit to each other leading to identical  magnetic and geometric \UC.
Hence 
the symmetry of
the QDJS is similar to that of the honeycomb lattice (see Figs.
\ref{t11_10_erg}b and
\ref{til3erg}).
The similarities between the spectra of the square and the CaVO lattice as
well as the SHD and the honeycomb lattice are obvious:   
The QDJS are well separated from the other states and follow eq.\ (\ref{eqtos}). 
There are no low-lying singlets within the spin gap.

Since the magnetic \UC\ contains 8 sites, the largest CaVO lattice we consider 
has $N=32$ sites and is defined by 
the edge vectors
$(2,-2);(2,2)$. 
It has 
GS energy per bond $E_0/\mathrm{bond}=-0.372903$, spin gap $\Delta=0.281788$ 
and square of the order parameter 
$(m^+)^2=0.178018$. The two non-equivalent NN correlations functions for
$N=32$ are $\langle {\bf S}_i {\bf S}_j\rangle_{J}= -0.311103$ for $J$ bonds 
belonging to
squares and $\langle {\bf S}_i {\bf S}_j\rangle_{J'}=-0.403803$ for 
$J'$ dimer bonds (cf.\ Fig.\ \ref{cavo1}).

The largest SHD lattice considered has $N=36$ sites and is defined by the 
edge vectors
$(2,1);(-1,1)$. 
It has 
GS energy per bond $E_0/\mathrm{bond}=-0.373118$, spin gap 
$\Delta=0.270929$ and square of the order parameter 
$(m^+)^2=0.163243$. 
The three non-equivalent NN correlations functions for
$N=36$ are $\langle {\bf S}_i {\bf S}_j\rangle_{SH} = -0.414324$ for 
NN bonds belonging to 
squares and hexagons, 
$\langle {\bf S}_i {\bf S}_j\rangle_{SD}=-0.395046$ for NN bonds belonging
to squares and dodecagons and 
$\langle {\bf S}_i {\bf S}_j\rangle_{HD}=-0.309984$ for NN bonds belonging
to hexagons and dodecagons.

The finite-size extrapolation for the CaVO and even more for the SHD lattice
suffers from the restriction to a small number of unit cells in the accessible 
finite lattices. Hence the  extrapolation is  particularly
uncertain and should be taken with extra care.   
For the finite-size extrapolation of the GS energy 
(Fig.\ \ref{vgl_ex_e12}a),
and the order parameter 
(Fig.\ \ref{vgl_ex_m12}a) according to 
formulae (\ref{gs_scale}), 
(\ref{m_scale}), (\ref{gap_scale}) we use finite lattices of $N=16,24,32$ (CaVO)
and $N=12,24,36$  (SHD).
The extrapolation leads to the following 
results for the CaVO lattice: 
\bit
\item
GS energy per bond: $E_0/\mathrm{bond}=-0.3689$ \\ 
(for comparison: 
linear SWT \cite{ueda96dd}: $E_0/\mathrm{bond}=-0.3584$);
\item
spin gap:
$\Delta=0.1149$  
(for comparison: 
QMC \cite{troyer96}: $\Delta \sim 0$);
\item
order parameter:
$m^+=0.2303 \sim  0.461 \; m^+_{\mathrm{class}}$ \\ 
(for comparison: 
linear SWT \cite{ueda96dd}: $m^+=0.212$; 
QMC \cite{troyer96}: $m^+ = 0.178$)
\eit
and for the SHD lattice: 
\bit
\item
GS energy per bond: $E_0/\mathrm{bond}=-0.3713$ \\ 
(for comparison: 
variational (Huse-Elser)\cite{tomczak99}: $E_0/\mathrm{bond} =-0.3605$; variational
(resonating valence bond (RVB))\cite{tomczak01}: $E_0/\mathrm{bond}=-0.3688$);
\item
spin gap:
$\Delta=0.1435$; 
\item
order parameter:
$m^+=0.2126 \sim   0.425 \; m^+_{\mathrm{class}}$ \\ 
(for comparison: 
 variational
(RVB)\cite{tomczak01}: $m^+=0.2546$).
\eit
Due to the competition between the bonds the order parameters for the CaVO
and the SHD lattice are smaller than for the honeycomb lattice.
Nevertheless we find convincing  evidence that the GS is semi-classically
\Neel ordered. This conclusion is well supported by other methods
\cite{troyer96,tomczak01,ueda96dd,albrecht96a,gelfand96,manuel98,tomczak99}.
However, the quantum competition between  non-equivalent bonds leads to a
tendency to form local singlets either on neighboring bonds or along polygons.
In connection with the observed   
rotationally invariant quantum paramagnetic GS with a finite 
spin gap in CaV$_4$O$_9$ for the CaVO lattice a $J-J'$- HAFM with different
strengths
of NN bonds $J$ and $J'$ is considered, where $J$ is the NN bond belonging
to a square and $J'$ is the NN bond  not belonging to a dimer (Fig.\
\ref{cavo1}). Within this model a quantum phase transition 
between the semi-classical \Neel ordered phase and a quantum paramagnetic
rotationally invariant singlet phase with gapped excitations is obtained.
We will discuss this $J-J'$- HAFM and its quantum phase transition 
in more detail in section~\ref{qpt}. 

\subsection{Semi-classical LRO on frustrated  lattices} 
\label{canted}
The classical GS for non-bipartite frustrated lattices may be
collinear (weak frustration) or non-collinear (strong frustration)
and depends on the special features of the lattice. 
The frustration may enhance the effect of quantum fluctuations 
so that the magnetic order may be stronger weakened than for the bipartite
lattices. Thus the frustrated HAFM on 2D lattices is an interesting 
candidate
for a magnetic system with a quantum paramagnetic GS.

\subsubsection{The triangular lattice (T1)}
\label{secDreieck}
The triangular lattice is strongly frustrated but has largest coordination
number $z=6$ (see Fig.\ \ref{fig_ov1}). 
Already in the 70ties Anderson and Fazekas \cite{anderson73dd,fazekas74} 
considered the spin half HAFM on the triangular lattice. They argued that
the GS for the 2D triangular lattice might be similar to that for the 1D HAFM
and
proposed a spin-liquid
like rotationally invariant resonating valence bond GS instead of a
semi-classical \Neel state.
Starting in the late eighties several authors found, however, more and more
evidence for a \Neel ordered GS 
(see e.g. 
\cite{bernu92,bernu93,bernu94,huse88,jolicoeur89,miyake92,deutscher93,trumper99,capriotti99,capriotti01,farnell01}).

\begin{figure}

\vspace{0.9cm}
\begin{center}
 \setlength{\unitlength}{1mm}
 \begin{picture}(55,70)
 \put(0,20){\myframe{\hspace{-3.4cm}\includegraphics[height=4.2cm]{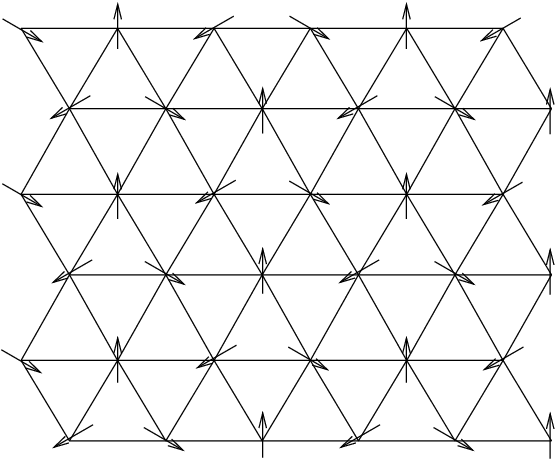}} }
 \put(5,3){\hspace{-2.cm}\makebox(0,0)[lc]{a}}
 \put(2,8){ \myframe{\hspace{+2.2cm}\includegraphics[height=7.5cm]{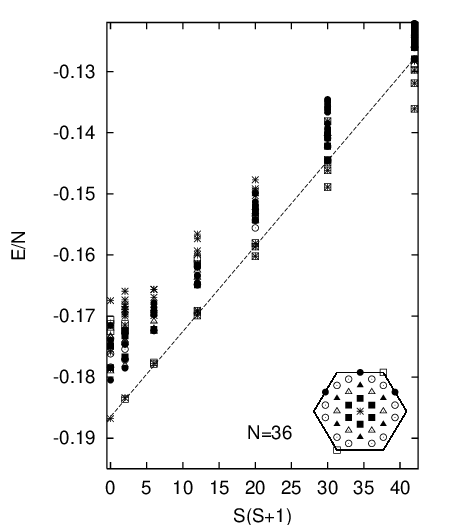}} }
 \put(5,3){\hspace{+7.2cm}\makebox(0,0)[lc]{b}}
 \put(+44,70){ \myframe{\includegraphics[height=0.8cm]{pict0.eps}}}
 \end{picture}
\end{center}
\caption[Energiespektrum des Dreiecksgitters]{ \label{t1_cl_gs}
The HAFM on the triangular lattice (T1)\\
a: Classical GS   \hspace{0.3cm}
b: Low-energy spectrum for $N=36$ 
(the inset shows the {\bf k} points in the Brillouin zone). 
}
\end{figure}

The classical GS is a three-sublattice \Neel state with an angle of
$120^\circ$
 between the spins of different sublattices (Fig.\ \ref{t1_cl_gs}a). It
breaks the translational symmetry of the
lattice. The energy per bond is $E_0^{\mathrm{class}}/\mathrm{bond}=-s^2/2=-0.125$  and the order parameter 
is $m^+_{\mathrm{class}}=\frac{1}{2}\sqrt{2/3}=0.40825$. 
 
The magnetic \UC\ is three times as large as the geometric one
and thus the QDJS belong to vectors 
${\bf Q}_1=(0,0)$, ${\bf Q}^+_2=(+4\pi/3,0)$ and ${\bf Q}^-_2=(-4\pi/3,0)$.
Low-lying states have been tabulated in \cite{bernu94}, however for
$N=36$ only in the sector ${\bf Q}_1=(0,0)$. Fig.~\ref{t1_cl_gs}b shows
our results for the low-lying states on the $N=36$ lattice. Apparently, 
the QDJS are well separated from the other states and follow eq.~(\ref{eqtos})
for $S \lesssim 4$. The lowest singlet excitation energy is above the first
triplet excitation. 
A special feature of the $E(S)$ behavior of the QDJS is a deviation 
from the linearity starting  in the vicinity of   
$S=N/6$. This comes from the Ising part of the Hamiltonian 
and is connected with distinguished Ising states having two spins up and 
one spin down per triangle \cite{miyashita86,hon99}  and 
results in a plateau in the magnetization versus external magnetic field 
curve (for a more detailed discussion, see section \ref{m_h}). 
However, this peculiarity emerging around $S(S+1)=42$ in Fig.\ \ref{t1_cl_gs}b 
is relevant only if $\sqrt{N} \lesssim N/6$, i.e.\ for small $N$.

The largest lattice considered has $N=36$ sites and is defined by the 
edge vectors
$(6,0);(0,6)$. It has GS energy per bond $E_0/\mathrm{bond}=-0.186791$, 
spin gap $\Delta =0.344211$ and square of the order parameter 
$(m^+)^2=0.124802$ (cf.\ \cite{bernu94}). 
A detailed discussion of the spectra can be found in
\cite{bernu92,bernu93,bernu94}.

For the finite-size extrapolation of the GS energy 
(Fig.\ \ref{vgl_ex_e12}b),
the spin gap  and the order parameter 
(Fig.\ \ref{vgl_ex_m12}b) we use only even finite lattices  of size
$N=24,30,36$.
The extrapolation according to formulae (\ref{gs_scale}), 
(\ref{m_scale}), (\ref{gap_scale}) leads to the following results: 
\bit
\item
GS energy per bond: $E_0/\mathrm{bond}=-0.1842$ \\ 
(for comparison: RVB \cite{anderson73dd}: $E_0/\mathrm{bond}
= -0.154$; SWT \cite{miyake92}: $E_0/\mathrm{bond}= -0.1823$;
former exact diagonalization \cite{bernu94}: $E_0/\mathrm{bond}=-0.1815$, 
Green's function Monte
Carlo \cite{capriotti99}: $E_0/\mathrm{bond}= -0.1819$; 
CCM \cite{farnell01}: $E_0/\mathrm{bond}= -0.1835$);
\item
spin gap:
$\Delta= 0.1293$; 
\item order parameter:
$m^+=0.1577 \sim 0.386  \; m^+_{\mathrm{class}}$ \\ 
(for comparison: 
sublattice magnetization $m^{sl}=\langle  S_i^z \rangle$
in linear SWT \cite{miyake92}:
$m^{sl}=0.2387=0.4774\;m^{sl}_{\mathrm{class}}$; 
Green's function Monte
Carlo \cite{capriotti99}: $m^{sl}=0.205=0.41\;m^{sl}_{\mathrm{class}}$; 
CCM \cite{farnell01}: $m^{sl}=0.2107=0.4214\;m^{sl}_{\mathrm{class}}$). 
\eit
Obviously, the extrapolated gap is quite large, whereas the order parameter
is smaller than that obtained by other means. This suggests stronger
finite-size effects than for bipartite lattices. Nevertheless, the
existence of semi-classical \Neel LRO is not in question.
\label{mSlSWT}

\subsubsection{The maple-leaf (T4) and the bounce (T7) lattices}
\label{secMaple}
\begin{figure}

\begin{center}
\myframe{\includegraphics[height=7.5cm]{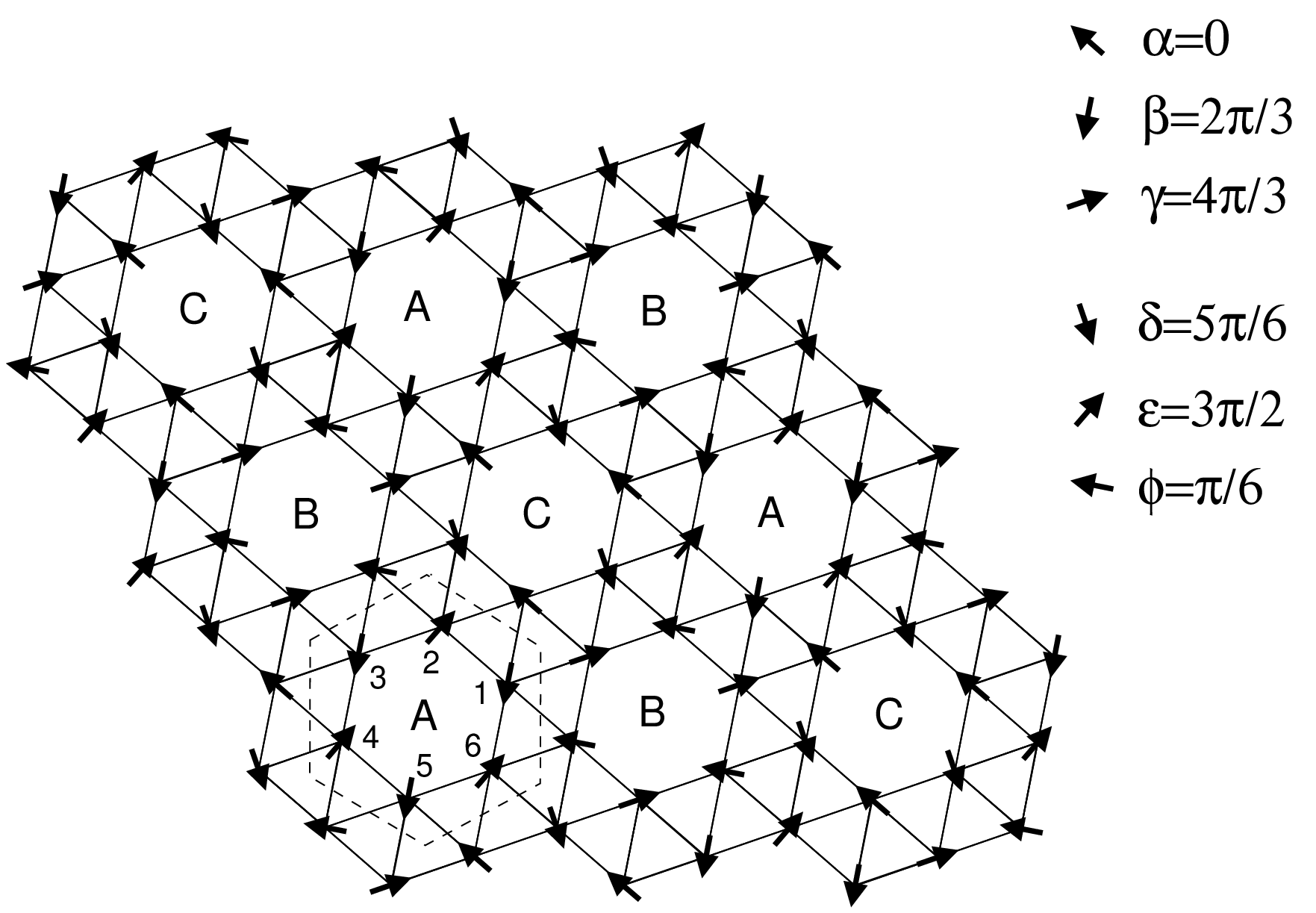}}
\end{center}
\caption[cl GS til4]{ \label{t4_cl_gs}
The classical GS of the HAFM on the maple-leaf lattice (T4). The geometric
\UC\ is shown with dashed lines. The magnetic \UC\ contains 
three geometric {\UC}s labeled by A,B,C.
}
\end{figure}

The maple-leaf lattice \cite{betts95} is obtained from the triangular 
lattice by a  1/7 depletion of sites. 
Its geometric \UC\ contains six sites and the underlying
Bravais lattice is a triangular one (cf.\ Fig.\ \ref{t4_cl_gs}).
It is also strongly frustrated  but has lower coordination number
($z=5$) than the triangular lattice. 
Furthermore, it has three non-equivalent NN
bonds (solid, dashed and dotted lines in Fig.\ \ref{t347}). 
Thus the quantum fluctuations might be more
important and the HAFM on the maple-leaf lattice was considered as a
candidate for a  quantum paramagnet \cite{schmalfuss02}.

The bounce lattice is related to the maple-leaf lattice. It can be obtained
from the maple-leaf lattice by bond depletion as described in section
\ref{relation} (see Fig.\ \ref{t347}). It has also a 
geometric \UC\ with 6
sites, an underlying triangular Bravais lattice and contains 
two non-equivalent NN bonds.
The coordination number 
$z=4$ is lower than for the maple-leaf lattice but it is
less frustrated, since the omitted bond was a frustrating one.
As far as we know no antiferromagnetic material has, as yet,
been synthesized with the lattice structure of tilings 4 or 7.

The classical GS of the maple-leaf lattice 
is a six-sublattice \Neel state shown in Fig.\ \ref{t4_cl_gs} with 
$E_0^{\mathrm{class}}/\mathrm{bond} = -s^2(\sqrt{3}+1)/5=-0.13660$ and $m^+_{\mathrm{class}}=0.39434$.
The less frustrated bounce lattice has also a six-sublattice \Neel GS 
with a $120^\circ$
structure on each triangle and a collinear up-down structure on each
hexagon leading to 
$E_0^{\mathrm{class}}/\mathrm{bond} = -2s^2/3=-0.16667$ and $m^+_{\mathrm{class}}=1/\sqrt{6}=0.40825$.
Both classical GSs 
break the translational symmetry of the
lattice,
the corresponding magnetic \UC\ is three times as large as the geometric one 
and contains 18 sites. Therefore the applicability of finite-size calculations
is particularly limited.

\begin{figure}

\vspace{0.9cm}
\begin{center}
 \setlength{\unitlength}{1mm}
 \begin{picture}(55,70)
 \put(2,8){\myframe{\hspace{-3.9cm}\includegraphics[height=7.5cm]{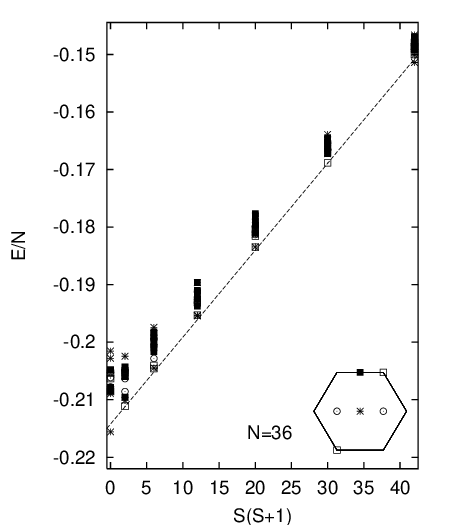}} }
 \put(-20,69){ \myframe{\includegraphics[height=0.8cm]{pict4.eps}}}
 \put(5,3){\hspace{-2.cm}\makebox(0,0)[lc]{a}}
 \put(2,8){ \myframe{\hspace{+2.2cm}\includegraphics[height=7.5cm]{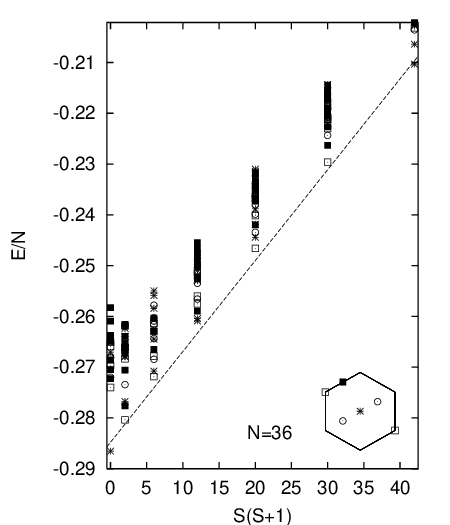}} }
 \put(5,3){\hspace{+7.2cm}\makebox(0,0)[lc]{b}}
 \put(+44,69){ \myframe{\includegraphics[height=0.8cm]{pict7.eps}}}
 \end{picture}
\end{center}
\caption[Energiespektrum T4 und T5]{ \label{t4_7_erg}
Low-energy spectrum  for HAFM on the maple-leaf (T4) and on the bounce (T7) 
lattice 
(the insets show the {\bf k} points in the Brillouin zone). 
a: maple-leaf lattice with $N=36$. 
\hspace{0.3cm} 
b: bounce lattice  with $N=36$. 
}
\end{figure}

The low-lying spectra for both lattices with $N=36$ sites 
are shown in  Fig.\ \ref{t4_7_erg}. The
lowest states in each sector of $S$ are QDJS
belonging to appropriate symmetries 
${\bf Q}_1=(0,0)$ and  ${\bf Q}^+_2=(+4\pi/3,0)$, 
${\bf Q}^-_2=(-4\pi/3,0)$.
They  follow eq. (\ref{eqtos}). 
The lowest singlet excitation energy is above
the first triplet excitation. 

The largest finite  lattices considered have
$N=36$ sites and are  defined by the edge 
vectors 
$(3,0);(1,2)$ for both tilings.
Note that these finite lattices do not have the full
symmetry of the corresponding infinite lattices.
%
%
%
%
%
The $N=36$ maple-leaf lattice has GS energy per bond $E_0/\mathrm{bond}=-0.215589$, 
spin gap $\Delta=0.400009$ and square of the order parameter 
$(m^+)^2=0.106101$. 
A picture of this lattice and a table of the correlation functions are given
in \cite{schmalfuss02}. 
The non-equivalent NN bonds lead to different NN correlations:
$\langle {\bf S}_i{\bf S}_j\rangle_T = -0.1777$ (belonging to a dashed line
in Fig.\ \ref{t347}, middle),
$\langle {\bf S}_i{\bf S}_j\rangle_H=-0.3656$ (belonging to a solid line 
in Fig.\ \ref{t347}, middle) and $\langle {\bf S}_i{\bf S}_{j}\rangle_D
=0.0086$ (belonging to a dotted line
in Fig.\ \ref{t347}, middle).\footnote{Note that these values and the
corresponding values for the bounce lattice are averaged values,
since the $N=36$ lattices do not have all lattice symmetries of the 
infinite lattice. As a result one has to average over 
three different values for a certain correlation function.}
It appears that the correlation functions  of the quantum system 
fit quite well to the classical GS.

The $N=36$ bounce lattice  
has GS energy per bond $E_0/\mathrm{bond}=-0.286540$, 
spin gap $\Delta=0.445138$ and square of the order parameter 
$(m^+)^2=0.119073$. 
The non-equivalent NN bonds lead to different NN correlations:
$\langle {\bf S}_i{\bf S}_j\rangle_T = -0.1723$ (belonging to a dashed line
in Fig.\ \ref{t347} right) and 
$\langle {\bf S}_i{\bf S}_j\rangle_H=-0.4008$ (belonging to a solid line 
in Fig.\ \ref{t347} right). The correlation function of the omitted bond 
(see Fig.\ \ref{t347}) is
$\langle {\bf S}_i{\bf S}_{j}\rangle_D = 0.1116$.
It is obvious that the NN correlations $\langle {\bf S}_i{\bf S}_j\rangle_D$
and $\langle {\bf S}_i{\bf S}_j\rangle_H$ are  enhanced by 
omitting the frustrating bond, whereas $\langle {\bf S}_i{\bf S}_j\rangle_T$
remains almost the same.

\label{p:maple-leaf}

We use finite maple-leaf lattices and bounce lattices of size $N=18$ and $36$
for the finite-size extrapolation of the GS energy
(Fig.\ \ref{vgl_ex_e12}b), the spin gap and the order parameter
(Fig.\ \ref{vgl_ex_m12}b).
By using formula (\ref{gs_scale}), (\ref{m_scale}) and (\ref{gap_scale})
we obtain for the maple-leaf lattice: 
\bit
\item GS energy per bond: $E_0/\mathrm{bond}=-0.2137$ \\ 
 (for comparison: SWT \cite{schmalfuss02}: $E_0/\mathrm{bond}
 = -0.20486$; variational \cite{schmalfuss02}: $E_0/\mathrm{bond}= -0.1988$);
\item
 spin gap: $\Delta=0.2548$; 
\item
 order parameter: $m^+=0.0860 \sim 0.218 \; m^+_{\mathrm{class}} $ \\ 
(for comparison:
sublattice magnetization $m^{sl}=\langle  S_i^z \rangle$ in
SWT \cite{schmalfuss02}:
 $m^{sl}= 0.154=0.308\;m^{sl}_{\mathrm{class}}$).
\eit
An extrapolation of the gap based on a variational approach 
was presented in
\cite{schmalfuss02} and leads to $\Delta = 0.0180$.

The corresponding extrapolation for the bounce lattice yields:
\bit
\item
 GS energy per bond: $E_0/\mathrm{bond}=-0.2837$; 
\item
 spin gap: $\Delta=0.2926$; 
\item order parameter:
 $m^+=0.1095 \sim 0.268 \; m^+_{\mathrm{class}} $. 

\eit
Obviously, the extrapolated order parameters are small but finite. 
The fact that the order parameter for the bounce lattice is larger than for the 
maple-leaf lattice seems to be related to the lower frustration.
Taking into consideration 
results of the \SWT\ and the variational approach presented for the maple-leaf
lattice in \cite{schmalfuss02}
we conclude that the semi-classical  six-sublattice \Neel LRO survives for both
lattices.
However, this statement needs 
confirmation by further studies.

\subsubsection{The trellis lattice (T5)}
\label{secTrellis}

\begin{figure}

\vspace{0.9cm}
\begin{center}
 \setlength{\unitlength}{1mm}
 \begin{picture}(55,70)
 \put(0,8){ \myframe{\hspace{-3.3cm}\includegraphics[height=5.6cm]{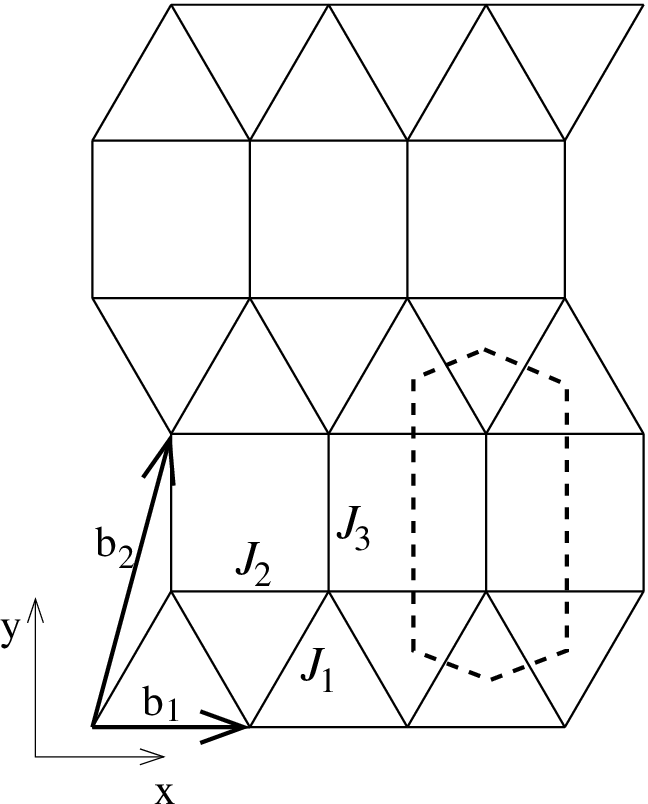}} }
 \put(5,3){\hspace{-2.cm}\makebox(0,0)[lc]{a}}
 \put(2,8){ \myframe{\hspace{+2.2cm}\includegraphics[height=7.5cm]{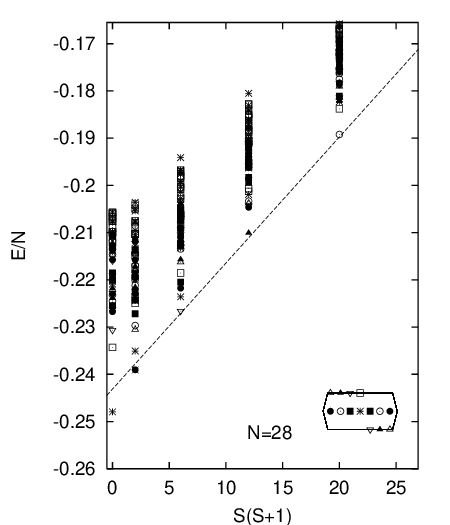}} }
 \put(5,3){\hspace{+7.2cm}\makebox(0,0)[lc]{b}}
 \put(+44,68){ \myframe{\includegraphics[height=0.8cm]{pict5.eps}}}
 \end{picture}
\end{center}
\caption[Trellis]{ \label{trellis}
The trellis lattice (T5).\\
a: Illustration of the lattice with basis vectors  ${\bf b}_1$
and  ${\bf b}_2$, geometric \UC\ (dashed) and  the non-equivalent NN bonds
$J_1$, $J_2$ and $J_3$. In the classical GS the spins form a spiral along the
zigzag chains ($J_1$, $J_2$ bonds) whereas the spins along a $J_3$ bond are 
antiparallel.   \\
b: Low-energy spectrum for  $N=28$ (the inset shows the {\bf k} points 
in the  Brillouin
zone).
}
\end{figure}

The trellis lattice is to some extent exceptional since its structure
corresponds to  a system of coupled ladders or alternatively of coupled zigzag
chains. 
Its geometric \UC\ contains 2 sites (cf.\ Fig.\ \ref{trellis}a).
It has the same coordination number
$z=5$ as the maple-leaf lattice but its frustration is slightly smaller (cf.
Fig.\ \ref{fig_ov1}). 
Furthermore, it has three non-equivalent NN
bonds, labeled by $J_1$, $J_2$ and $J_3$ in Fig.\ \ref{trellis}a.  

The  HAFM on the trellis lattice is related to  the 
magnetism of 
SrCu$_2$O$_3$,
CaV$_2$O$_5$ and MgV$_2$O$_5$ \cite{normand97,miyahara98dd}. However, the
$J_1$, $J_2$ and $J_3$ bonds are not of equal strength in these
materials (for instance in SrCu$_2$O$_3$ the zigzag $J_1$ coupling 
is weak leading to a quasi-1D ladder structure).    
The classical GS is a \Neel state for $J_2 < J_1/4$, and is
an incommensurate  spiral state along the zigzag 
chains ($x$-direction) for $J_2 > J_1/4$, 
where the  angles  between neighboring bonds are $\alpha_2
=2\arccos{(J_1/4J_2)}$ ($J_2$ bond), $\alpha_1=\pi+\alpha_2/2$ 
($J_1$ bond) and $\alpha_3=\pi$ ($J_3$ bond). This leads
to a classical GS with pitch angles $\alpha_1= \pi+\arccos{(1/4)}=1.41957\pi$; 
$\alpha_2= 2\arccos{(1/4)}=0.83914\pi$, GS energy per bond 
$E_0^{\mathrm{class}}/\mathrm{bond} =-0.65s^2= -0.1625$ and
$m^+_{\mathrm{class}}=0.39894$ for the perfect lattice ($J_1=J_2=J_3$).

The incommensurability of the classical GS creates additional difficulties
applying exact diagonalization for finite lattices since 
the classical pitch angles
$\alpha_1$ and $\alpha_2$ may be in conflict with periodic 
boundary conditions.
In order to minimize this boundary effect we consider only 
finite lattices of $N=20$, $28$ and $36$ sites  
having pitch angles $\alpha_2^{(N)}$ deviating by not more than 6\% from the
true values $\alpha_2$. The $N=36$ lattice is defined by the 
edge vectors $(9,0);(-1,2)$ 
and has a pitch angle $\alpha_2^{(36)} = 1.059 \alpha_2$.
Its GS energy per bond is $E_0/\mathrm{bond}=-0.247578$, spin gap 
$\Delta=0.605227$ and square of the order parameter 
$(m^+)^2=0.109897$.
The three non-equivalent NN correlations functions for
$N=36$ are 
$\langle {\bf S}_i {\bf S}_j\rangle_{J_1}= -0.098835$, 
$\langle {\bf S}_i {\bf S}_j\rangle_{J_2}= -0.283938$ and 
$\langle {\bf S}_i {\bf S}_j\rangle_{J_3}= -0.472341$
(cf.\ Fig.\ \ref{trellis}a).

In Fig.\ \ref{trellis}b the QDJS are shown.  
Although the boundary conditions are not perfect it can be seen that the QDJS
 are separated from the other states and follow approximately eq. 
(\ref{eqtos}). 
The lowest singlet excitation is above 
the first
triplet excitation. 
The translational symmetry of the QDJS is more complex than in the other
lattices. It is connected with the ${\bf q}$ vector of
the spiral state.
We find $Q_y=0, \pi$ and  $Q_x=6\pi(N/2-S)/7 \mbox{ mod } 2\pi$ for
$N=28$ and
$Q_x=8\pi(N/2-S)/9 \mbox{ mod } 2\pi$ for $N=36$.

For the finite-size extrapolation of the GS energy 
(Fig.\ \ref{vgl_ex_e12}b),
the spin gap  and the order parameter 
(Fig.\ \ref{vgl_ex_m12}b) we use finite lattices  of size
$N=20,28,36$.
The extrapolation according to formulae (\ref{gs_scale}), 
(\ref{m_scale}), (\ref{gap_scale}) leads to the following results: 
\bit
\item GS energy per bond: $E_0/\mathrm{bond}=-0.2471$; 
\item
spin gap:
$\Delta= 0.49$; 
\item
order parameter:
$m^+=0.0885 \sim 0.222 \;  m^+_{\mathrm{class}}$. 
\eit
Although our data do not allow a secure conclusion the results are in
favor of a spiral long-range ordered phase. This conclusion is in 
agreement
with the findings in \cite{normand97} based on a Schwinger boson technique and 
linear \SWT.

%
\subsubsection{The SrCuBO lattice (T6)}
\label{secSrCuBO}

\begin{figure}

\vspace{+1.7cm}
\begin{center}
 \setlength{\unitlength}{1mm}
 \begin{picture}(55,70)
 \put(0,46){ \myframe{\hspace{-2.9cm}\includegraphics[height=3.4cm]{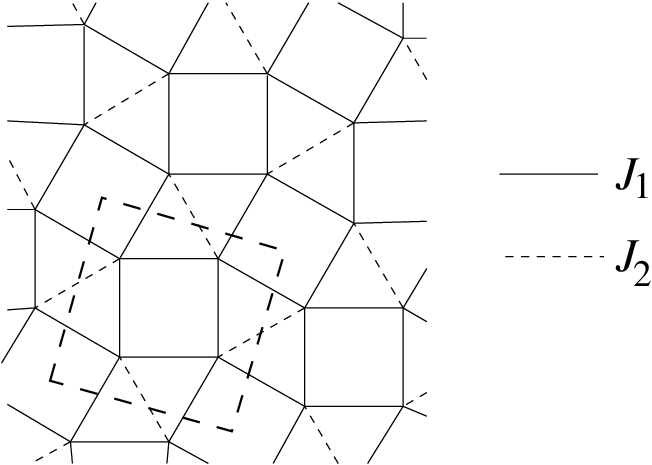}} }
 \put(0,10){ \myframe{\hspace{-2.9cm}\includegraphics[height=3.4cm]{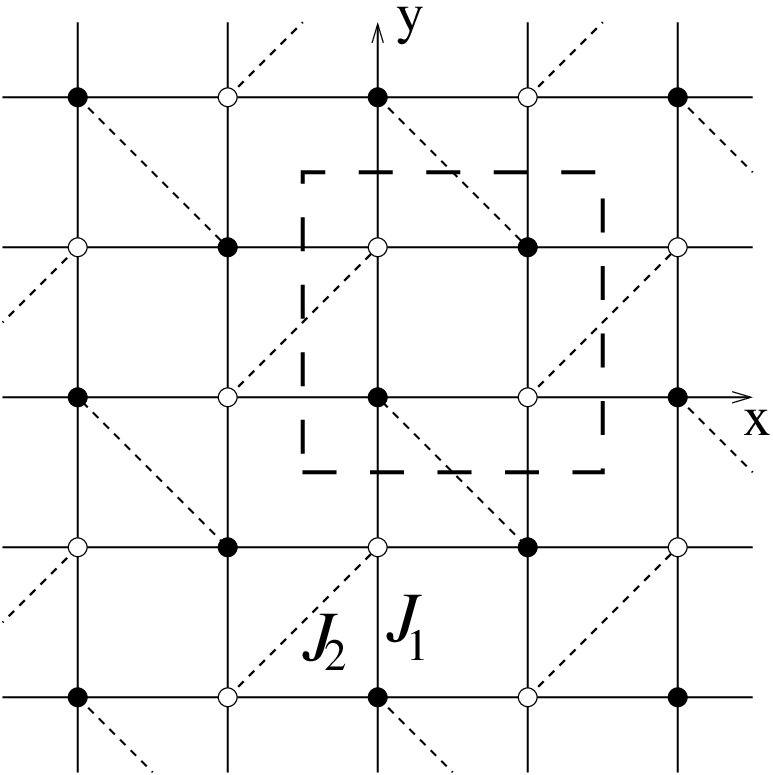}}}
 \put(5,3){\hspace{-2.cm}\makebox(0,0)[lc]{a}}
 \put(2,8){ \myframe{\hspace{+2.2cm}\includegraphics[height=7.5cm]{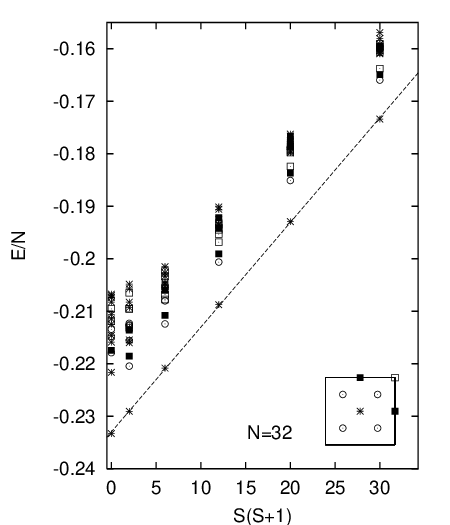}} }
 \put(5,3){\hspace{+7.2cm}\makebox(0,0)[lc]{b}}
 \put(+44,65){ \myframe{\includegraphics[height=0.8cm]{pict6.eps}}}
 \end{picture}
\end{center}

\caption[Struktur des SrCuBo-Gitters]{ \label{srcubo1}
The SrCuBO lattice (T6).
a:  Comparison of the SrCuBO lattice (above) 
and the Shastry-Sutherland model (below).
The \UC\ is illustrated by the
dashed square. \\
b: Low-energy spectrum for $N=32$ 
(the inset shows the {\bf k} points in the  Brillouin zone). 
}
\end{figure}

The SrCuBO lattice is weakly frustrated, has four sites in the geometric \UC\
and
two non-equivalent NN bonds $J_1$ and $J_2$ (see Fig.\ \ref{srcubo1}a, top).  
It can be transformed by an appropriate distortion 
to a square lattice with one diagonal bond
in each second square (see Fig.\ \ref{srcubo1}a, bottom). 
This frustrated square lattice is known as Shastry-Sutherland
model \cite{shastry81dd,albrecht96}
introduced in the 80ties as a 2D spin half HAFM  with an exactly known
quantum GS. 
Indeed for
large frustrating $J_2$ the GS is a so-called orthogonal dimer
product state with dimer
singlets on each $J_2$ bond.
Although the Shastry-Sutherland model initially was understood as
a `toy model' it has attracted much renewed attention as
it provides a representation of the magnetic properties of the recently 
discovered 2D spin gap system SrCu$_2$(BO$_3$)$_2$
\cite{kageyama99dd,miyahara99}.
The experimental findings stimulated a series of 
theoretical studies for the spin half HAFM on the SrCuBO
lattice with varying bonds $J_1$, $J_2$, see Refs. 
\cite{weihong99dd,hartmann00,momoi00,koga00a,misguich01,chung01,weihong01,carpentier02,lauchli02} 
and the recent review \cite{miyahara03dd}.
We will discuss the GS phase diagram in the $J_1-J_2$ plane 
below in section \ref{qpt}. 
In this section we consider $J_1=J_2$, only. In this case
the classical GS is the two-sublattice \Neel state with energy per bond
$E_0^{\mathrm{class}}/\mathrm{bond}=-0.6s^2=-0.15$  and with order parameter 
$m^+_{\mathrm{class}}=0.5$.
The geometric \UC\ of the SrCuBO lattice contains four sites and the
translational symmetry of the lattice and of the classical \Neel GS fit to
each other. The spectrum of the SrCuBO lattice (Fig.\ \ref{srcubo1}b) is
therefore comparable with that of the honeycomb lattice (Fig.\ \ref{til3erg}). 
The QDJS are well separated from the other states and follow eq.~(\ref{eqtos}). 
The lowest singlet excitation is 
above the first triplet.

The largest lattice considered with $N=36$ sites is defined by 
the edge vectors
$(3,0);(0,3)$
and has 
GS energy per bond $E_0/\mathrm{bond}=-0.233410$, spin gap  
$\Delta=0.319735$ 
and square of the order parameter 
$(m^+)^2=0.169048$, that is 80\% of the order parameter of the corresponding
square lattice.
The two non-equivalent NN correlations functions for
$N=36$ are $\langle {\bf S}_i {\bf S}_j\rangle_{J_1}= -0.332886$ (almost the
same as for the square lattice)
and $\langle {\bf S}_i {\bf S}_j\rangle_{J_2}=0.164493$.

For the finite-size extrapolation of the GS energy (Fig.\ \ref{vgl_ex_e12}b),
the spin gap 
 and the order parameter 
(Fig.\ \ref{vgl_ex_m12}b) we use finite lattices  of size
$N=20$, $32$ and $36$. 
The extrapolation according to formulae (\ref{gs_scale}), 
(\ref{m_scale}), (\ref{gap_scale}) leads to the following results: 
\bit
\item
GS energy per bond: $E_0/\mathrm{bond}=-0.2310$ \\ 
(for comparison: series expansion \cite{weihong99dd}: $E_0/\mathrm{bond}=-0.231$;
Schwinger boson mean field  \cite{albrecht96}: $E_0/\mathrm{bond}=-0.231 $;  
CCM \cite{richter03}:
$E_0/\mathrm{bond}=0.2311 $);
\item
spin gap:
$\Delta=0.0927$; 
\item
order parameter:
$m^+=0.2280 \sim   0.456 \;  m^+_{\mathrm{class}}$ \\ 
(for comparison: series expansion \cite{weihong99dd}: $m^+=0.200$;
Schwinger boson mean field \cite{albrecht96}: $m^+=0.203 $; CCM 
\cite{richter03}: $m^+=0.211 $).
\eit
Due to frustration the order parameter is only about 70\% of that of the
square lattice but it is the largest one of all frustrated lattices. 
There is no doubt of semi-classical GS \Neel order for this lattice. 
This conclusion is in agreement with 
several other studies like series expansion
\cite{weihong99dd,hartmann00,koga00a} and 
bosonic representations \cite{albrecht96,lauchli02}. 
However, the \Neel LRO is destroyed by further increasing the
frustrating bond $J_2$ (see section \ref{qpt}).

\subsection{Absence of semi-classical LRO 
on frustrated  lattices - the \kagome (T8) and the star (T9) lattices} 
\label{seckagome}

\begin{figure}
\begin{center}
\myframe{\includegraphics[height=5.5cm]{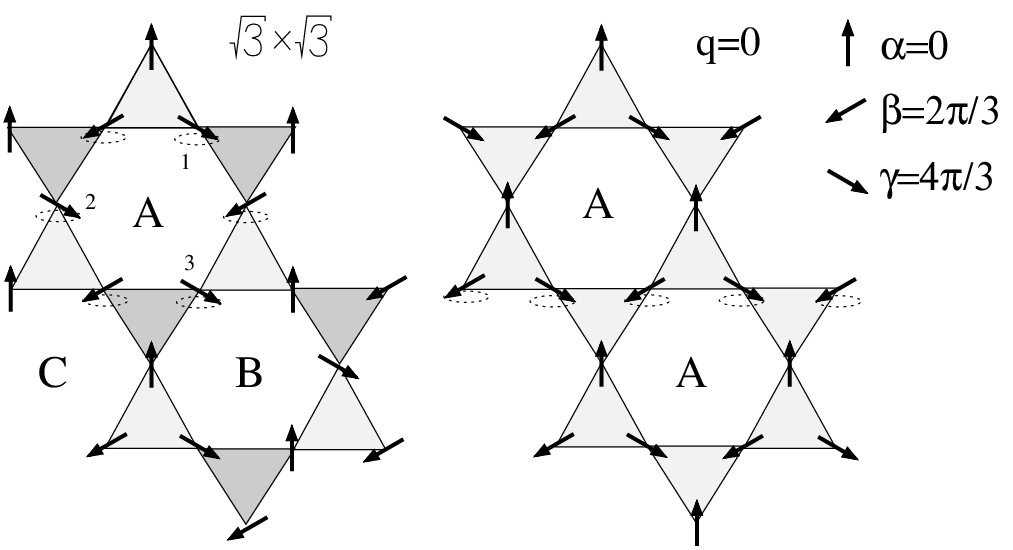}}
\end{center}
\caption[klassischer Grundzustand des \kagome-Gitters]{ \label{t8_cl_gs}
Two variants of the GS 
of the classical HAFM  on the \kagome lattice (T8): the \stateX$ $ state (left)
and the \stateO$ $ state (right). The dotted ellipses show further
degrees of freedom of the highly degenerate  classical GS.
The Wigner-Seitz geometrical \UC\ contains three sites (A1, A2, A3).
The light and gray triangles symbolize different chiralities.
 
}
\end{figure}

\begin{figure}

\vspace{1cm}
\begin{center}
\myframe{\includegraphics[height=7.2cm]{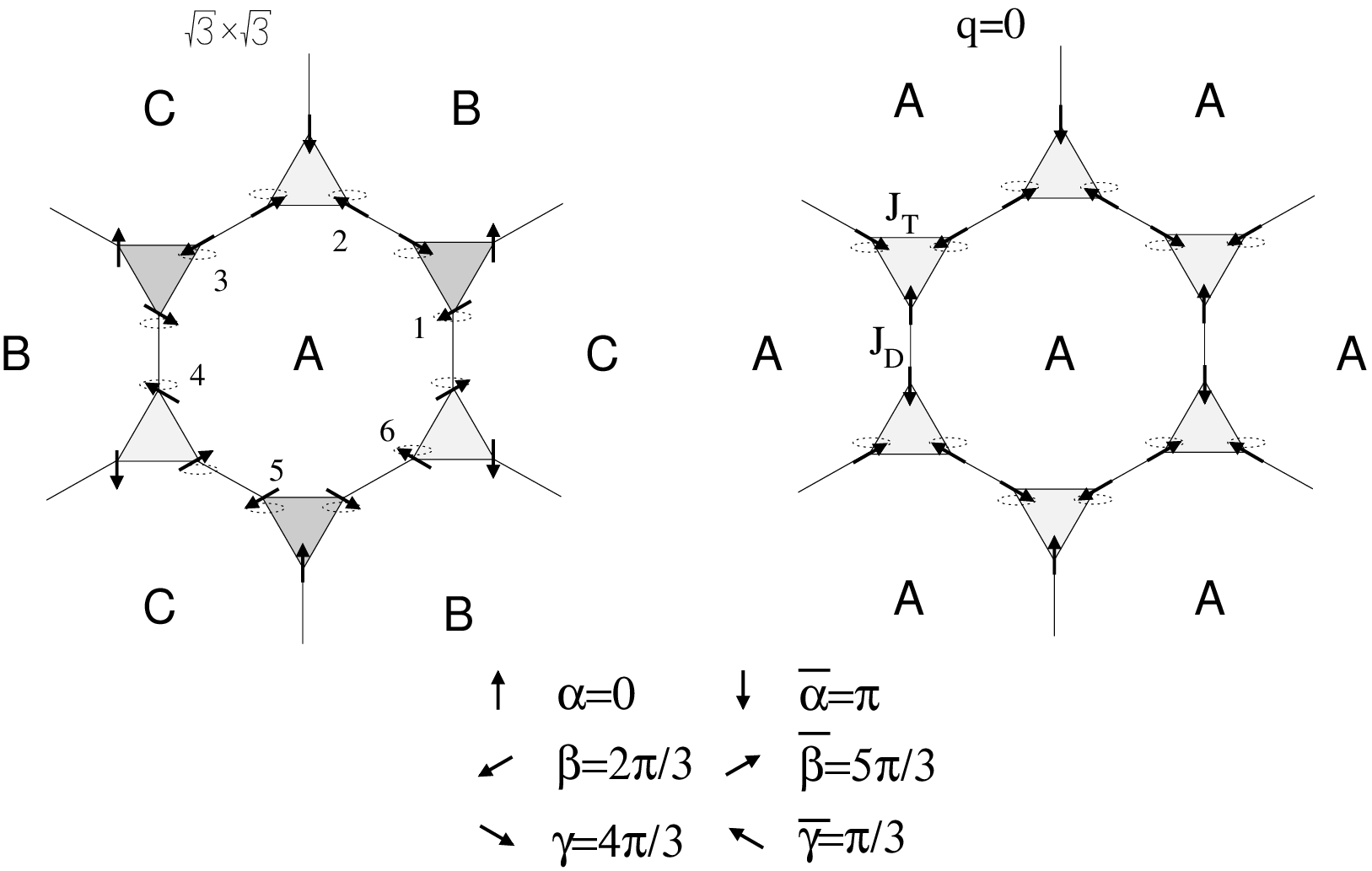}} 
\end{center}
\caption[klassischer Grundzustand des  star-Gitters]{ \label{t9_cl_gs}
Two variants of the GS 
of the classical HAFM  on the star lattice (T9): the \stateX$ $
state (left) and the \stateO$ $ state (right).
 The dotted 
ellipses show further
degrees of freedom of the highly degenerate  classical GS. 
The Wigner-Seitz geometrical \UC\ consists of six sites located on 
a hexagon (sites A1,..., A6).
The light and gray triangles symbolize different chiralities.
}
\end{figure}

Among the non-bipartite frustrated lattices the \Kagome\footnote{The name
\kagome stems from the Japanese language and means a bamboo-basket woven
pattern \cite{kago}.}  and the star 
lattice
play an exceptional role. The \kagome lattice is strongest frustrated (as
strong as the triangular lattice) and has low coordination number $z=4$, cf.\
Fig.\ \ref{t8_cl_gs}. 
It can be obtained by a 1/4 site depletion or alternatively by a 1/3 bond
depletion (with an appropriate subsequent distortion)
of the triangular lattice.  
Whereas the triangles in the \kagome lattice are corner
sharing, they are separated by a dimer in the star lattice.
Its degree of frustration  is 
less than for the \kagome lattice but it has an even lower  coordination
number $z=3$ and two non-equivalent NN bonds $J_D$ and $J_T$, cf.\ Fig.\
\ref{t9_cl_gs}. As indicated in Fig.\ \ref{fig_relat}, the star lattice  
can be obtained by a 2/5 bond depletion of the maple-leaf (T4) 
or alternatively by a 1/4 bond depletion of the bounce lattice 
(T7) with an appropriate subsequent distortion.
Both the \kagome and the star lattices are
characterized by  strong quantum fluctuations.  

After realizing in the early nineties that the quantum GS of the HAFM on the 
triangular lattice is \Neel ordered the
HAFM on the \kagome lattice came into the focus of interest as a hot 
candidate for a 2D quantum spin system with an exotic non-\Neel ordered GS 
\cite{bernu93,lecheminant97,waldtmann98dd,farnell01,zeng90,harris92,chubukov92,leung93,yang93,mila98dd,waldtmann00,yu00,mambrini00dd,shankar00,sindzingre00,maleyev02,canals02}. 
Indeed, most of the recent investigations are in favor of
a quantum paramagnetic GS,
although its nature is far from being well understood. 
A possible physical realization of the  \kagome HAFM is SrCrGa oxide,
which is, however, a layered \kagome HAFM with spin $3/2$
\cite{ramirez94dd,ramirez00dd}. A novel spin-1/2 \kagome like HAFM has been
found recently in volborthite Cu$_3$V$_2$O$_7$(OH)$_2 \cdot 2$H$_2$O
\cite{hiroi01dd}. By contrast,
the spin half HAFM on the star lattice has not been considered in the
literature so far, nor is a physical realization currently known. However, we
mention that a projection of the three-dimensional non-frustrated
magnetic compound green dioptase Cu$_6$Si$_6$O$_{18}\cdot 6$H$_2$O 
has the shape of the star lattice \cite{gros02dd}.

The geometric \UC\ of the \kagome (star) lattice  contains three (six) 
sites and the underlying Bravais lattice is a triangular one (cf.\ Figs.\
\ref{t8_cl_gs} and \ref{t9_cl_gs}).
The classical GS for the \kagome lattice was studied in  
\cite{chalker92,huse92,reimers93}. 
In analogy to the
triangular lattice the angle between neighboring spins    
is 120$^\circ$. Its energy per bond is $E_0^{\mathrm{class}}/\mathrm{bond}=-s^2/2=-0.125$. 
However, in contrast to the triangular lattice 
there is a non-trivial infinite degeneracy of the classical GS typical for
a classical HAFM with corner-sharing triangles.

In the classical GS of the  star lattice the two non-equivalent NN bonds carry
different NN spin correlations: the angle between neighboring spins    
on dimer bonds $J_D$ is 180$^\circ$, whereas the angle 
on triangular bonds $J_T$ is 120$^\circ$. Its energy per bond is 
$E_0^{\mathrm{class}}/\mathrm{bond}=-2s^2/3=-0.1667$. 
Although the star lattice is not built by corner-sharing triangles,  
the classical GS for this lattice also exhibits 
a non-trivial infinite degeneracy very similar to that of the \kagome
lattice.

Two particular variants of the classical GS characterized by
a certain wave vector 
are shown in Figs.\ \ref{t8_cl_gs} and \ref{t9_cl_gs}. 
The states on the left side of Figs. \ref{t8_cl_gs} and \ref{t9_cl_gs} 
exhibit  the same
symmetry as  the classical GS for the triangular lattice having a magnetic
\UC\ three times as large as the geometric \UC\ (so-called
\stateX $\mbox{}$ state).
The states on the right side of Figs. \ref{t8_cl_gs} and \ref{t9_cl_gs} 
have the same translational symmetry as the lattice (so-called
$q=0$ state) and therefore the magnetic and the geometric \UC\ are
identical. Both states are highly degenerate as indicated by the
dotted elliptic lines at the top of spins.  
%

Let us consider the order parameter (\ref{mdef}) for the classical GS.
If we take the perfect ordered planar \stateX $\mbox{ }$ and  \stateO 
$\mbox{ }$ state, then we get for both lattices
$m^+_{\mathrm{class},\sqrt{3}\times\sqrt{3}}=
m^+_{\mathrm{class},q=0}=\frac{1}{2}\sqrt{2/3}=0.40825$.
However, one has to take into account the high degeneracy of the GS.
In order to average over these degenerate states we performed numerical
calculations of the ground states for classical systems of up to $N=432$ sites.
The numerical results lead to the conclusion that for large $N$ we have
$m^+_{\mathrm{class,averaged}}=0.25$ for both lattices. 
This corresponds to a GS phase with decoupled spins for larger 
spin-spin separations.

In the quantum case the largest \kagome lattice considered 
has $N=36$ sites and is defined by  the edge vectors 
$(4,2);(2,4)$. 
It has GS energy per bond $E_0/\mathrm{bond}=-0.219188$, spin gap  
$\Delta=0.164190$  and square of the order parameter $(m^+)^2= 0.059128$.
We mention that the result for $E_0/\mathrm{bond}$ was already given in
\cite{waldtmann98dd,leung93}.
\label{kag36orderParam}

The largest  star lattice considered has $N=36$ sites and is defined by 
the edge vectors
$(2,0);(1,3)$. 
It has GS energy per bond $E_0/\mathrm{bond}=-0.310348$, spin gap  
$\Delta=0.480343$ and square of the order parameter $(m^+)^2=0.082299$.
Note that the value of the spin gap is particularly large. The only
$N=36$ lattice having a larger spin gap is the trellis lattice for which,
however, the large spin-gap is most likely a finite-size artifact due to the
incommensurate structure of the states.
The two non-equivalent NN correlation functions for
$N=36$ are $\langle {\bf S}_i {\bf S}_j\rangle_{J_T}= -0.170339$ 
(that is weaker than for the \kagome and the triangular lattice)
and $\langle {\bf S}_i {\bf S}_j\rangle_{J_D}=-0.590367$.\footnote{Note that
these values are averaged values,
since the $N=36$ star lattice does not have all lattice symmetries of the 
infinite lattice. As a result one has to average over 
two different values.} 
We mention that the NN correlation 
$\langle {\bf S}_i {\bf S}_j\rangle_{J_D}$
is the strongest correlation we found in all lattices, thus indicating a strong
tendency to form local singlets on the $J_D$ bonds.

\begin{figure}
\vspace{0.9cm}
\begin{center}
 \setlength{\unitlength}{1mm}
 \begin{picture}(55,70)
 \put(0,8){ \myframe{\hspace{-3.9cm}\includegraphics[height=7.5cm]{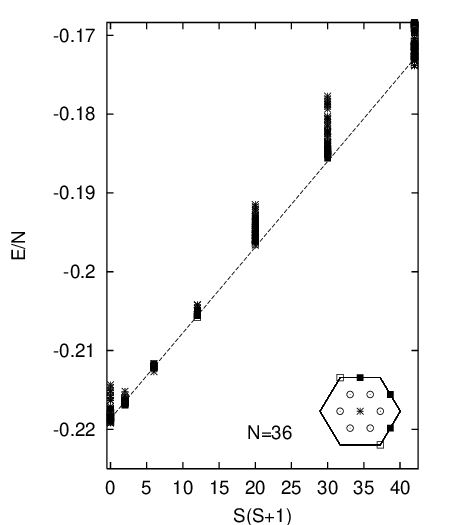}} }
 \put(5,3){\hspace{-2.cm}\makebox(0,0)[lc]{a}}
 \put(-20,65){ \myframe{\includegraphics[height=0.8cm]{pict8.eps}}}
 \put(2,8){ \myframe{\hspace{+2.2cm}\includegraphics[height=7.5cm]{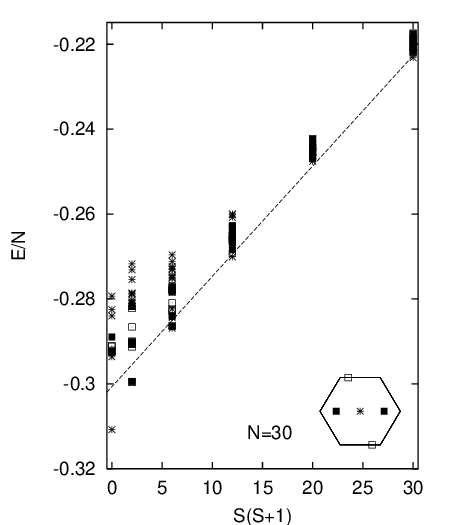}} }
 \put(5,3){\hspace{+7.2cm}\makebox(0,0)[lc]{b}}
 \put(+44,65){ \myframe{\includegraphics[height=0.8cm]{pict9.eps}}}
 \end{picture}
\end{center}
\caption[sq32 Pisa]{ \label{t8_9_erg}
Low-energy spectrum for HAFM on the \kagome (T8) and on the  star (T9) lattice 
(the insets show the {\bf k} points 
in the  Brillouin
zone).\\
a: \kagome with  $N=36$. \hspace{0.3cm}
b:  star with $N=30$. 
}
\end{figure}

The spectra of both lattices are shown Fig.\ \ref{t8_9_erg}. For both spectra
it is obvious
that the lowest states $E_{\min}(S)$ are not well described by 
eq.\ (\ref{eqtos}). In particular, the lowest states belonging to $S=0$ and
$S=1$ deviate significantly from a straight line. 
We do not see well separated QDJS as well as 
spin-wave excitations. Furthermore, the symmetries of the lowest states in
each sector of $S$ cannot be attributed to the classical \stateX$ $ 
or \stateO$ $ ground states in general.
The \kagome lattice is an exceptional case in 
that a large number of non-magnetic singlets fill the singlet-triplet gap. 
For instance for $N=27$ there are 153 \cite{waldtmann98dd}
and for $N=36$ on finds 210 \cite{lhuillier00dec} 
non-magnetic excitations within the spin gap 
and in the thermodynamic limit possibly a gapless singlet continuum. 
This unusual number of low-lying singlets 
is attributed to the non-trivial huge degeneracy of the classical GS.  
By contrast, the star lattice does not show low-lying singlets.
This can be attributed to the special property of the quantum GS to form
local singlets on the $J_D$ bonds which somehow makes the singlet GS of
the star lattice exceptional. As a consequence, the quantum GS of the star
lattice has lowest energy per bond among all frustrated lattices and is 
well separated from the excitations. Especially the first singlet excitation
has comparably high energy.

We mention that a detailed discussion of the spectrum for the \kagome lattice
was given in \cite{lecheminant97,waldtmann98dd}.

For the finite-size extrapolation of the GS energy (Fig.\ \ref{vgl_ex_e12}b),
the gap and the order parameter
(Fig. \ref{vgl_ex_m12}b) we use finite lattices  of size
$N=12$, $18$, $24$, $30$ and $36$ (\Kagome)
 and of $N=18$, $24$, $30$ and $36$ (star). 
The extrapolation leads to the following results for the \kagome lattice: 
\bit
\item GS energy per bond: $E_0/\mathrm{bond}= -0.2172 $ \\
(for comparison: SWT \cite{harris92}:  $E_0/\mathrm{bond}  = -0.2353$;  
 former exact diagonalization ($N=9,\ldots,21$) \cite{zeng90}:  $E_0/\mathrm{bond}=
-0.217$;
 CCM \cite{farnell01}: $E_0/\mathrm{bond}= -0.2126$; Green's function decoupling
\cite{yu00,canals02}:
$E_0/\mathrm{bond}= -0.215$);
\item
spin gap:
$\Delta= 0.0397$;
\item
order parameter:
$m^+= 0.000  \sim 0.0 \; m^+_{\mathrm{class}}$.
\eit
In fact, the extrapolation gives the unphysical value $m^+=-0.0146<0$ (cf.
Fig.\ \ref{vgl_ex_m12}b). We interpret this as vanishing order parameter.

For the  star lattice we obtain: 
\bit
\item GS energy per bond: $E_0/\mathrm{bond}= - 0.3093 $; 
\item spin gap:
$\Delta= 0.3809$;
\item
order parameter:
$m^+= 0.0385  \sim 0.094 \ldots 0.150 \; m^+_{\mathrm{class}} $ 
(the first value corresponds to $m^+_{\mathrm{class}}=0.40825$ of the 
perfect ordered planar \stateX $\mbox{ }$ and  \stateO 
$\mbox{ }$ classical GS, see Fig.\ 
\ref{t9_cl_gs}, whereas the second value corresponds to
$m^+_{\mathrm{class,averaged}}=0.25$ obtained by averaging over all
degenerate classical ground states).
\eit
The extrapolated spin gap for the \kagome lattice is small but finite and 
corresponds  to the values reported in the literature 
(see e.g. \cite{lhuillier01sep}), but we should remark  that the 
existence of a spin gap at
all is not a fully secure statement.

%

For both lattices the exact diagonalization data yield indications for a
quantum paramagnetic GS. For the \kagome lattice this statement is
known from detailed studies by
C.~Lhuillier, H.-U.~Everts and coworkers as well as other 
groups published over the last 10 years. However, the star lattice represents 
a new example for a quantum HAFM  on a
uniform 2D lattice without semi-classical GS ordering. 
We emphasize that the quantum paramagnetic GS for the star lattice is
different in nature to the quantum GS for the \kagome lattice. 
The quantum GS for the star lattice is characterized by an extremely strong NN
correlation on the dimer bonds (more than 60\% larger than the NN correlation 
of the honeycomb lattice having the same coordination number $z=3$)
and a weak NN
correlation on the triangular bonds (only about 30\% of the NN dimer
correlation and significantly weaker than the triangular NN correlation 
of the \kagome and the triangular lattices).
The singlet-triplet spin gap is particularly large (about ten times larger
than that for the \kagome lattice). 
Although the classical GS exhibits a huge non-trivial degeneracy, remarkably 
one does not find low-lying singlets within this large spin gap,  
rather the first singlet excitation  is well above the lowest triplet state.
The low-lying spectrum as a whole resembles the spectrum of weakly coupled 
dimers \cite{phd}.
All these features support the conclusion that the quantum GS of the HAFM on
the star lattice is dominated  by local singlet pairing and represents a
so-called valence-bond crystal state (see also section \ref{summary_1}).

\subsection{Summary and comparison}
\label{summary_1}
Based on extensive exact diagonalization studies and on available results in
the literature we 
discuss the GS ordering of the spin half HAFM
on the 11 uniform Archimedean tilings in two dimensions. 
Of course we are not able to
clarify all aspects of the GS properties of these quantum many-body
systems. Nevertheless the comparative discussion of the 11 lattices leads to
conclusions on the influence of lattice structure on GS magnetic ordering in
two dimensions and this way on the existence or absence of 
semi-classical LRO in these systems.    

The HAFM has been already studied
intensively in the literature for some of these lattices 
(square (T2), triangular (T1), honeycomb (T3),
\kagome (T8), SrCuBO (T6), CaVO (T11)) and the physical 
picture seems to be more or less clear for those lattices.  
For some other lattices (SHD (T10),  maple-leaf (T4) and  trellis (T5)) 
only a few results are available
in the literature so far and 
the conclusions on the GS ordering are less reliable.
The HAFM on the star lattice (T9) as well as on the bounce lattice (T7)
has not been studied till now. 

\begin{figure}
\begin{center}
 \setlength{\unitlength}{1mm}
 \begin{picture}(110,80)
 \put(-10,5){\myframe{\includegraphics[height=7.5cm]{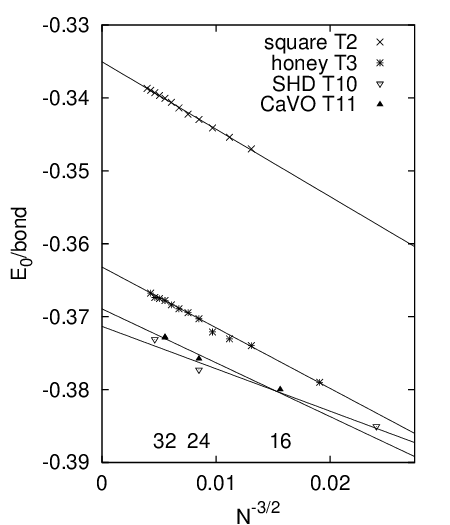}} }
 \put( 10,0){\makebox(0,0)[lc]{a}}
 \put( 55,5){\myframe{\includegraphics[height=7.5cm]{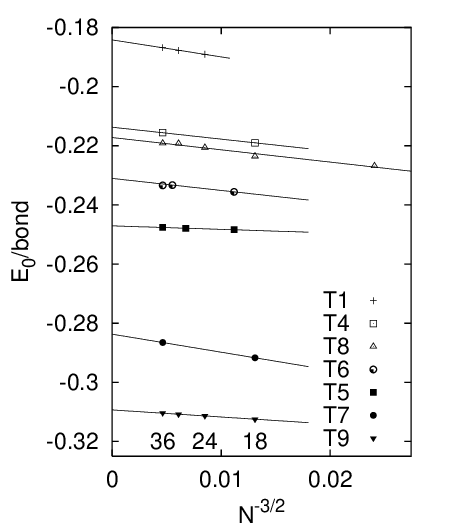}} }
 \put( 75,0){\makebox(0,0)[lc]{b}}
 \end{picture}
\end{center}
\caption{ \label{vgl_ex_e12}
 Finite-size extrapolations of GS energy per bond $E_0/\mathrm{bond}$,
 a -  bipartite lattices, b - frustrated lattices.
}
\end{figure}

\begin{figure}
\begin{center}
 \setlength{\unitlength}{1mm}
 \begin{picture}(110,80)
 \put(-10,5){\myframe{\includegraphics[height=7cm]{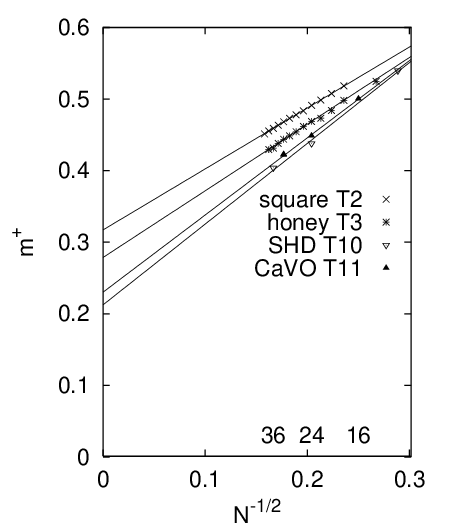}} }
 \put( 10,0){\makebox(0,0)[lc]{a}}
 \put( 55,5){\myframe{\includegraphics[height=7cm]{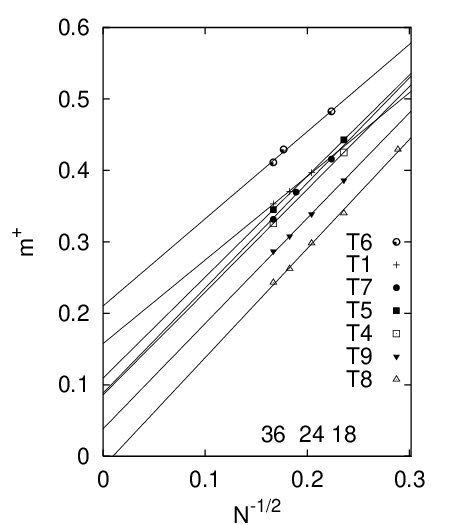}} }
 \put( 75,0){\makebox(0,0)[lc]{b}}
 \end{picture}
\end{center}
\caption{ \label{vgl_ex_m12}
 Finite-size extrapolations of $m^+$,
 a -  bipartite lattices, b - frustrated lattices.
}
\end{figure}

Let us summarize  the results of the preceding sections:   
The GS of the spin  half HAFM on the bipartite (i.e. non-frustrated) lattices  
is semi-classically \Neel ordered. The reduction of the order parameter by
quantum fluctuations  depends on the coordination number and on the
competition of non-equivalent NN bonds (cf. table \ref{tabmag}).
The low-energy spectra exhibit
some typical features for magnetic systems with semi-classical order, 
namely well separated 
quasi-degenerate joint states (QDJS) 
with symmetries belonging to the classical GS
ordering.
Another indication for semi-classical ordering is the disappearance of the 
spin gap in the thermodynamic limit. We find, at least for the lattices
with not too large unit cells, indications for a vanishing spin gap. 
However, the
finite-size extrapolation of the spin gap is less reliable than for the
magnetization (see section \ref{scaling})   and therefore we do not consider
the spin gap as a main criterion for the existence of semi-classical LRO. 
The comparison of the finite-size behavior of the GS energy 
shown in Fig.\ \ref{vgl_ex_e12} shows that the  
extrapolation coefficient $A_3$ (cf.\ eq.\ (\ref{gs_scale})) 
for the bipartite lattices is largest in
agreement with long-ranged spin-spin correlations. We mention that a
suppression of semi-classical LRO in bipartite lattices can appear in systems
with NN bonds of different strength this way 
increasing the competition of non-equivalent NN bonds (see section \ref{qpt}).

\begin{table}
\caption[Vergleich Magnetisierung]{ \label{tabmag}
Comparison of the ground-state energy per bond $E_0/\mathrm{bond}$ and the 
order parameter $m^+$ (eq.\ (\protect{\ref{mdef}})) of the spin half HAFM
obtained by finite-size extrapolation (see text). In order to see the effect of
quantum fluctuations, we present $m^+$ scaled by its corresponding classical
value $m^+_{\mathrm{class}}$. Furthermore, we show the coordination number $z$
and indicate, whether all NN bonds are equivalent or not by EQ and NEQ,
respectively. For the star lattice (last row) the first value corresponds 
to $m^+_{\mathrm{class}}$ of the 
perfect ordered planar \stateX $\mbox{ }$ and  \stateO 
$\mbox{ }$ classical GS, see Fig.\ 
\ref{t9_cl_gs}, whereas the
second value corresponds to $m^+_{\mathrm{class}}$ averaged over all
degenerate classical ground states.

}
\begin{center}
\begin{tabular}{l|c|c|c|c} 
 tiling                   & $z$            & $\;$ NN bonds $\;$   &$E_0/\mathrm{bond}$   &$m^+/m^+_{\mathrm{class}}$    \\ \hline
{\it bipartite}           & \hspace{1cm}   & \hspace{1cm}   &        &            \\ 
 square (T2)              & 4              & EQ             &$-0.3350$  & 0.635      \\ 
 honeycomb (T3)           & 3              & EQ             &$-0.3632$  & 0.558      \\ 
 CaVO (T11)               & 3              & NEQ            &$-0.3689$  & 0.461      \\ 
 SHD  (T10)               & 3              & NEQ            &$-0.3713$  & 0.425      \\ \hline
 {\it frustrated}         &                &                &        &            \\ 
 SrCuBO  (T6) $\;$        & 5              & NEQ            &$-0.2310$  & 0.456      \\ 
 triangular (T1)          & 6              & EQ             &$-0.1842$  & 0.386      \\ 
 bounce (T7)              & 4              & NEQ            &$-0.2837$  & 0.286      \\ 
 trellis (T5)             & 5              & NEQ            &$-0.2471$  & 0.222      \\ 
 maple-leaf (T4)          & 5              & NEQ            &$-0.2171$  & 0.218      \\ 
 \kagome (T8)             & 4              & EQ             &$-0.2172$  & 0.000      \\ 
 star (T9)                & 3              & NEQ            &$-0.3093$  & 0.094 \ldots 0.150      \\ 
\end{tabular}
\end{center}
\end{table}

The situation for the frustrated lattices is more complex. Some of the   
criteria for semi-classical LRO might be weaker pronounced.  
For the HAFM on the \kagome and on the star lattice we find evidence for
a quantum  paramagnetic GS whereas for the other
frustrated lattices  there are indications for semi-classical LRO.
Although the order parameter $m^+$ is additionally weakened by the interplay of
quantum fluctuations and frustration the extrapolated values of $m^+$
remain finite (between 22\% and 45\% of the classical values) for 
the tilings T1,T4,T5,T6,T7. It vanishes however for the \kagome lattice and
is at least very small for the star lattice (see table \ref{tabmag}).    
Except for the \kagome and the star lattices the low-energy spectra exhibit
some typical features for magnetic systems with semi-classical order, 
namely well separated QDJS with symmetries belonging to the classical GS
ordering.
The comparison of the finite-size behavior of the GS energy 
(Fig.\ \ref{vgl_ex_e12}) shows the   
smallest extrapolation coefficient $A_3$ for the \kagome and the star
lattice being in agreement with short-range
spin-spin correlations. Although the extrapolation coefficient $A_3$ is very 
small for 
the trellis lattice, too, we interpret this as a particular finite-size effect
due to the incommensurate structure of the classical GS.

%
%
%

\begin{figure}
\begin{center}
\myframe{\includegraphics[height=5.5cm]{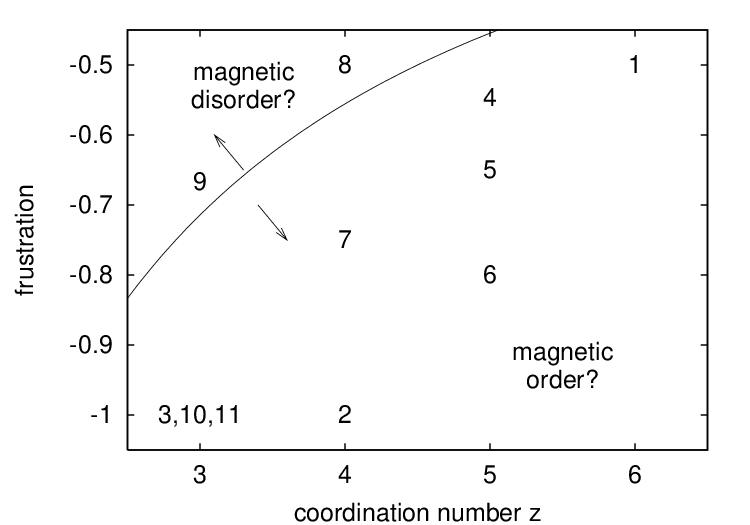}}
\end{center}
\caption{ \label{vgl_ex_e0}
Border line between semi-classical magnetic order and quantum magnetic
disorder in a parameter space spanned by 
 frustration (classical GS energy per bond divided by $s^2$, see section
\ref{geometry})  and 
coordination number $z$. The numbers  indicate the location of the 
Archimedean tilings  in this parameter space.
}
\end{figure}

We conclude that the interplay of lattice structure and quantum fluctuations
may lead to a non-classical quantum paramagnetic singlet GS for frustrated
lattices with low coordination number and strong frustration, 
i.e.\ for the \kagome and the star lattice (see Fig.\ \ref{vgl_ex_e0}). 
Although extensive studies have been performed for the \kagome lattice
\cite{bernu93,lecheminant97,waldtmann98dd,farnell01,zeng90,harris92,chubukov92,leung93,yang93,mila98dd,waldtmann00,yu00,mambrini00dd,shankar00,sindzingre00,maleyev02,canals02},
the spin half HAFM on the star lattice is considered in this article for the
first time. By contrast with all the other lattices, the
\kagome and the star lattice show a huge non-trivial degeneracy of the
classical GS due to strong frustration.

Although there is no semi-classical GS order for both lattices,
the nature of both quantum ground states is basically different
in the quantum case.
We argue that the origin for this difference lies in the existence of 
non-equivalent NN
bonds in the star lattice whereas all NN bonds in the \kagome lattice are
equivalent.
That leads also to significant differences in the low-lying spectrum of both
lattices. 
The \kagome lattice has probably a finite spin gap, but within this spin gap
a large number (increasing exponentially with system size) 
of low-lying singlets appear\cite{lecheminant97,waldtmann98dd,mila98dd} 
which seem to be a remnant of the non-trivial classical GS degeneracy.
However, the HAFM on the star lattice has a particularly 
large spin gap but also
a well pronounced singlet-singlet gap (even larger than the spin gap) 
which is in accord with a GS 
dominated by local singlet pairing on non-equivalent NN $J_D$ bonds assisted by
frustration. As a consequence, the huge classical GS degeneracy has no 
remnant in the spectrum of the quantum model.
We mention that the checkerboard (planar pyrochlore) lattice is another
example, where the non-trivial classical GS degeneracy   
does not lead to a continuum of low-lying singlets and the ground-state is
most likely a valence bond crystal (see e.g.\ \cite{FMSL03,BrHo02}).
Furthermore, examples are known that many low-lying non-magnetic
excitations within the spin gap may appear although the classical GS
is not non-trivially degenerate \cite{sierp01,LiMMSL00}.

For all the other bipartite and frustrated lattices the quantum fluctuations
seem to be not strong enough to destroy the classical order.
However, we should again emphasize that our conclusions about
semi-classical LRO possesses some uncertainty, in particular for the trellis,
maple-leaf and bounce lattices. 

The above presented study provides some criteria for the appearance
of novel quantum ground states in 2D spin systems.
Although only for a few of the lattices under consideration 
direct realizations in real materials have been found till now,  
in several cases slightly modified models, e.g.  models with NN couplings
of non-equal strength or with inclusion of next-nearest neighbor couplings, 
are appropriate  for the description of real magnetic substances. 

At the end of this paragraph we will classify the magnetic ordering 
on the 11 Archimedean tilings using the four basic 
types of low-energy physics in 2D isotropic quantum 
antiferromagnets proposed and described 
recently by  Lhuillier, Sindzingre, Fouet and  Misguich
\cite{lhuillier00dec,lhuillier01sep,lhuillier01oct,lhuillier02dec,lhuillier03}.
The first type of GS phases is the semi-classical LRO (collinear or
noncollinear). Most of the lattices belong this class, namely all
bipartite lattices (T2,T3,T10,T11) but also the frustrated tilings
(T1,T4,T5,T6,T7). The GS of the 
HAFM on these lattices breaks the $SU(2)§$
symmetry. The low-lying excitations are gapless Goldstone modes (magnons).     
As discussed above, the order parameter is reduced by quantum fluctuations.
The three other types of GS phases, namely the so-called valence bond crystal, 
type I spin liquid and type II spin liquid are purely quantum.

The so-called valence bond crystal is a phase characterized by the formation
of local singlets with high binding energy 
built by  an even number  of spins (most likely by two or four spins)
connected by  NN bonds (singlet `valence bonds').
The correlation between the singlets
is weak leading to a fast exponential decay of the spin pair correlation to zero.
The GS is a rotationally invariant singlet of the total spin without
$SU(2)$ symmetry breaking. However, breaking of translational symmetry of the
lattice is possible but not necessary.
The valence bond crystal possesses long-range 
singlet-singlet (dimer-dimer or plaquette-plaquette) 
correlations.
All excitations above the GS are gapped leading to an exponential (i.e.
thermally activated)
low-temperature behavior of the specific heat $c$ and of 
the susceptibility
$\chi$.
A candidate for such a GS phase is the
HAFM on the star lattice. For this lattice the possible  
dimer-dimer LRO would fit to the lattice geometry. Another candidate is  
 the
$J_1-J_2$ model on the square lattice, widely discussed in the literature
(see \cite{capriotti01,kotov00dd} and references therein), where the 
valence bond crystal phase would break the translational symmetry of the
lattice. 
The type I spin liquid has some similarity to the valence bond crystal. 
It has  
also  a rotationally invariant singlet GS without
$SU(2)$ symmetry breaking, it has a fast exponential decay of the 
spin pair correlation to zero and gapped excitations leading to 
thermally  activated 
low-temperature behavior of $c$ and $\chi$. However,  
the GS does not possess singlet-singlet 
long-ranged correlations but is likely
to be characterized by short ranged resonating valence bonds.
There is no good candidate for this phase
among the  Archimedean tilings. But this phase might be realized in the
$J_1-J_2$ model on the honeycomb lattice \cite{fouet01}. 
The type II spin liquid  has  
also  a rotationally invariant singlet GS, 
a fast exponential decay of the 
spin pair correlation to zero and no long-ranged singlet-singlet 
correlations. 
The spin gap $\Delta$ to the first triplet
excitation is finite giving rise to a  
thermally activated 
low-temperature behavior of the susceptibility $\chi$. However, there is a
gapless continuum of singlets which could be described by a family of 
short-ranged valence bond states 
\cite{mambrini00dd}   the number of which is exponentially growing 
with size $N$.
This gapless continuum implies that the system has a zero-temperature 
residual
entropy and that the low-temperature specific heat is not thermally
activated.
The best candidate for this  type of spin liquid is 
the spin half HAFM on the \kagome lattice.

\section{Quantum phase transitions in 2D HAFM - the 
 CaVO $J-J'$ model and the Shastry-Sutherland
model} \label{qpt}

Phase transitions have been a subject of great interest to physicists over
many decades. Besides thermal phase transitions, 
the so-called quantum phase transitions (or zero-temperature
transitions) have started to attract a lot of attention (see chapter by
\ChapQPT\ in this book).
For zero-temperature order-disorder transitions 
we basically need the interplay between the interparticle interactions and
quantum fluctuations. 
Canonical models to discuss quantum phase transitions are quantum spin 
models. As discussed above the HAFM on  most of the  2D lattices possesses
semi-classical LRO in the GS, but the interplay of quantum fluctuations and 
strong competition  between bonds may suppress this order. 
The competition may appear 
either as frustration or by non-equivalent NN bonds or a
combination of both. Indeed, the strength of this 
competition may serve as the
control parameter of a zero-temperature  order-disorder transition.
It can be tuned by changing the relative magnitude of
non-equivalent NN bonds or by introducing next-nearest neighbor bonds.
The Archimedean tilings therefore represent a wide playground for the
investigation of zero-temperature transitions.  

 A generic model of a frustrated HAFM widely discussed in the literature 
(see, e.g.,
Refs.~\cite{capriotti01,kotov00dd,doucot88,schulz92dd,ri93,sushkov94,oitmaa96,Bishop6,sorella00,sushkov01})
is the spin-half
$J_1-J_2$ model on the square lattice,
where the frustrating $J_2$ bonds plus quantum fluctuations are believed to
lead to a second-order transition from a N\'eel-ordered state to a
quantum paramagnetic state at about $J_2 \approx 0.38J_1$.
The properties of the latter state are still far from
being understood. One favored quantum phase for $J_2 \sim 0.5 J_1$ is
a valence bond crystal.
However, there are examples where frustration leads to a
first-order transition in quantum spin systems in contrast to a second-order
transition in the corresponding classical model (see, e.g., 
Refs.~\cite{krueger00,koga00a,xian95,richter98dp,koga00dd,lcdipla02}).

The competition between non-equivalent NN bonds melts the
semi-classical \Neel order by formation of local singlets. 
By contrast to frustration, which yields
competition in quantum as well as in classical systems, the local singlet
formation is a pure quantum effect. 
Both mechanisms may of course be mixed as, for instance, in CaV$_4$O$_9$ or
in SrCu$_2$(BO$_3$)$_2$ (see, e.g., Refs.~\cite{ueda96dd,albrecht96b,koga00a}).

\begin{figure}

\vspace{0.9cm}
\begin{center}
 \setlength{\unitlength}{1mm}
 \begin{picture}(55,70)
 \put(0,8){
\myframe{\hspace{-3.5cm}\includegraphics[height=6.4cm]{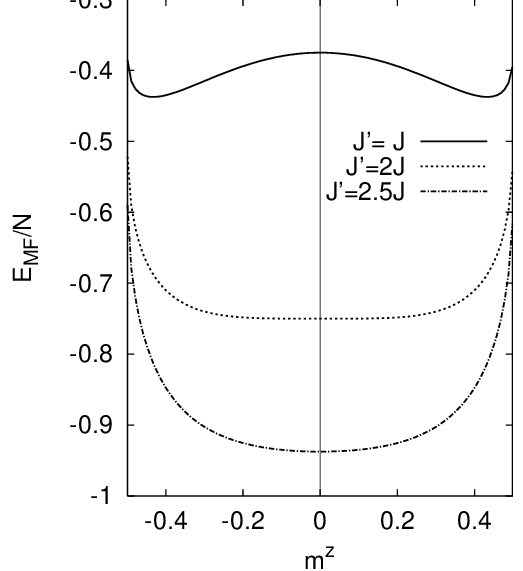}} }
 \put(5,3){\hspace{-2.cm}\makebox(0,0)[lc]{a}}
 \put(2,8){ \myframe{\hspace{+2.4cm}\includegraphics[height=6.4cm]{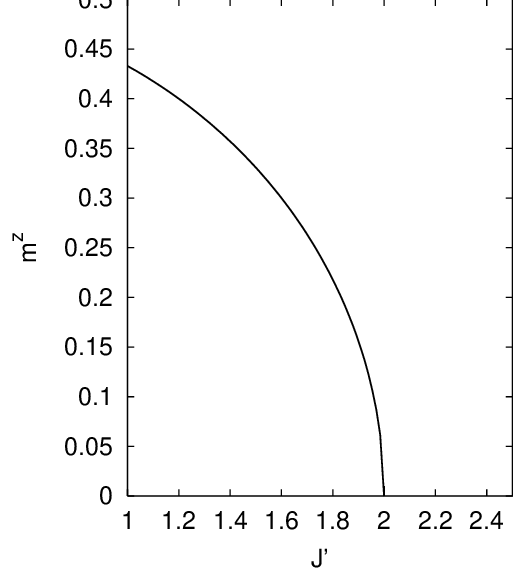}} }
 \put(5,3){\hspace{+7.2cm}\makebox(0,0)[lc]{b}}
 \end{picture}
\end{center}
\caption[t6_32]{ \label{landau} Mean field results 
for the $J-J'-$HAFM on the CaVO lattice. \\
a: Energy versus order parameter  \hspace{0.3cm} 
b: Order parameter versus $J'$
 \hspace{0.3cm}
}
\end{figure}

Let us first discuss a mean-field like approach to describe the 
continuous quantum phase
transition driven by local singlet formation.
To that end we study the HAFM
on the CaVO lattice having  two
non-equivalent NN bonds $J$ and $J'$, see Fig.\ \ref{cavo1}.
The uncorrelated mean-field state for N\'eel LRO is the two-sublattice 
N\'eel state
$\;  |\phi_{MF_1}\rangle  = | \hspace{-0.15cm} \uparrow \rangle|\hspace{-0.15cm}\downarrow \rangle 
|\hspace{-0.15cm}\uparrow \rangle|\hspace{-0.15cm}\downarrow \rangle |\hspace{-0.15cm}\uparrow \rangle
|\hspace{-0.15cm}\downarrow \rangle
\ldots  \;$ and for the dimerized singlet state it is
the rotationally invariant 
product state of local singlets of the two spins belonging to a $J'$ bond
$|\phi_{MF_2}\rangle  = 
\prod_{\{i,j\}_{J'} } \left\{|\uparrow_{i} \rangle|\downarrow_{j} \rangle
-|\downarrow_{i} \rangle|\uparrow_{j} \rangle \right\} /\sqrt{2} \; ,$ 
where 
$i$ is a site in the sublattice $A$ and $j$ a site in sublattice $B$.
In order to describe the transition between both states, we consider 
an uncorrelated
product state interpolating between $|\phi_{MF_1}\rangle$ and 
$|\phi_{MF_2}\rangle$ of the form\cite{krueger00,gros95}
\begin{equation} \label{psi_mf} 
|\Psi_{MF}(t)\rangle = \prod_{\{i,j\}_{J'} }
\frac{1}{\sqrt{1+t^2}} \left\{|\uparrow_i\rangle |\downarrow_{j}\rangle
-t |\downarrow_i\rangle |\uparrow_{j}\rangle\right\}.
\end{equation}
We have 
 $ |\Psi_{MF}(t=0)\rangle = |\phi_{MF_1}\rangle$  and 
 $ |\Psi_{MF}(t=1)\rangle = |\phi_{MF_2}\rangle$.
The minimal value of the energy is given by 
\begin{equation}
\label{en_{MF}}
   \frac{E_{MF}}{N} = 
\frac{ \langle\Psi_{MF}|H| \Psi_{MF}\rangle}{N} 
= \left\{\begin{array}{ll}
        -\frac{3J'}{8}-\frac{1}{16J}(2J-J')^2  & \quad J'\le 2 J\\
        -\frac{3J'}{8}            & \quad J'>   2 J.\\
    \end{array}\right.
\end{equation}
Furthermore, it is found that the sublattice magnetization $m^z$ has the
following form
\begin{equation} \label{m_{MF}}
  m^z=\langle
\Psi_{MF}|S_{i \in A}^z|\Psi_{MF}\rangle = \left\{\begin{array}{ll}
         \frac{1}{4J}\sqrt{(2J-J')(2J+J')} & \quad J'\le 2 J\\
         0                  & \quad J'>   2 J.\\
    \end{array}\right.
\end{equation}
Note that $m^z$ vanishes at a critical point $J'_c=2 J$, and that the critical 
index is the mean-field index $1/2$ (see Fig.\ \ref{landau}b). 
Using the relation between the variational parameter $t$ and the 
sublattice magnetization $m^z$ we find the relation 
$ E_{MF}/N  =      -\frac{1}{8}J'\hspace{0.1cm}-
\hspace{0.1cm}\frac{1}{4}J' 
\sqrt{1\hspace{0.1cm}-\hspace{0.1cm}4\left (m^z \right )^2} 
\hspace{0.1cm}-
\hspace{0.1cm} J\hspace{0.1cm}\left (m^z \right )^2 \hspace{0.1cm}
$ showing the typical behavior of a second-order transition, 
see Fig.\ \ref{landau}a.   
We can expand $E_{MF}$ up to the fourth order in $m^z$ near the critical point 
and find a Landau-type expression, given by 
$
 E_{MF}/N  \;=\;  -\frac{3}{8}J'  
\hspace{0.05cm}+\hspace{0.05cm}\frac{1}{2} 
\left (J' - 2J\right) \hspace{0.05cm} \left (m^z \right )^2 
\hspace{0.05cm} +
\hspace{0.05cm} 
\frac{1}{2} J'\hspace{0.1cm} \left (m^z \right )^4 \;$.

Although this mean-field like description  gives some qualitative insight
into the physics of the quantum phase transition for the CaVO lattice
 more elaborated
investigations \cite{troyer96,ueda96dd,gelfand96,troyer97} show that the 
quantum phase transition  to a rotationally invariant gapped dimerized GS 
phase takes 
place at $J'/J
\approx 1.7 $ and to the plaquette singlet GS phase  at $J'/J
\approx 0.9$.
The critical exponents of quantum
phase transitions driven by the competition of non-equivalent NN bonds 
in 2D quantum HAFMs are not the mean field exponents 
but  those of the
three-dimensional classical Heisenberg model \cite{troyer97,tomczak01a}.

%

Another interesting example for quantum phase transitions in spin systems
appears in
the Shastry-Sutherland model, i.e.\ the $J_1-J_2$ HAFM on the SrCuBO lattice
(T6). We will use in this section the Shastry-Sutherland representation 
(frustrated square lattice, see Fig.\ \ref{srcubo1}a, lower part).    
The classical GS of this model has two phases: The collinear \Neel
phase for $J_2 \le J_1$ and a spiral phase for $J_2 > J_1$ (cf.\ Fig.~1 in
Ref.~\cite{albrecht96}).
The transition between the two classical 
phases is of second order. 

\begin{figure}

\begin{center}
\myframe{\includegraphics[width=9 cm]{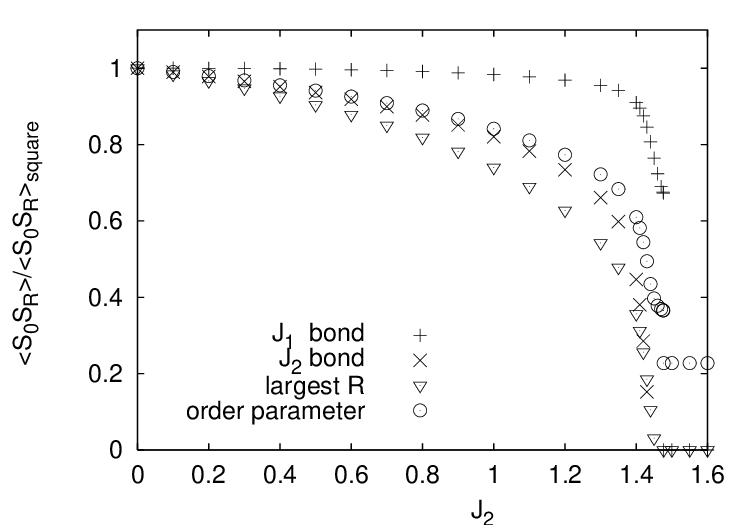}}
\end{center}
\caption[t6_32]{ \label{ZiZj}
Spin-spin correlation and square of sublattice magnetization (order
parameter) scaled by their
values for the square lattice ($J_2=0$)
for the Shastry-Sutherland model ($N=32$).
}
\end{figure}

For $J_2 \le J_1$ the physics of the quantum model is similar to that of 
the classical model, i.e., we have semi-classical \Neel order (see 
section \ref{secSrCuBO}). However, the quantum model exhibits new features
for stronger frustration $J_2 > J_1$.
Firstly, one finds that the collinear \Neel phase in the quantum model 
can survive into the region where classically it is already unstable
\cite{albrecht96,koga00a,lauchli02,weihong01}.
This effect is known as {\it order from disorder} \cite{villain,shender} and
is widely observed in quantum spin systems (see, e.g.
\cite{krueger00,schulz92dd,chubukov92}).

Secondly, one knows already from  Shastry and
Sutherland \cite{shastry81dd} that for large enough $J_2$ the
quantum GS is a
 rotationally invariant 
product state of local pair singlets 
$|\phi\rangle  = 
\prod_{\{i,j\}_{J_2} } [|\uparrow_{i} \rangle|\downarrow \rangle_{j}
-|\downarrow_{i} \rangle|\uparrow \rangle_{j}]/\sqrt{2}\quad$ 
(so-called orthogonal dimer state), 
where 
$i$ and $j$ correspond to those sites which cover the $J_2$ bonds. 
This  orthogonal dimer phase sets in at around 
$J_2^c \approx (1.45 \dots 1.48) J_1$
\cite{weihong99dd,miyahara99,hartmann00,koga00a,weihong01,lauchli02}. 
The nature of the transition to the dimer phase is still a matter of
discussion, although there are arguments that the transition is probably
of first order \cite{albrecht96,koga00a}.
In the region $1.2 J_1 \lesssim J_2 \lesssim 1.45 J_1 $ the main 
challenging question is
whether the system has an intermediate phase.  Candidates are quantum
spiral phases or more favorable a plaquette RVB like phase. 
Despite numerous investigations, a definite picture concerning the existence
and nature of an intermediate phase has not yet emerged.

We illustrate
such behavior discussed above by finite-lattice results ($N=32$) 
for the spin-spin correlation along the NN $J_1$ bond, along the diagonal
$J_2$ bond and for the largest separation $R=4$ available in the finite
$N=32$ Shastry-Sutherland 
lattice as well as for the square of 
sublattice magnetization ${\bar m}^2$ (cf.\ eq.~(\ref{ord_square})) 
shown in Fig.~\ref{ZiZj}. We have scaled the correlation functions and
the sublattice magnetization by their corresponding values for the square
lattice ($J_2=0$) for better comparison. 
The small changes in the correlation functions and
the sublattice magnetization are in agreement with
the survival of the collinear \Neel ordering up to about $J_2 \sim 1.2 J_1$.
Beyond $J_2 \sim 1.2 J_1$ the correlation functions change
drastically up to $J_2=1.4785J_1$, where
for $N=32$ the rotationally invariant orthogonal dimer state becomes
the GS. At this point the correlation functions and
the sublattice magnetization jump to their values of the orthogonal dimer
state. The behavior in the region $1.2 J_1 \lesssim J_2 < 1.4785J_1$
preceding the transition to the orthogonal dimer state
seems to be in accordance with the existence of an intermediate phase.

\def\abs#1{\left\vert #1 \right\vert}
\def\vec#1{{\bf #1}}
\def\Zed{\mathbb{Z}}
\def\DeltaI{\Delta_I}
\def\DeltaIc{\Delta_{I,c}}

\section{Magnetization process}

\label{m_h}

\begin{figure}[tb]
\begin{center}
\includegraphics[width=7 cm]{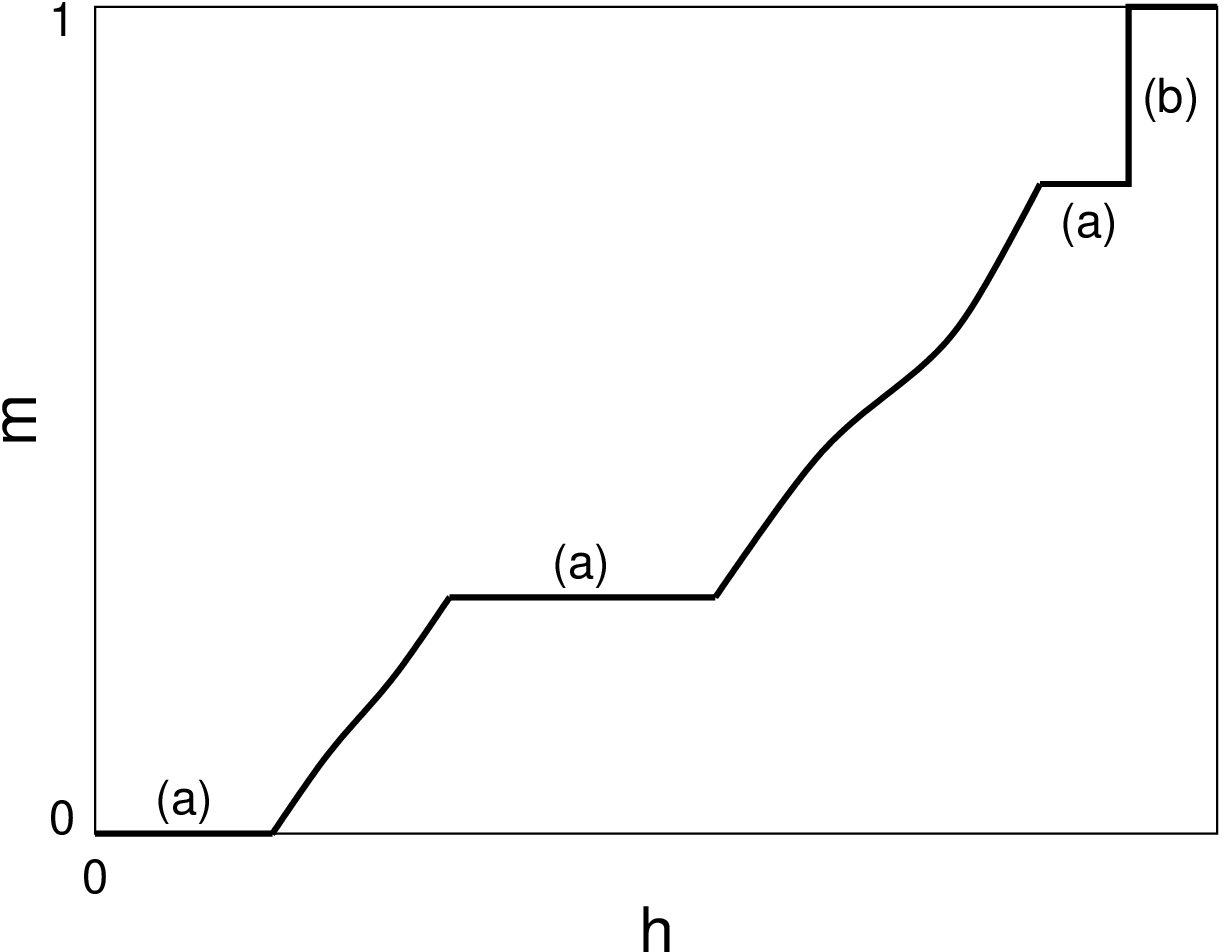}
\end{center}
\caption{
\label{figMschem}
Schematic magnetization curve illustrating some
plateaux (a) and a jump below saturation (b).}
\end{figure}

In this final section we discuss the effect of a uniform external magnetic
field on the models discussed so far. Once a small but finite magnetization is
created by the external field, spins can no longer align completely
antiparallel in the classical ground state even for a bipartite lattice.
Since this is similar to the effect of geometric frustration, one can regard
the magnetic field as introducing or enhancing frustration. One may
therefore expect that a strong external field can induce further interesting
quantum effects. In particular, we will discuss the quantum
phenomena which are sketched in Fig.~\ref{figMschem}:
\begin{itemize}
\item[(a)] Plateaux have a fixed magnetization $m$
   in a region of the applied magnetic field $h$.
   Note that a plateau with magnetization $m = 0$
   corresponds to a spin gap at zero magnetic field $h=0$. \\
   On a plateau, the magnetization $m$ typically assumes a
   (simple) rational fraction of its saturation value.
\item[(b)] Also some examples of jumps associated with a special
   degeneracy in the spectrum will be discussed in section \ref{secImag}.
\end{itemize}

Specifically we consider the Heisenberg antiferromagnet (\ref{ham}) in a uniform
external magnetic field $h$
\be
 H= \sum_{<i,j>} J_{ij} \vec{S}_i \vec{S}_j - h \sum_i S_i^z \, .
\label{hamField}
\ee
In the following we will focus on the zero-temperature magnetization process
of the Heisenberg antiferromagnet (\ref{hamField}) on the 11 Archimedean and
some related lattices. Some further aspects of two-dimensional $s=1/2$
antiferromagnets in an external field have been summarized e.g.\ in
\cite{lhuillier01sep,habil}.

In the present context it will sometimes be useful to allow for general
length $s$ of the local spin. One can also introduce an $XXZ$ anisotropy
as a prefactor $\DeltaI$ multiplying the $z$-$z$ interaction term in
(\ref{ham}).
Note that a magnetic field $h \ne 0$ already breaks the symmetry from
$SU(2)$ down to $U(1)$ such that in contrast to the case $h=0$, there
is no reason for the Heisenberg point $\DeltaI = 1$ to be special.
We will nevertheless concentrate mainly on $s=1/2$ and $\DeltaI = 1$.

An important observable is the magnetization
\begin{equation}
m = {1 \over s\,N} \sum_{i} S^z_{i}
\label{defM}
\end{equation}
which we normalize to saturation value $m=1$
(recall that $N$ is the total number of spins in the system).
The magnetization (\ref{defM}) is a conserved quantity
for the Hamiltonian (\ref{hamField}): $[H,m] = 0$.
One can therefore replace the operator (\ref{defM}) by its expectation
value and by slight abuse of notation we will use the same symbol
for both. The conservation of $m$ is also technically useful
for computing the magnetization curve since one can
relate energies with a field $E(h)$ to the energies $E(S^z,h=0)$ for fixed
total $S^z$ at $h=0$
\be
E(h) = E(S^z,h=0) - h \, S^z \, . 
\label{eField}
\ee
This implies that the GS 
energies in the sectors $S^z$ and $S^z + 1$ cross at the magnetic field
\be
h = E(S^z+1,h=0) - E(S^z,h=0) 
\label{finiteDiff} 
\ee
i.e.\ at this value of $h$ the magnetization increases by $1/s\,N$.
The ground states with a given total spin $S$
typically carry the maximal possible $S^z$ and hence $S=S^z$ holds
for them. In such a situation, $E(S^z,h=0) = E(S)$
of the preceding sections and
(\ref{finiteDiff}) implies that the $h(m)$ curve
is obtained by (discrete) differentiation of the $E(S)$
curve at $h=0$ with respect to $S \sim m$.
In particular, if there is a regime with a quadratic dependence of $E$
on $S$ like in (\ref{eqtos}), the magnetization curve $m(h)$ becomes
linear in this regime.

If $E(S)$ has a downward cusp, one obtains two different fields
$h_1$ and $h_2$ when approaching the associated value of $m$ from
below and above, respectively, and one finds a plateau in $m(h)$.
In one dimension, the appearance of plateaux is governed by
the following quantization condition on the magnetization $m$
\cite{AOY} (see also \cite{habil} for a more detailed discussion)
\begin{equation}
s V \left(1 - m\right) \in \Zed \, .
\label{condM}
\end{equation}
Here $V$ is the number of spins in a translational unit cell of the
{\bf ground state} (i.e.\ the lowest state for a given $m$) which
can be larger than (namely an integer multiple of) the unit
cell of the Hamiltonian if translational symmetry is spontaneously broken.

In two dimensions, there is no proof yet that the condition (\ref{condM})
is a necessary one.
Nevertheless, the condition (\ref{condM}) should apply to those
cases where plateau states are ordered (e.g.\ valence bond crystals)
and it is therefore at least a useful guide also in two dimensions.

\begin{figure}
\begin{center}
\begin{tabular}{lll}
\includegraphics[height=0.31\columnwidth]{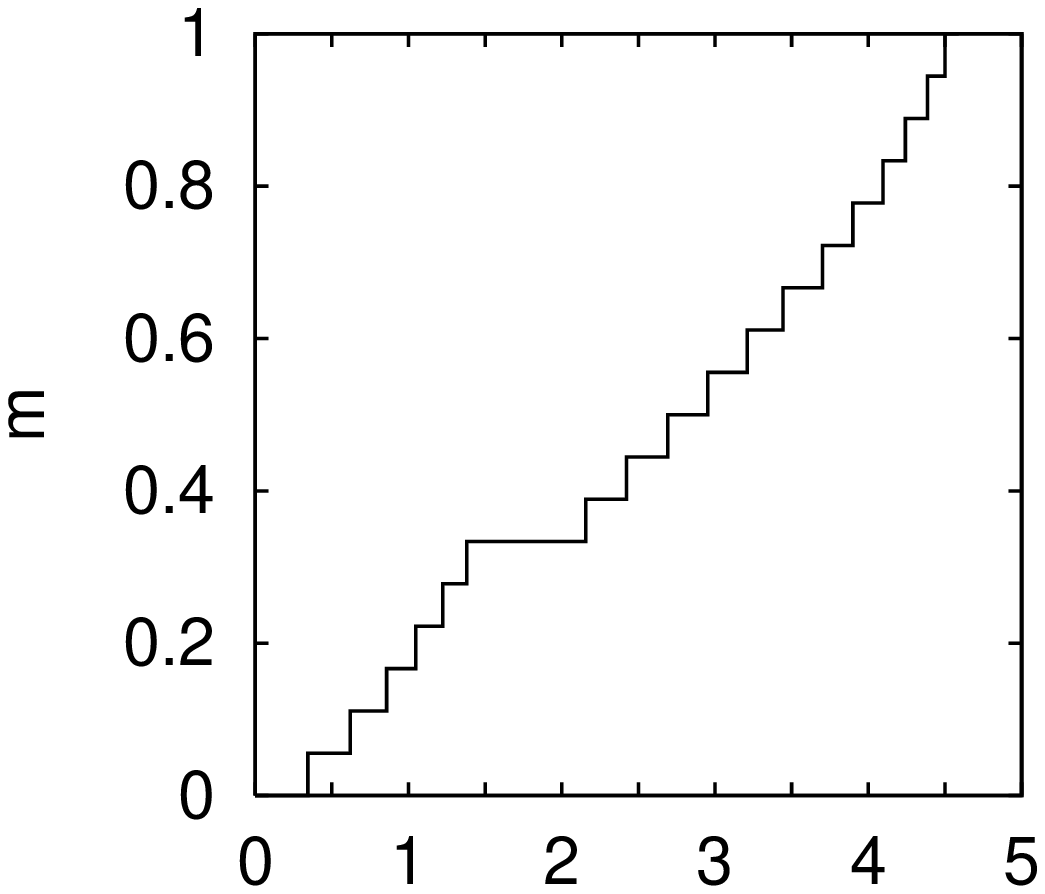} &
\hspace*{-4mm}
\includegraphics[height=0.31\columnwidth]{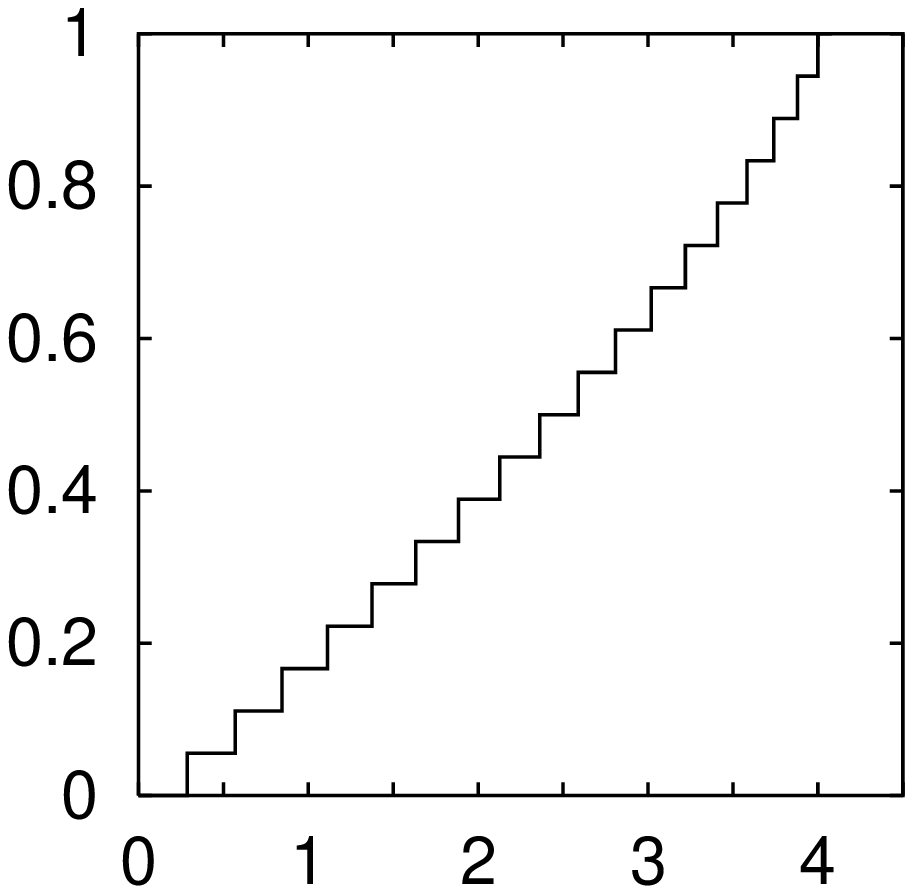} &
\hspace*{-4mm}
\includegraphics[height=0.31\columnwidth]{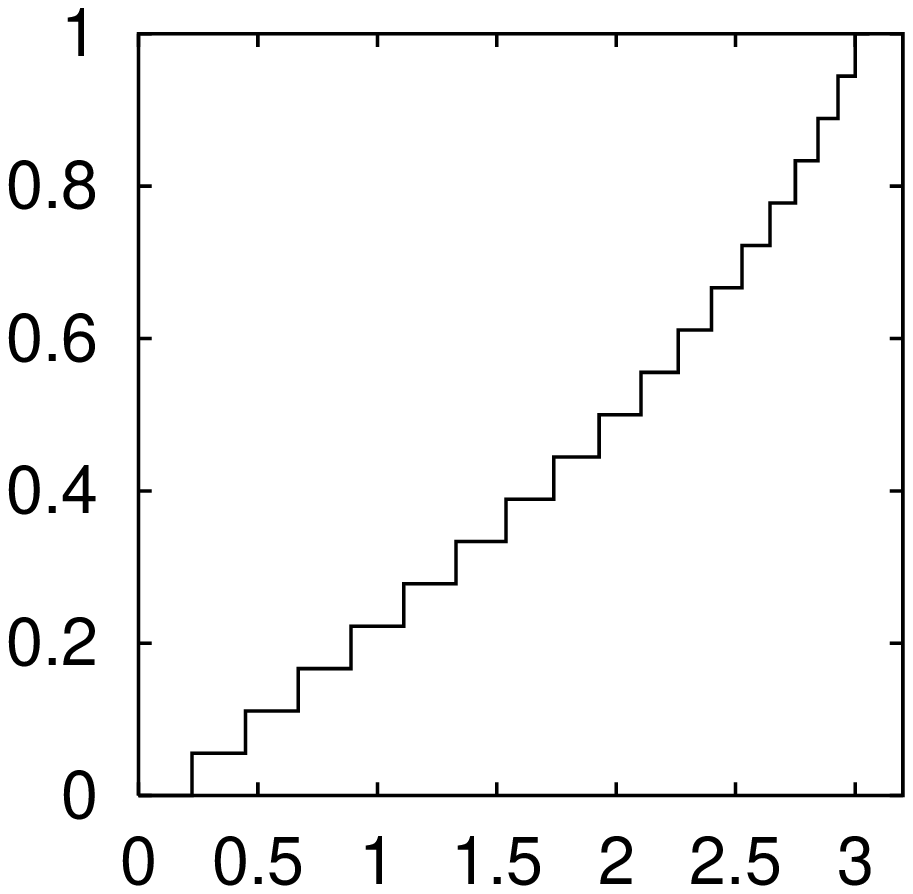} \\
\qquad\quad\, T1 = triangular & \qquad T2 = square  & \qquad T3 = honeycomb \\
\includegraphics[height=0.31\columnwidth]{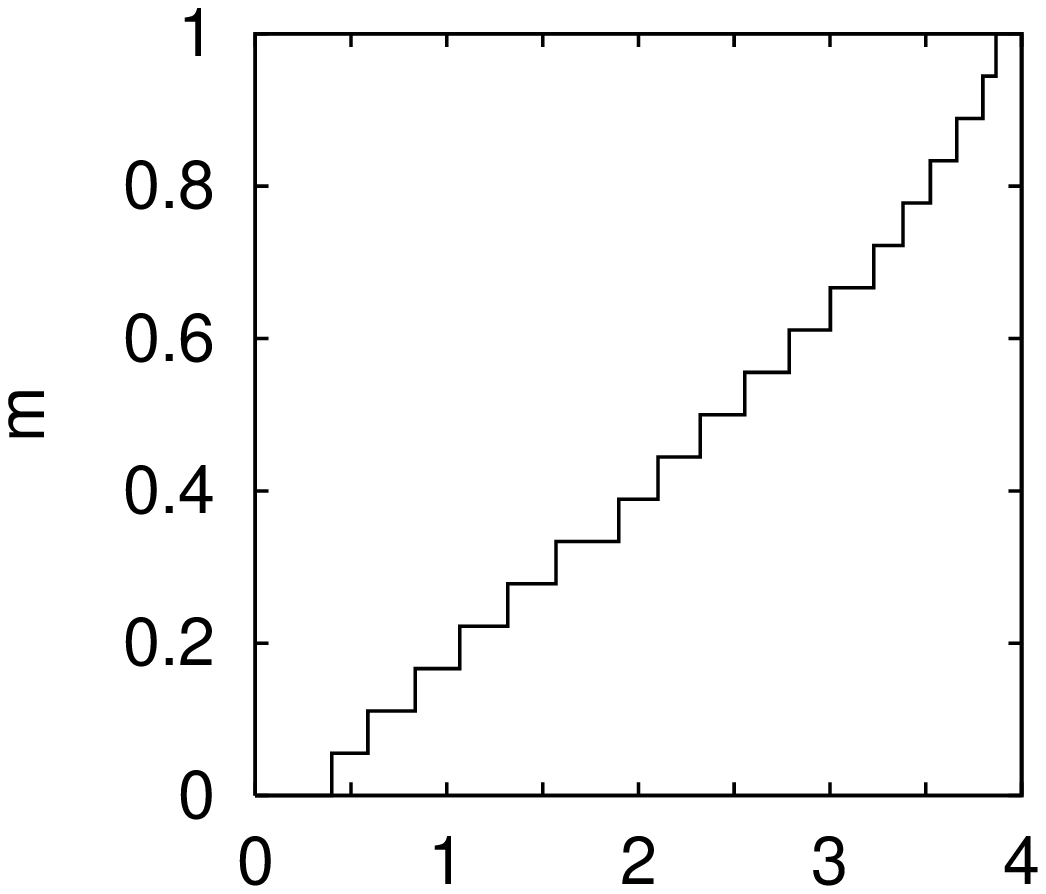} &
  &
\hspace*{-4mm}
\includegraphics[height=0.31\columnwidth]{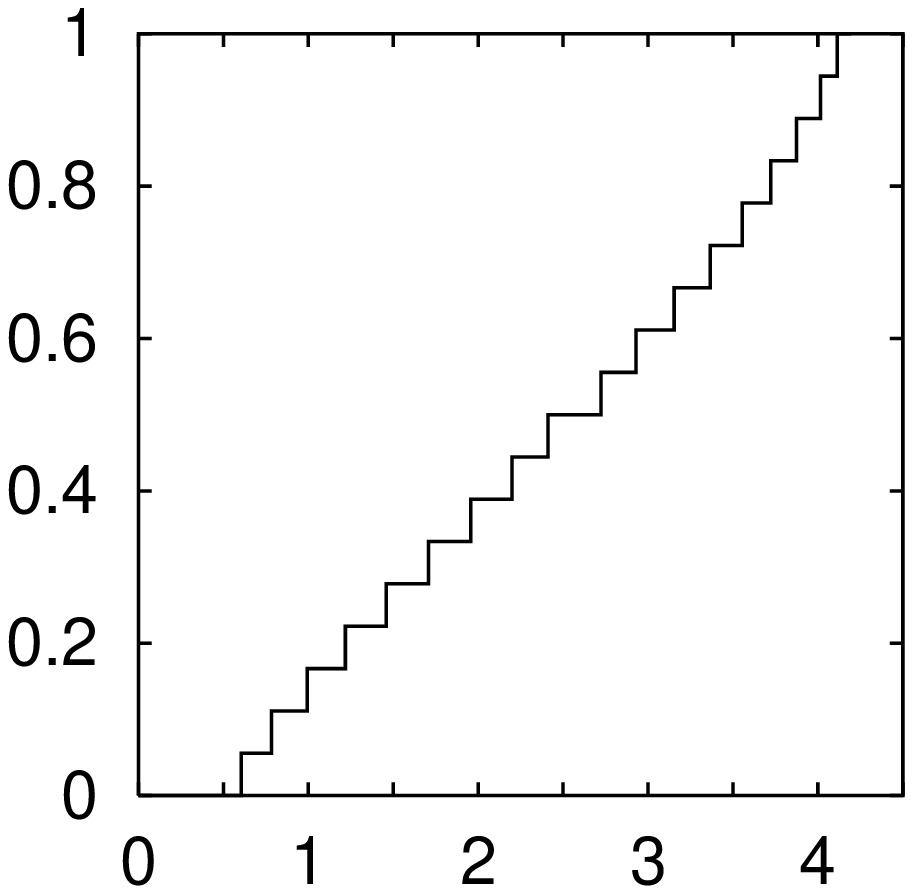} \\
\qquad\quad\, T4 = maple leaf & & \qquad T5 = trellis \\
\includegraphics[height=0.31\columnwidth]{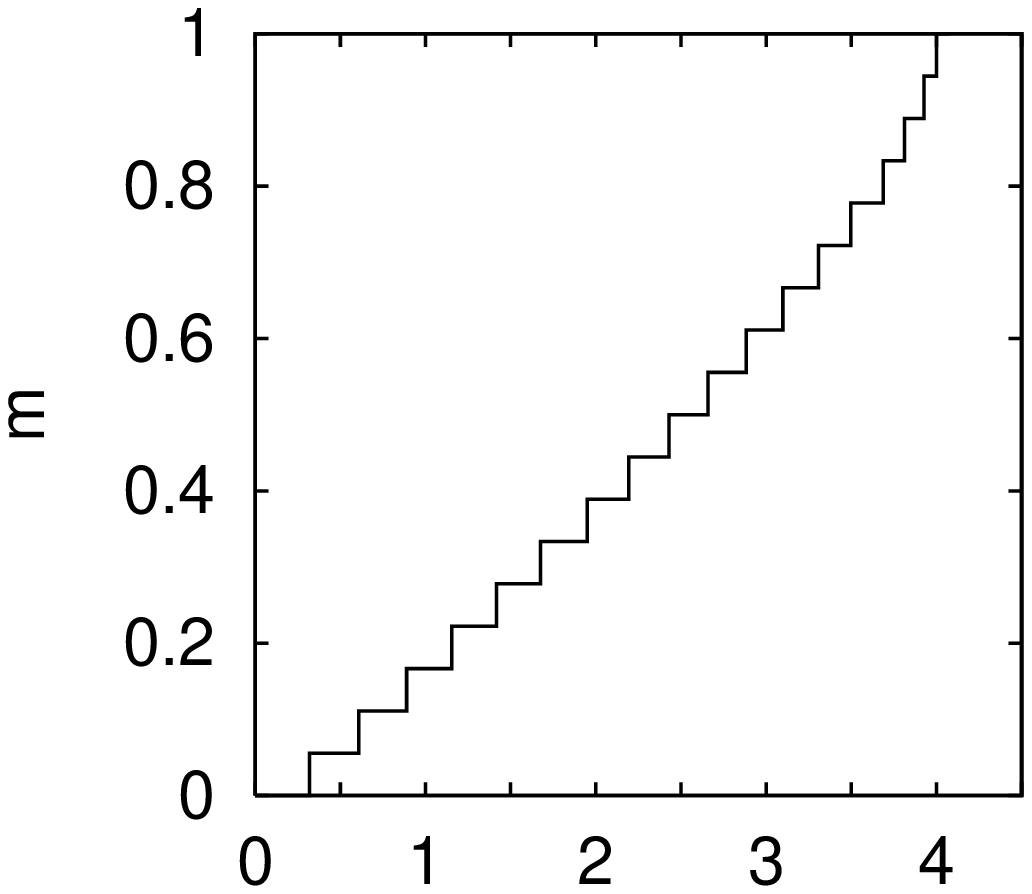} &
\hspace*{-4mm}
\includegraphics[height=0.31\columnwidth]{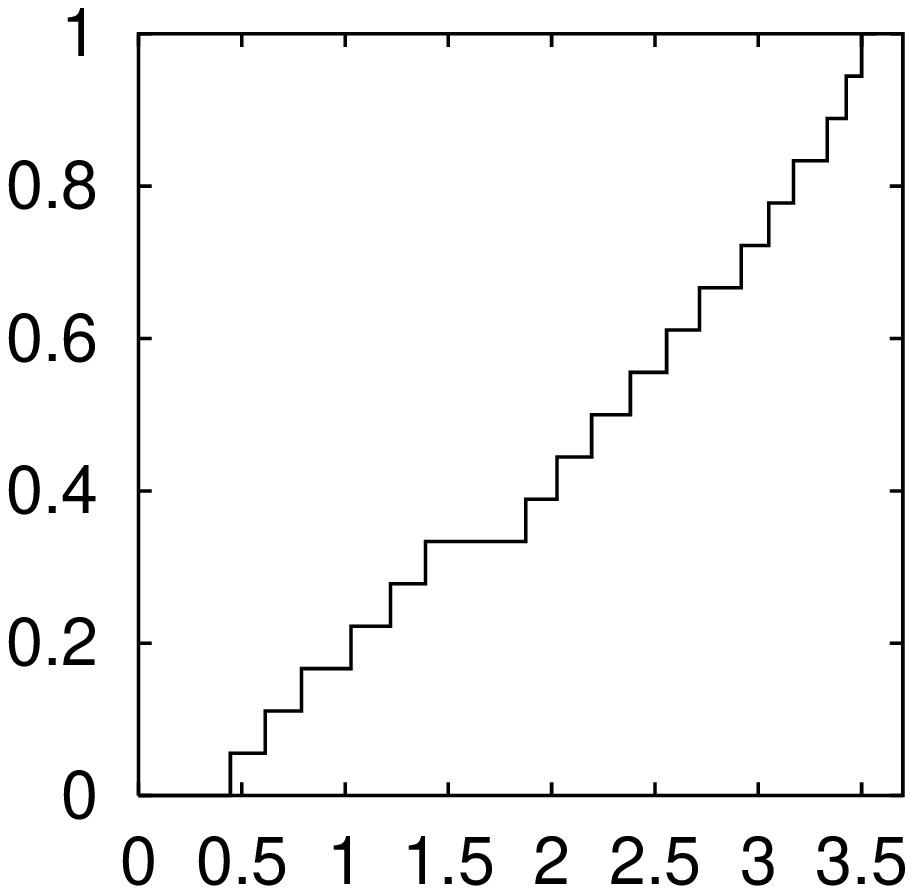} &
\hspace*{-4mm}
\includegraphics[height=0.31\columnwidth]{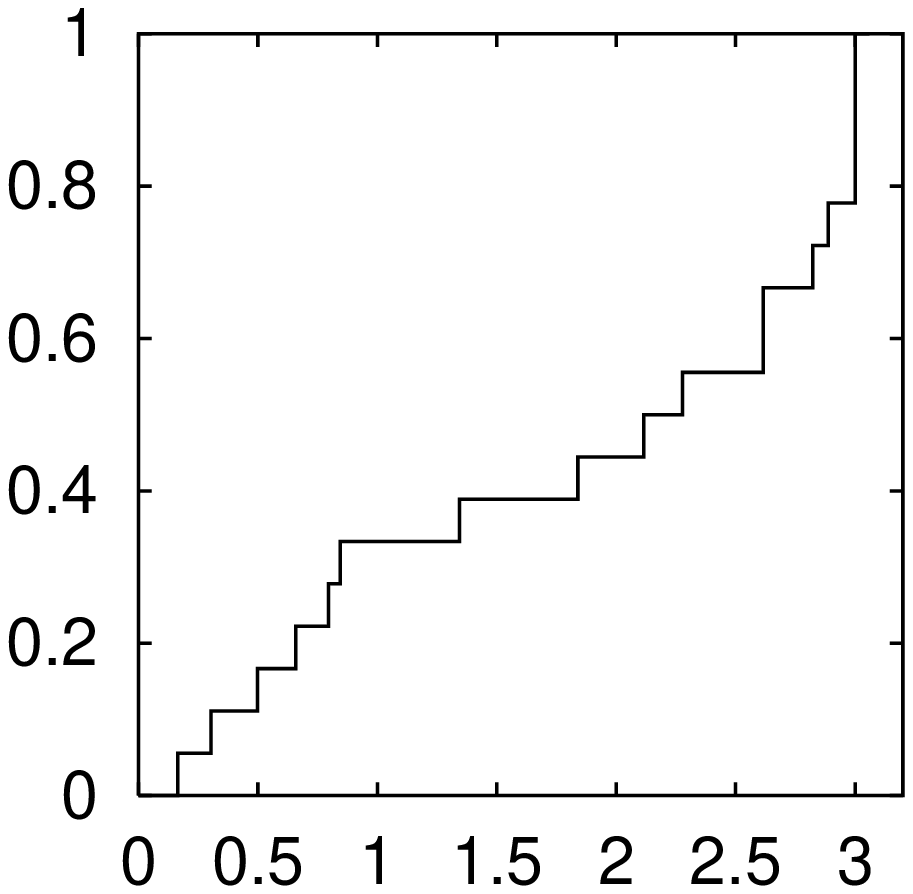} \\
\qquad\quad\, T6 = SrCuBO & \qquad T7 = bounce & \qquad T8 = \kagome \\
\includegraphics[height=0.3315\columnwidth]{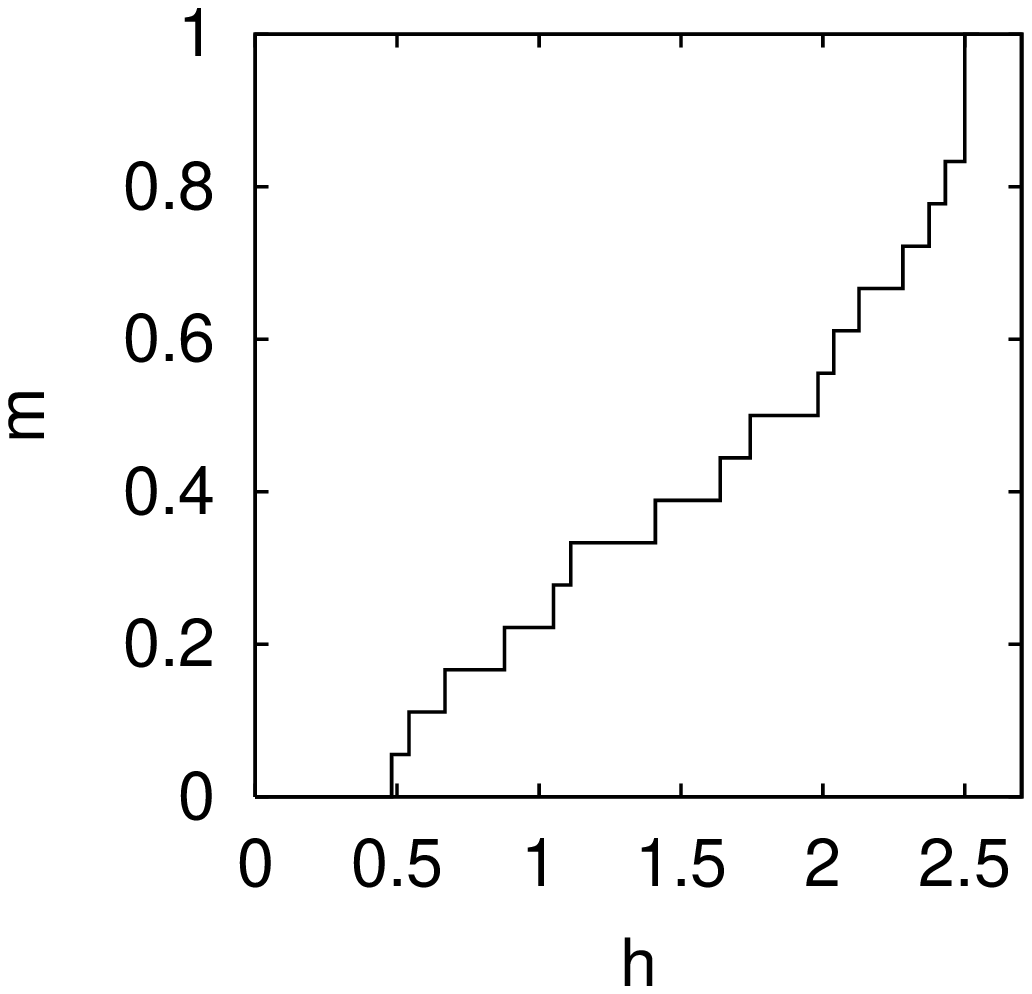} &
\hspace*{-4mm}
\includegraphics[height=0.3315\columnwidth]{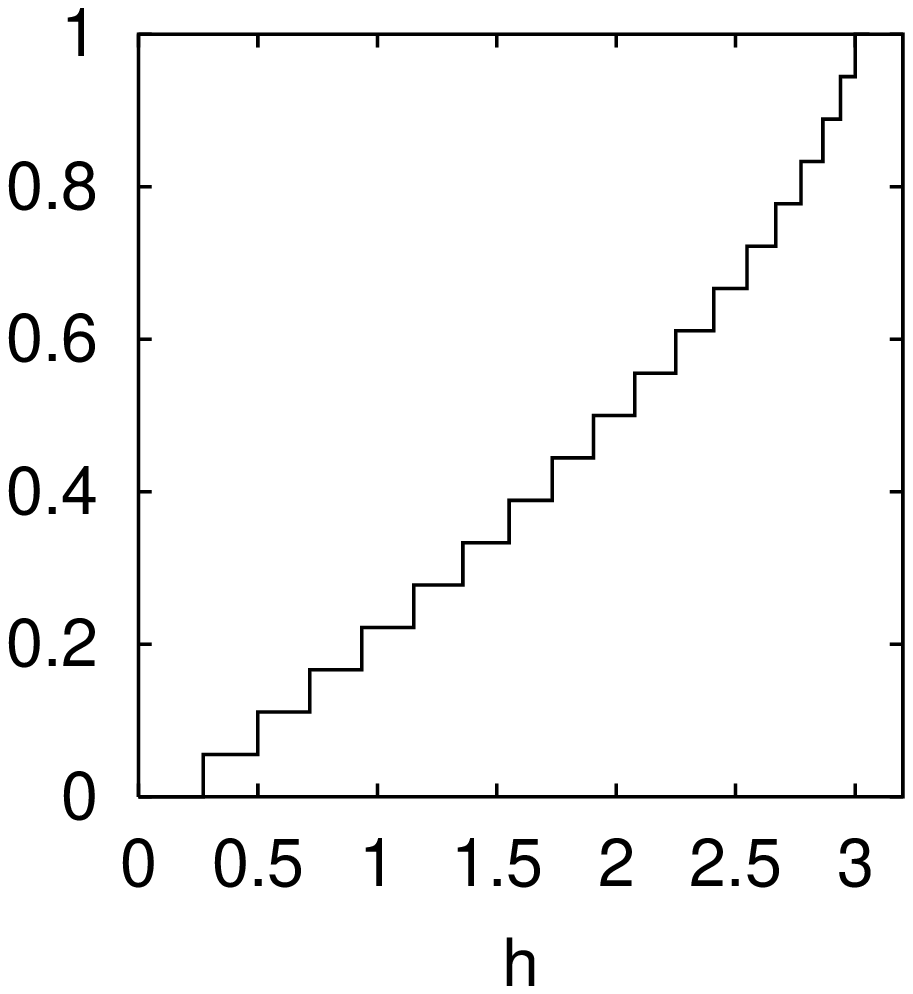} &
\hspace*{-4mm}
\includegraphics[height=0.3315\columnwidth]{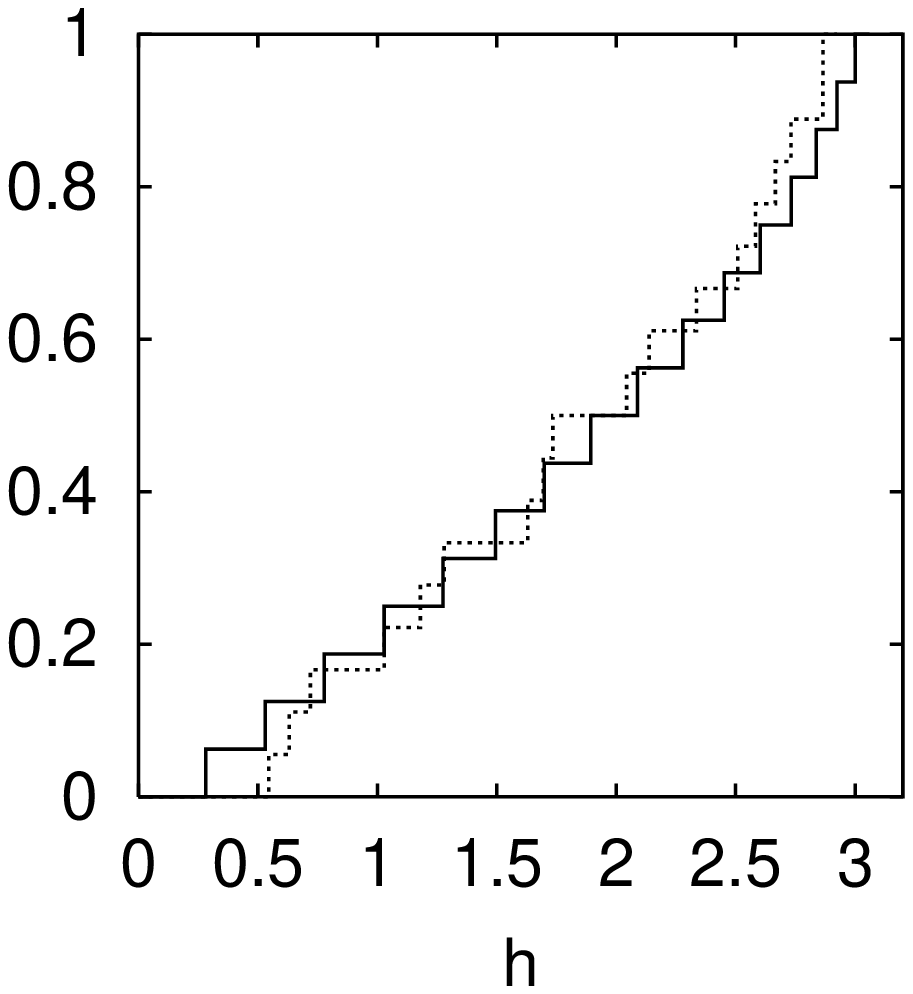} \\
\qquad\quad\, T9 = star & \qquad T10 = SHD & \qquad T11 = CaVO
\end{tabular}
\end{center}
\caption{\label{mArch}
 Magnetization curves of the $s=1/2$ Heisenberg antiferromagnet with $J=1$
 on all 11 Archimedean tilings. Results are for $N=36$ sites
 except for the tiling T11 (CaVO) where the full curve shows
 a result for $N=32$ which should be more representative than
 $N=36$ (shown as the dashed curve). For further details
 compare the text.
}
\end{figure}

Fig.\ \ref{mArch} shows results for magnetization curves of all 11
Archimedean lattices. With one exception, all these curves have been
computed for finite lattices with $N=36$ sites. Since for $s=1/2$
only the discrete values $S^z=0$, $1$, $\ldots$, $N/2$ are allowed
for a given $N$, one finds step-like curves on a finite lattice.
The task is then to determine which parts of these curves will become
smooth in the thermodynamic limit $N \to \infty$ or where anomalies like
plateaux or jumps remain in this limit.
 
Clearly, the behavior in a magnetic field is
even richer than the $h=0$ properties and we will therefore not
aim at a complete analysis.
Before we proceed with a discussion of some selected aspects,
we would like to add
some remarks on two tilings that we will not discuss further.
Firstly, on the trellis lattice (T5) the
ground states in a magnetic field carry incommensurate momenta.
They correspond in the $x$-direction to the twist angle
$\alpha_2$ discussed in section \ref{secTrellis} (although for $m<1$
the GS momenta in a 
quantum system are in general different
from the classical twist angle). 
Since irrational momenta are not realized for any finite lattice,
one obtains additional finite-size effects.
However, we have checked that these effects are sufficiently
small for the $N=36$ lattice which we have used
to render the result in Fig.\ \ref{mArch} qualitatively representative.

Secondly, the ground state on the CaVO lattice (T11) has a unit cell with
           8 spins (see section \ref{sectil11}). Since this does not fit on
           a lattice with $N=36$ sites, one observes large
           finite-size artifacts 
           in this case. In fact, the CaVO lattice
           is the only one among the 11 Archimedean lattices where no good
           magnetization curve can be obtained for $N=36$. For completeness,
           we nevertheless show this result as the dotted curve in
           Fig.\ \ref{mArch}, but we also show a curve
           for $N=32$ (full line) which should be considered as representative.

In the following three sections we discuss the tilings T2 (square),
T1 (triangular) and T8 (\Kagome) in more detail.

\subsection{Square lattice}

\label{secSquare}

\begin{figure}[tb]
\begin{center}
\includegraphics[width=9.5 cm]{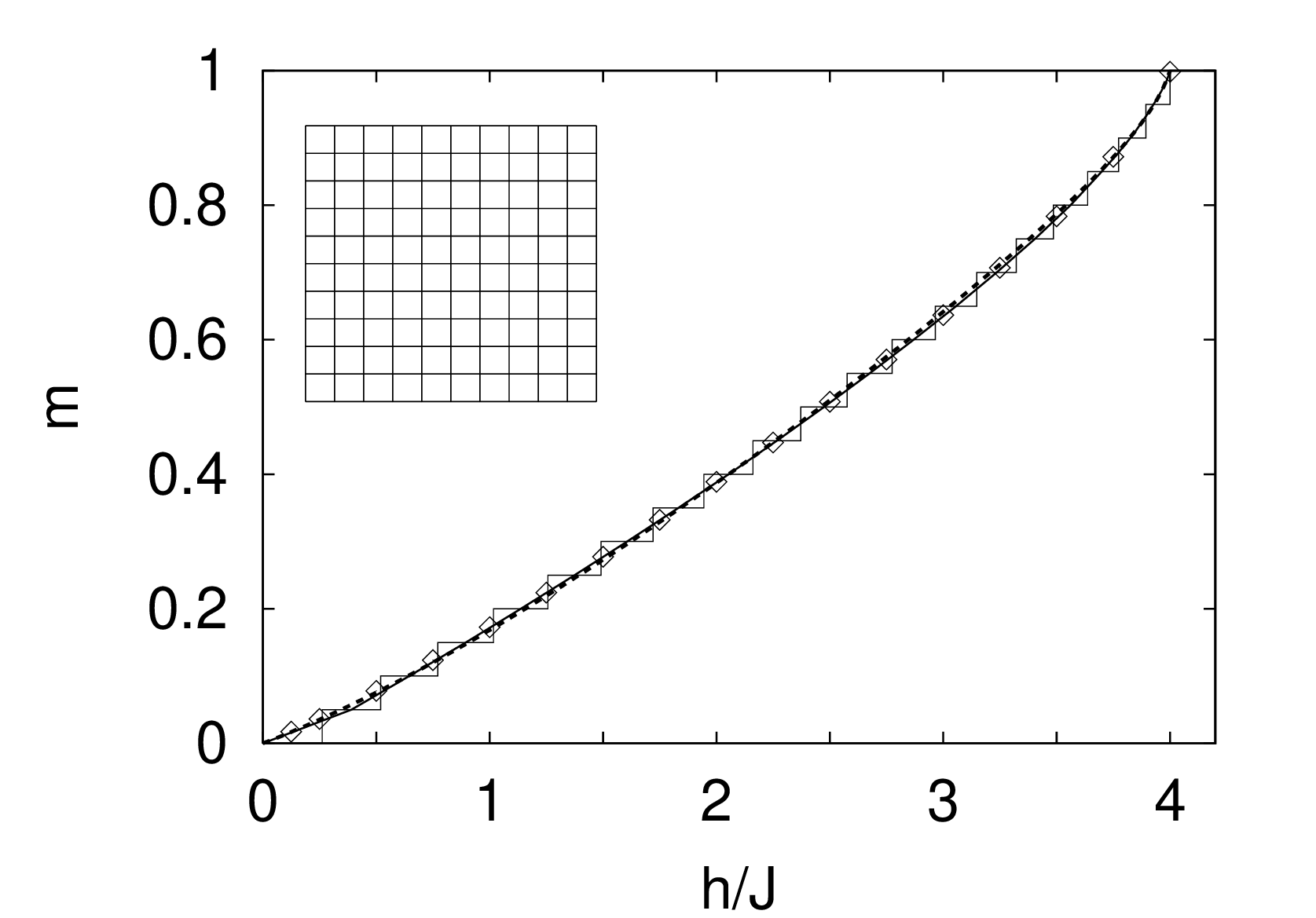}
\end{center}
\caption{
\label{figSquare}
Magnetization curve of the $s=1/2$ Heisenberg antiferromagnet on the
square lattice. The thin solid line is for $N=40$ sites,
the full bold line is an extrapolation to the thermodynamic limit.
A second-order spin-wave result \cite{ZhNi} (bold dashed line)
and QMC results (diamonds) are also shown.}
\end{figure}

Let us start with a brief discussion of the magnetization process of the square
lattice which is well understood and probably representative for the
non-frustrated Archimedean tilings.
Fig.~\ref{figSquare} shows the magnetization curve of the
$s=1/2$ square-lattice Heisenberg antiferromagnet obtained by
different approaches. Firstly, the thin full line shows the result obtained
by exact diagonalization for a finite lattice with $N=40$ sites \cite{HSR03}
(see also \cite{LoNo93,YaMue,hon99} for earlier exact diagonalization studies).
The full bold line denotes an extrapolation of the exact diagonalization data
to the thermodynamic limit which
is obtained by connecting the midpoints of the finite-size steps
at the largest available system size. One observes a smooth magnetization
curve with no peculiar features (in particular no plateaux) for $\abs{m} < 1$.
Note that close to saturation the extrapolated curve includes data at
large system sizes, which are not shown explicitly in Fig.~\ref{figSquare}
(the curve is based exclusively on finite lattices with at least $8 \times 8$
sites for $m \ge 0.84375$). The high-field part of the
magnetization curve is therefore particularly well controlled by exact
diagonalization.

The magnetization curve of a classical Heisenberg antiferromagnet would 
be just
a straight line for all fields up to saturation. Hence, the curvature
of the magnetization curve Fig.~\ref{figSquare} is due to quantum
effects. These quantum effects can also be studied by spin-wave theory;
a second-order spin-wave result \cite{ZhNi} is
shown by the bold dashed line in Fig.~\ref{figSquare}.

Finally, the magnetization process of the square lattice can also be studied
by quantum Monte Carlo (QMC) since this lattice is not frustrated,
We have generated some values of $m(h)$ on a $64\times 64$ lattice
(typically at $T=J/50$ which we have lowered to $T=J/200$ upon approaching
saturation) using the ALPS stochastic-series-expansion QMC application
\cite{AWT03,ALPS}.
These results are shown by the diamonds in Fig.~\ref{figSquare}
(statistical errors are much smaller than the size of the symbols) and agree
with available stochastic-series-expansion QMC results
\cite{sandvik}.

The quantitative differences of the results of all three approaches
are small, i.e., each approach yields a good description of
the $s=1/2$ HAFM on the square lattice. 
As the spin-wave approach \cite{ZhNi} is based
on a \Neel state, we may therefore conclude that \Neel order prevails in
the transverse components for $\abs{m} < 1$
(see also \cite{YaMue} for a discussion from the point of view of exact
diagonalization).

The same picture is probably also valid for the other bipartite
non-frustrated tilings, namely T3 (honeycomb), T10 (SHD) and T11
(CaVO). All these lattices are believed to be \Neel ordered at $h=0$
(see section \ref{summary_1}). Upon application of a magnetic field,
the \Neel vector first turns perpendicular to the field and then
the sublattice magnetizations are smoothly tilted towards the field
direction until full polarization is reached. At least the numerical
results for the magnetization curves shown in Fig.\ \ref{mArch}
for the lattices T3 (honeycomb -- see also \cite{hon99} for further
details and numerical data), T10 (SHD) and T11 (CaVO) are
consistent with a smooth magnetization curve. 

{}From an experimental point of view, one needs a sufficiently small
exchange constant $J$ to render the saturation field acccessible in
a laboratory. 
Successful synthesis and measurement of the magnetization process
of suitable $s=1/2$ square lattice antiferromagnets have been reported in
\cite{WAWLT02}. 

\subsection{Triangular lattice}

\label{secTriag}

The $s=1/2$ $XXZ$ model on the triangular
lattice is among the first models whose magnetization process was
studied by exact diagonalization \cite{NiMi}. These early studies already
found a plateau with $m = 1/3$, at least for Ising-like anisotropies
$\DeltaI > 1$. Due to the restriction to at most 21 sites, it was first not
completely clear whether the plateau persists in the
isotropic regime $\DeltaI \approx 1$.
The magnetization process of the Heisenberg antiferromagnet 
($\DeltaI = 1$) was analyzed
further using spin-wave theory \cite{ChuGo}. This study
provided evidence that the $m = 1/3$ plateau exists
also at $\DeltaI = 1$ and estimates for its boundaries were obtained.

\begin{figure}[tb]
\begin{center}
\includegraphics[width=9.5 cm]{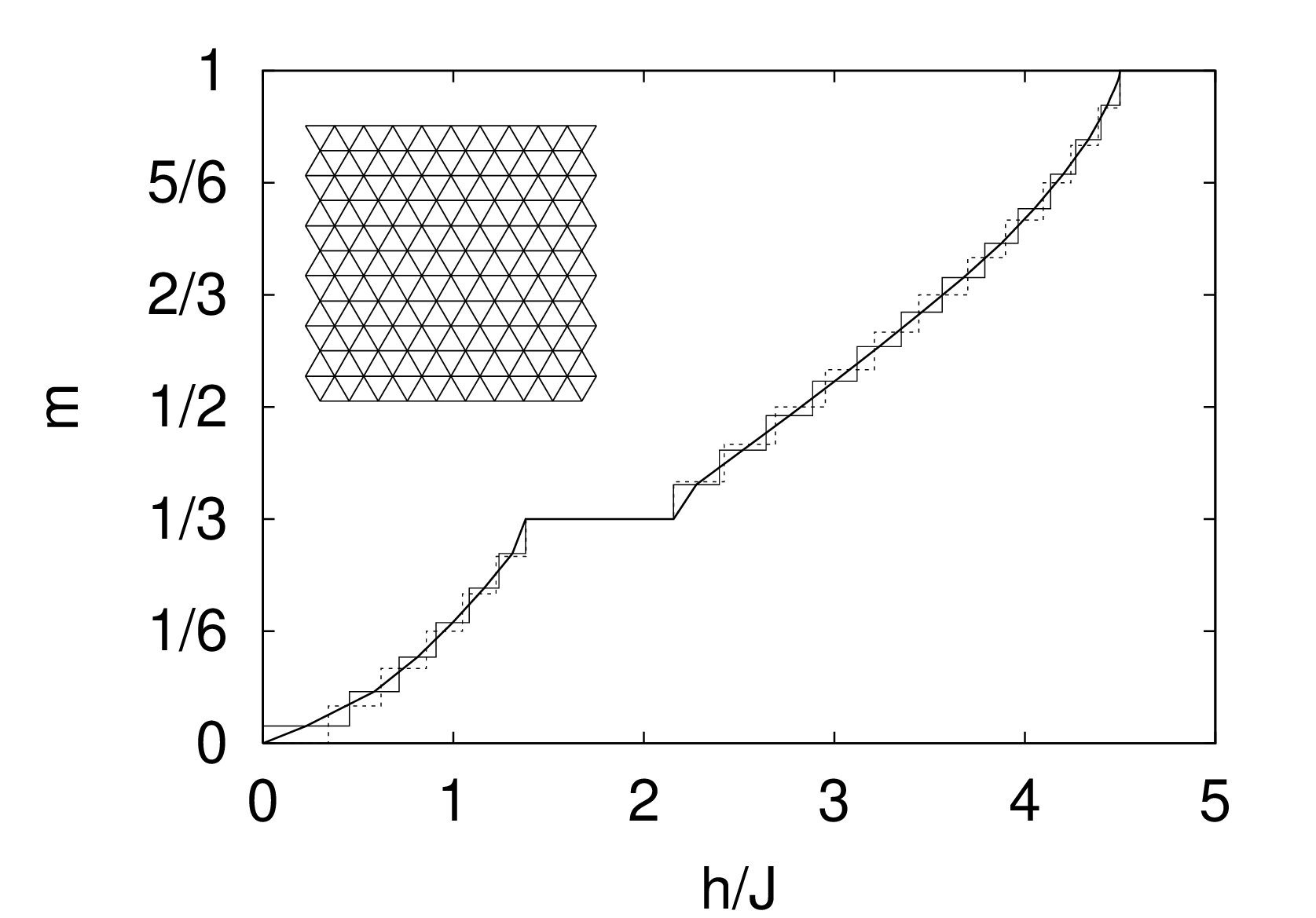}
\end{center}
\caption{
\label{figTriag}
Magnetization curve of the $s=1/2$ Heisenberg antiferromagnet on the
triangular lattice.
The thin dashed and solid lines are for $N=36$ and $39$ sites, respectively.
The bold line is an extrapolation to the thermodynamic limit.
}
\end{figure}

Fig.~\ref{figTriag} shows the magnetization curves obtained by
exact diagonalization for the $s=1/2$ Heisenberg antiferromagnet 
on finite lattices with $N=36$ and $39$ sites (thin lines).
There are small quantitative differences of the $N=36$ curve with exact
diagonalization results presented previously \cite{hon99,bernu94,habil}
whose origin is discussed in \cite{HSR03}.
Both curves in Fig.~\ref{figTriag} exhibit a clear plateau at $m = 1/3$ in an
otherwise smooth magnetization curve.
The spin-wave results for the magnetic fields
at the lower $h_1 = 3 \, (s - 0.084) \, J = 1.248 \, J$ and the upper boundaries
$h_2 = 3 \, (s + 0.215) \, J = 2.145 \, J$ of the $m = 1/3$
plateau \cite{ChuGo} are smaller by about $0.13 J$ (lower boundary)
and $0.01 J$ (upper boundary) than the exact diagonalization results
presented here for $N=39$ and $s=1/2$.

The full bold line in Fig.~\ref{figTriag} denotes an extrapolation of
the exact diagonalization data to the thermodynamic limit which is obtained by connecting
the midpoints of the finite-size steps at the largest available system size
(except for the boundaries of the $m = 1/3$ plateau where
corners were used). Close to saturation this includes again bigger
system sizes than those explicitly shown in Fig.~\ref{figTriag}.

The state of the $m = 1/3$ plateau can be easily
understood in the Ising limit $\DeltaI \gg 1$ \cite{miyashita86,hon99}.
Quantum fluctuations are completely suppressed in the limit $\DeltaI \to \infty$
and the $m = 1/3$ state is a classical state
where all spins on two of the three sublattices of the triangular lattice
point up and all spins on the third sublattice point down.
This state corresponds to an ordered collinear spin configuration.
It is threefold degenerate and breaks the translational symmetry.
One can then use perturbation theory in $1/\DeltaI$ to study the
$m = 1/3$ plateau of the $XXZ$ model \cite{hon99}.
However, the current best estimate of the point $\DeltaIc$ where
the $m = 1/3$ plateau disappears is obtained from a numerical computation
of the overlap of the Ising states and the $m=1/3$ wave function of the
full $XXZ$ model with $s=1/2$: $\DeltaIc = 0.76 \pm 0.03$ \cite{HSR03}.
This means that the $m=1/3$ plateau states of the Ising antiferromagnet
and the Heisenberg antiferromagnet on the triangular lattice are qualitatively
the same.

In the absence of a magnetic field, order persists in the Heisenberg 
antiferromagnet on the triangular lattice despite the geometric frustration
(see section \ref{secDreieck}). We have now seen that
the magnetic field enhances the frustration sufficiently
in the Heisenberg antiferromagnet on the triangular lattice
to open a spin gap and thus a plateau at $m=1/3$.

Among the other magnetization curves shown in Fig.\ \ref{mArch},
the one of the bounce lattice (T7) looks most similar
to the one of the triangular lattice. Indeed, also the tiling T7
consists of triangles and one may expect that also here an up-up-down
spin structure on each triangle gives rise to an $m=1/3$ plateau.
However, the covering of the complete lattice with up-up-down triangles
is not unique for the bounce lattice, indicating at least some differences
in the magnetization process of the triangular and bounce lattices.

\subsection{Kagom\'e lattice}

\label{secKag}

\begin{figure}[tb]
\begin{center}
\includegraphics[width=9.5 cm]{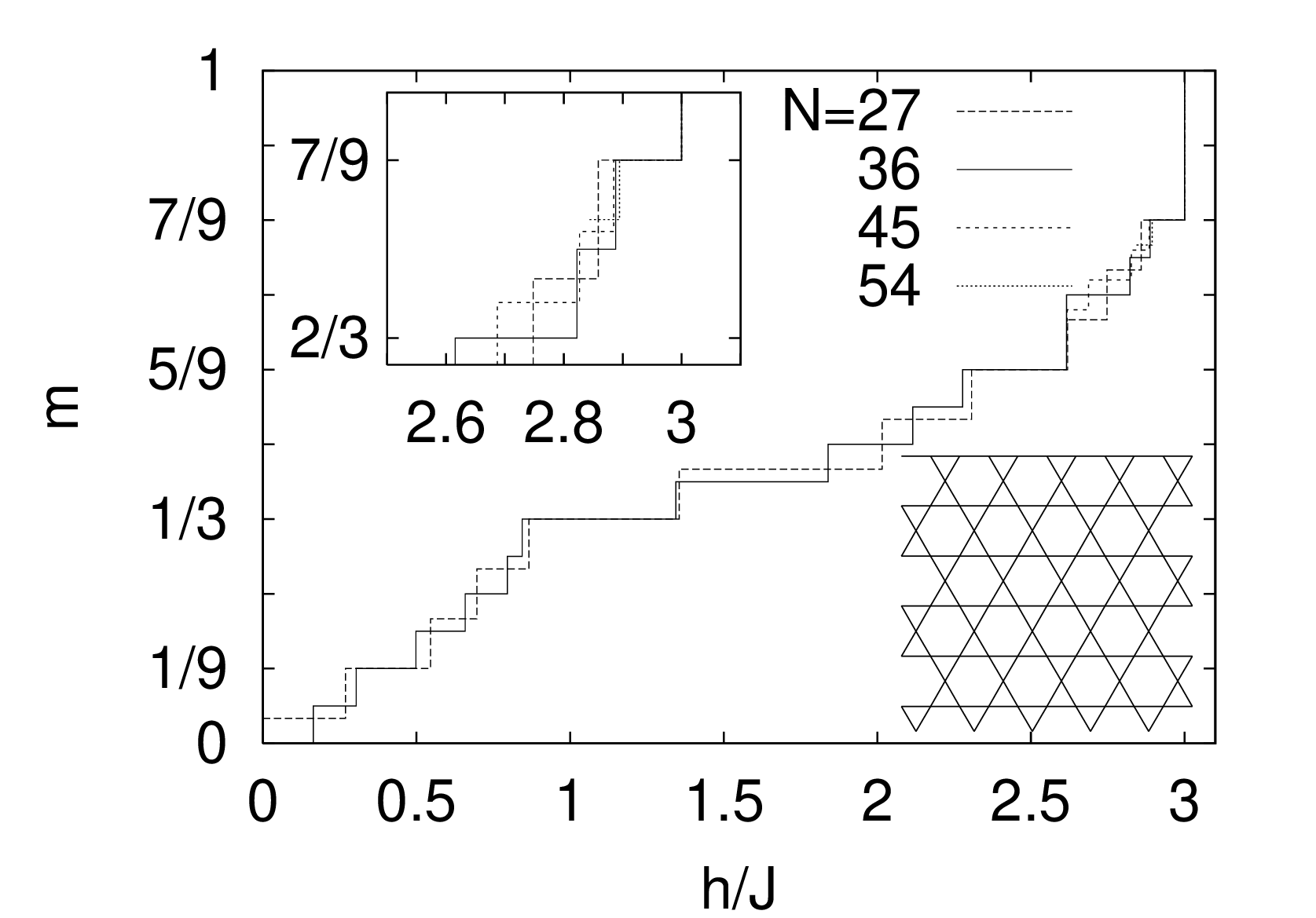}
\end{center}
\caption{
\label{figKag}
Magnetization curves of the $s=1/2$ Heisenberg antiferromagnet on the
\kagome lattice with $N=27$, $36$ (complete), $45$ and $54$ (partial).
The inset shows a magnified version of the region around $m=7/9$.
}
\end{figure}

Among the Archimedean lattices, the \kagome (T8) and star (T9) lattices
are characterized by the combination of strong frustration and low
coordination number. As discussed in sections
\ref{seckagome} and \ref{summary_1},
we believe that they give rise to a quantum paramagnetic ground state at $h=0$.
The $N=36$ magnetization curves in Fig.\ \ref{mArch} indicate that
these two lattices are presumably also those with the most complicated
and rich magnetization processes among all 11 Archimedean lattices.
Here we summarize the current understanding of the magnetization process
of the $s=1/2$ \kagome lattice and leave a detailed investigation of the
star lattice to the future.

Fig.\ \ref{figKag} shows complete magnetization curves for the \kagome
lattice with $N=27$ and $36$ sites as well as the high-field part of
$N=45$ and $54$ curves \cite{hida03,HSR03,SHSRS02}.
Firstly, there should be a plateau at $m=0$ associated
to the small spin gap above the quantum paramagnetic ground state.
However, this is difficult to recognize in Fig.\ \ref{figKag}.

\begin{figure}[tb]
\begin{center}
\includegraphics[width=5 cm]{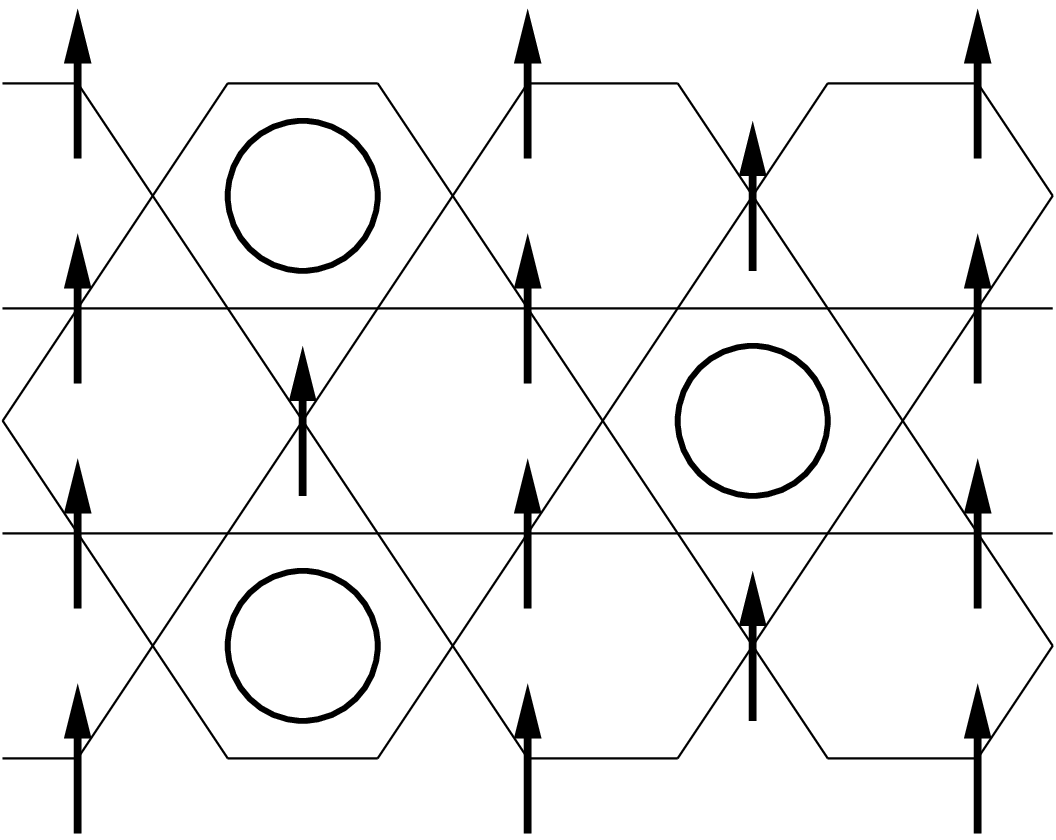}
\end{center}
\caption{
\label{figRoot3}
Part of the \kagome lattice with a $\sqrt{3} \times \sqrt{3}$ superstructure
indicated by the circles in certain hexagons. Arrows indicate spins
which are aligned with the magnetic field.
}
\end{figure}

A plateau at $m = 1/3$ may be better recognized
in Fig.\ \ref{figKag}. In fact, the presence of this plateau
at $m = 1/3$ in the $s=1/2$ Heisenberg 
antiferromagnet on
the \kagome lattice has been established previously by considering
also system sizes different from those shown in Fig.~\ref{figKag}
\cite{hida03,CGHP02}. The state of this plateau is, however, quite non-trivial.
For the classical Heisenberg antiferromagnet at $m = 1/3$,
thermal fluctuations select collinear states, but due to
the huge degeneracy of these states, there appears to be no real order
on the classical level at $m = 1/3$ \cite{zhitomirsky}
(see also \cite{ShHo}).
For $s=1/2$, it is useful to consider the $XXZ$ model. In the
Ising limit $\DeltaI \to \infty$ one can then first establish \cite{CGHHPRS03}
a relation to a quantum dimer model on the
honeycomb lattice which was argued \cite{MS2001,MSC2001} to give
rise to a valence bond crystal ground state with a
$\sqrt{3} \times \sqrt{3}$ order.
Fig.~\ref{figRoot3} shows a qualitative picture of this state.
In the present context the circles indicate
resonances between the two different
\Neel states on the surrounding hexagon.
The next step is to compute the overlap
of the $m = 1/3$ wave function of the $XXZ$ model
with that of the quantum dimer model as a function of $\DeltaI$
and one finds no evidence for a phase transition for $\DeltaI \ge 1$
\cite{CGHHPRS03}. This implies that also the $m = 1/3$
state of the $s=1/2$ Heisenberg 
antiferromagnet on the \kagome lattice is
an ordered state with features similar to the valence bond crystal.
There are many low-lying non-magnetic excitations above the lowest
$m=1/3$ state which can be considered as a remnant of the classical degeneracy.
However, the valence-bond-crystal-type order implies just three
degenerate $m=1/3$ ground states 
related by translational symmetry (see Fig.\
\ref{figRoot3} for illustration) and a gap to {\it all} excitations above this
three-fold degenerate ground state. Note that for the $s=1/2$ Heisenberg
antiferromagnet on the \kagome lattice
this non-magnetic gap in the $m=1/3$ sector turns out to be quite small
(estimates are of the order of $J/25$ \cite{CGHHPRS03}). 

There may be a further plateau at $m = 5/9$ in
Fig.\ \ref{figKag} although it is difficult to draw unambiguous conclusions
from the available numerical data in this region of magnetization values.

Finally, one can see a pronounced jump of height $\delta m = 2/9$
just below saturation and a plateau at $m = 7/9$ in the magnetization curve
of the $s=1/2$ \kagome lattice. Both features will be discussed in more
detail in the next section.

\subsection{Independent magnons and macroscopic magnetization jumps}

\label{secImag}

\begin{figure}[tb]
\begin{center}
\includegraphics[width=9 cm]{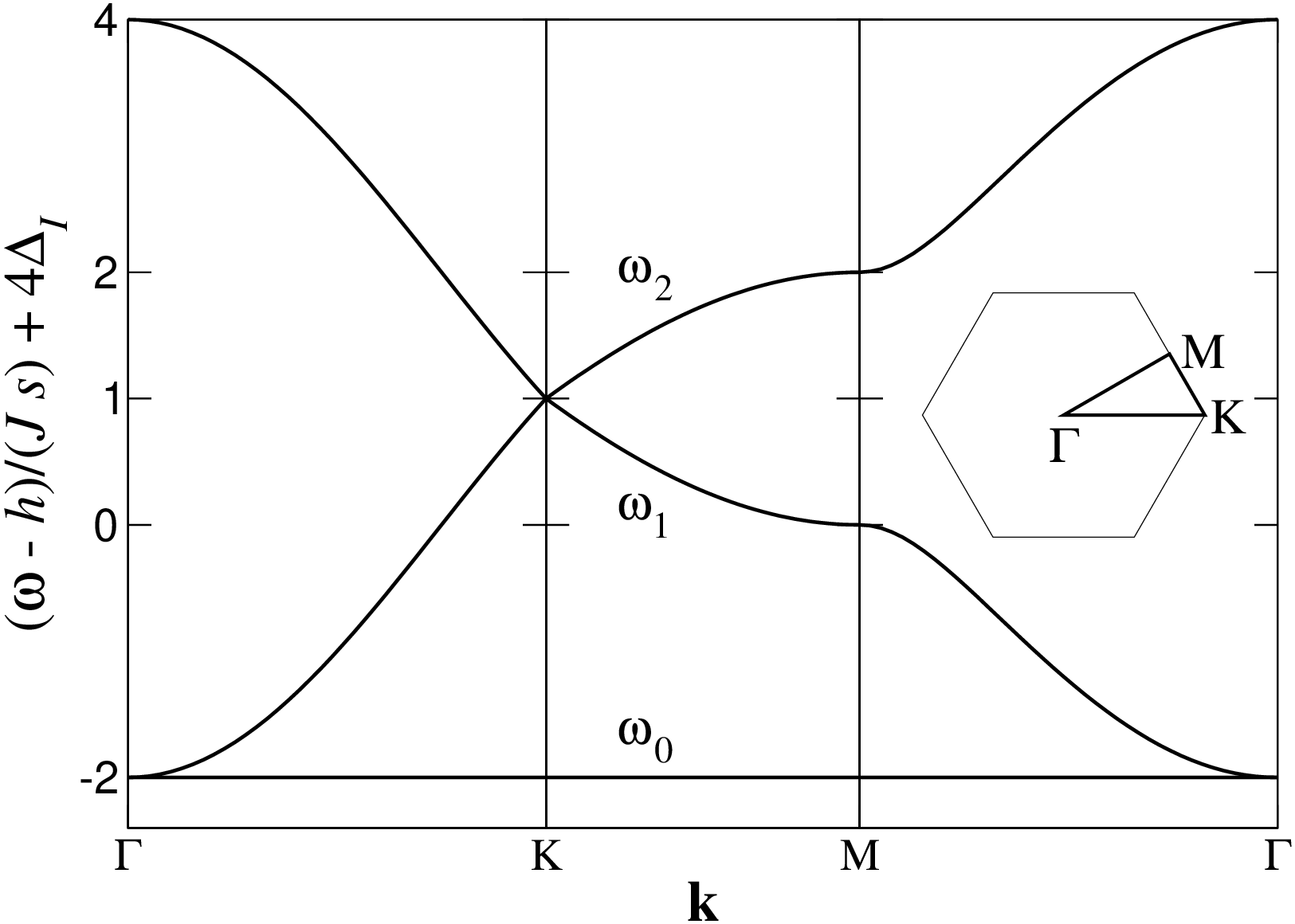}
\end{center}
\caption{
\label{figKagome}
The three branches $\omega_i(\vec{k})$ of one-magnon excitations above
the ferromagnetic background for the \kagome lattice along the path
in the Brillouin zone shown in the inset.
Note that $\omega_0(\vec{k})$ is completely independent of $\vec{k}$.}
\end{figure}

For the $s=1/2$ Heisenberg model on a given two-dimensional lattice it is a
very rare event that one can write down the ground state exactly. One such
exceptional case is the dimerized ground state arising in the two-dimensional
Shastry-Sutherland model \cite{shastry81dd} (see section \ref{qpt}).
It is therefore remarkable that in the high-field
region of some popular frustrated lattices such as the \kagome lattice
one can construct a macroscopic number of exact ground states.
We will discuss some aspects of the construction in more detail in this section,
focusing in particular on the \kagome lattice. Note that
similar constructions can be given for other
lattices \cite{SHSRS02,RSHSS03} and finite clusters \cite{SSRS01}
(for other points of view we also refer to \cite{SHSRS02,RSHSS03}).
We also wish to remark that the construction of exact eigenstates to be
described below works for models where no non-trivial
conservation laws are known. However, it
is restricted to the transition to saturation, since, as will become clear
in the following, it relies on the
knowledge of a reference state (namely the ferromagnetically  polarized state 
$\vert \!\! \uparrow \ldots \uparrow \rangle$ which is
a trivial eigenstate of the Hamiltonian) and an analytic
determination of the one-magnon excitations above it.

Now let us be more specific and,
as the first step, consider very high magnetic fields such that the
ground state is the ferromagnetically polarized state.
In highly frustrated spin models,
the lowest branch $\omega_0(\vec{k})$ of the one-magnon excitations above
the ferromagnetically polarized state often has some
flat directions (i.e.\ does not depend on some of the components $k_i$) or
is completely flat (i.e.\ independent of $\vec{k}$).
In the latter case, one finds a special type of jump just
below the saturation field as well as indications for a plateau below the
jump \cite{SHSRS02,RSHSS03}.

The explicit computation of the one-magnon spectrum above the ferromagnetically
polarized state depends on the model. For example, the \kagome lattice has a
unit cell containing three sites and the spectrum is obtained by diagonalization
of a $3 \times 3$ matrix. For the spin-$s$ $XXZ$ model one then finds the
three magnon branches $\omega_i(\vec{k})$ ($i=0$, $1$, $2$) which are shown
in Fig.\ \ref{figKagome}. Remarkably, the lowest branch
$\omega_0(\vec{k}) = h - (2 + 4 \DeltaI) \, J\, s$ is completely flat,
i.e.\ independent of $\vec{k}$. This property is a fingerprint
of the strong frustration caused by the triangles in the \kagome lattice.
In fact, the lowest magnon branch relative to the ferromagnetically polarized
state is also completely flat for some other popular highly frustrated lattices
including the pyrochlore lattice and its two-dimensional projection,
namely the checkerboard lattice \cite{SHSRS02}.

The one-magnon excitations can be localized in the real-space directions
corresponding to a flat direction in $\vec{k}$-space by using an inverse
Fourier transformation. If the dispersion is completely flat, one can
construct a magnon excitation that is localized in a finite volume.
For the \kagome lattice, these local magnon excitations are located
on the hexagons marked by circles in Fig.\ \ref{figRoot3}. Apart from
normalization, this state is given by
\begin{equation}
\vert 
 1 \rangle \sim \sum_{x} (-1)^x S_x^{-} \,
 \vert \!\! \uparrow \ldots \uparrow \rangle 
\end{equation}
where the sum runs
over the 6 corners of the hexagon. Localization can be verified since
each spin next to the hexagon is coupled to two spins in the hexagon
such that contributions of flipped spins propagating onto the exterior
site add with different signs and thus cancel. Therefore,
a localized magnon is an exact eigenstate of the $XXZ$ Hamiltonian on
the \kagome lattice.

Now one can create further localized magnon excitations. As long
as the local magnons are sufficiently well separated in space, they
do not interact and consequently the many-magnon state is still an
exact eigenstate. The non-trivial step is to verify that these
non-interacting localized magnon excitations are not only eigenstates
but in fact ground states in their respective magnetization subspaces.
This result is probably true for general $s$, general coupling
geometries with $J_{i,j} \ge 0$ and
$XXZ$ anisotropy $\DeltaI \ge 0$. In \cite{SHSRS02} the ground state property
was verified numerically for some cases and it has been shown rigorously
for certain subsets of the parameters, namely for $s=1/2$,
$\DeltaI \ge 0$ and all coupling constants $J_{i,j}$ equal \cite{SSRS01}
or for general $s$ and $J_{i,j} \ge 0$, but isotropic
interaction $\DeltaI = 1$ \cite{Schmidt02}.

If the localization region is finite, a macroscopic fraction of the
spins in the system can be flipped using local magnon excitations. Since
the energies of the individual excitations add without interaction terms,
one obtains a finite interval of the magnetization $m$
where the ground state energy $E(m)$ becomes a linear function.
Due to the relation (\ref{finiteDiff}), this linear behavior leads
to a finite jump in the magnetization curve $m(h)$ at the saturation
field $h_{\rm sat}$.

Inspection of Fig.\ \ref{figRoot3} shows that
at most $N/9$ local magnons fit on a finite \kagome lattice.
Therefore, a jump of
height $\delta m = 1/(9\,s)$ is predicted for the
\kagome lattice. For the $s=1/2$ Heisenberg 
antiferromagnet on the \kagome lattice
one indeed observes numerically a jump of height
$\delta m = 2/9$ which is independent of the system size
if boundary conditions are chosen appropriately (see Fig.\ \ref{figKag}).
Note that the height of the jump is
in general proportional to $1/s$ and vanishes in the classical limit
$s \to \infty$. Therefore, the macroscopic jump caused by independent
local magnons is a true macroscopic quantum effect.

The maximal number of local excitations is obtained for their closest possible
packing. The circles in Fig.\ \ref{figRoot3} indicate this state
for the \kagome lattice. This clearly is
an ordered (crystalline) state. According to general arguments
\cite{momoi00,oshikawa00}, one expects a gap above such a crystalline state
and consequently a plateau in the magnetization curve at the foot of
the jump. This conclusion is supported by the numerical magnetization curve
of the $s=1/2$ Heisenberg 
antiferromagnet on the \kagome lattice, Fig.\ \ref{figKag},
which exhibits a clear plateau at $m = 7/9$ with a width around $0.07 J$
\cite{RDS03}.

The excitation energy of a local magnon is exactly zero at the
saturation field $h_{\rm sat}$.
Hence, all independent magnon states are exactly degenerate at $h=h_{\rm sat}$.
The number of these states grows exponentially with $N$. This can be seen
by considering the subset of states where magnons sit only on
the positions of the crystalline state. Since the number of such positions is
proportional to $N$ and each position can be empty or occupied by a magnon,
one finds an exponentially growing lower bound on the number of independent
magnon states (this lower bound is $2^{N/9}$ for the \kagome lattice). 
In other words, the local magnon excitations give rise to a finite
zero-temperature entropy at $h=h_{\rm sat}$ for a quantum spin system ! 

\begin{figure}[tb]
\begin{center}
\includegraphics[width=9.5 cm]{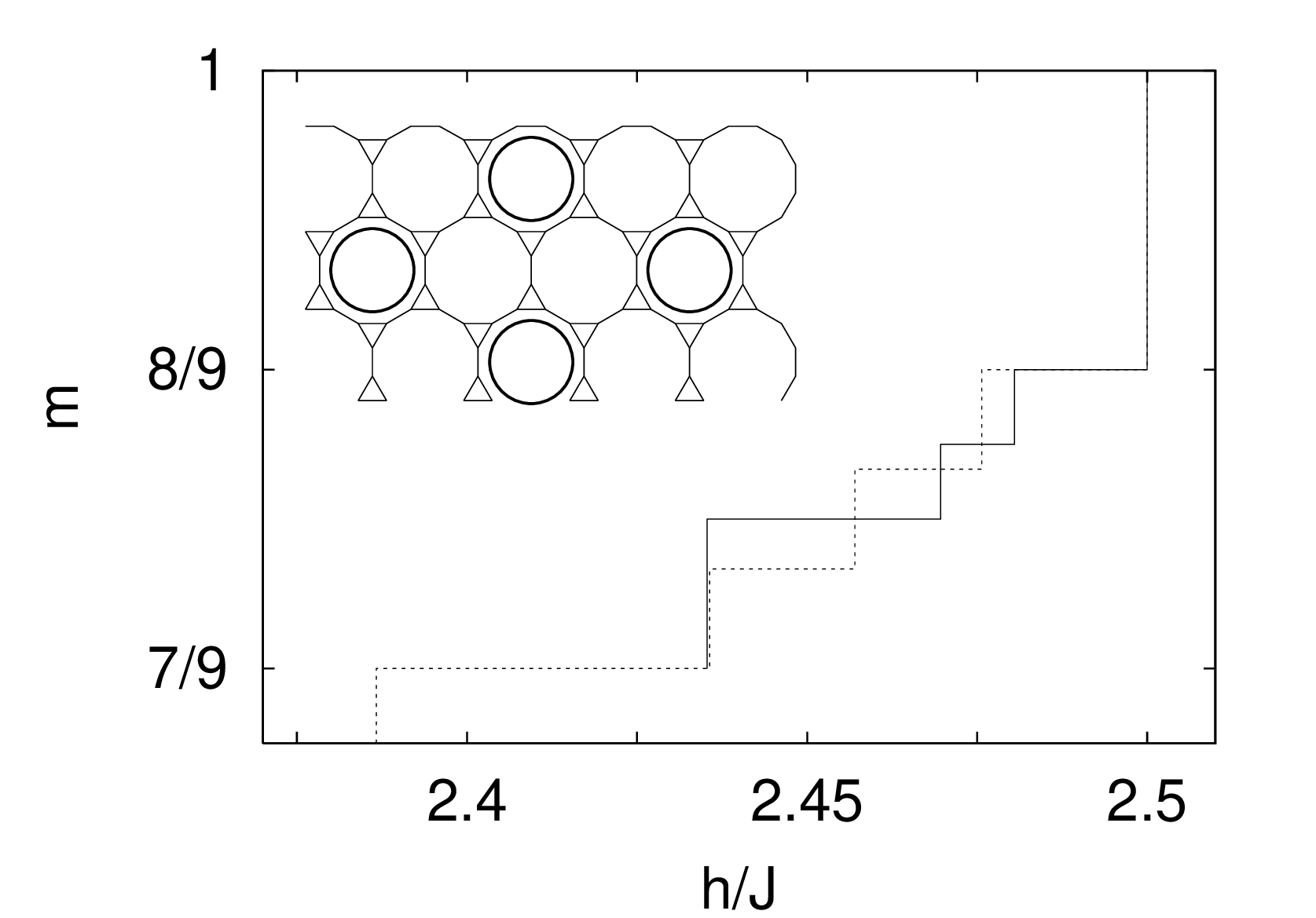}
\end{center}
\caption{
\label{figT9}
High-field part of the magnetization curves of the $s=1/2$ Heisenberg
antiferromagnet on the star lattice (T9) with $N=54$ (dashed line) and
$N=72$ sites (solid line).
The inset indicates the closest packing of local magnon excitations.
}
\end{figure}

The star lattice (T9) is the other Archimedean tiling which supports
local magnon excitations. In this case, the magnons
are localized around dodecagons as shown by the circles in the inset of
Fig.\ \ref{figT9}. The adjacent triangles again ensure localization via
destructive interference of hopping processes out of a dodecagon.
One can read off from the inset of Fig.\ \ref{figT9} that a finite
star lattice can in general accomodate at most $N/18$ local magnons.
This implies a jump of height $\delta m = 1/(18\,s)$ below saturation
with a plateau at $m = 1 - 1/(18\,s)$ corresponding to the crystalline
pattern of local magnon excitations sketched in the inset of Fig.\ \ref{figT9}.
The main panel of Fig.\ \ref{figT9} shows that a jump of the expected height
$\delta m = 1/9$ is indeed present in the magnetization curves of the
$s=1/2$ model on lattices with $N=54$ and $72$.
Note that the $N=36$ lattice whose magnetization curve is shown in
Fig.\ \ref{mArch} is not generic, but an exception from the point of view
of local magnons. Due to its small linear extent, it has more
and shorter cycles wrapping around the boundary than present in the
infinite system, namely of length eight while the dodecadons yield cycles
with length twelve. This $N=36$
lattice then supports not only two but three local magnons and therefore
the jump is higher than in the generic situation.
Note further that a plateau is expected below this jump, i.e.\ at $m=8/9$
for $s=1/2$. However, the $N=54$ and $72$ curves in Fig.\ \ref{figT9} do not
allow an unambiguous confirmation of the presence of such a plateau.  

The checkerboard and a square-\kagome lattice are further two-dimensional
lattices supporting local magnon excitations \cite{SHSRS02,RDS03}.
On the checkerboard lattice, a magnon is localized around a square.
This leads to a jump of size $\delta m= 1/(8\,s)$, as one can verify
numerically for $s=1/2$ \cite{RSHSS03}.

We would like to mention in passing that there are instabilities towards
lattice deformations. However, it can be argued that
the most favorable instability is one which preserves
the local magnon excitations as exact eigenstates and the
associated degeneracy \cite{RDS03}.

A related but different situation arises in two dimensions if
the minima of the one-magnon excitations form a one-dimensional manifold.
One example is the two-dimensional Shastry-Sutherland lattice \cite{MoTo00}
whose magnetization process will be discussed in the next section,
another one is the frustrated square lattice mentioned in section
\ref{qpt} at $J_2 = J_1/2$ \cite{gluzman94,frusH01}.
In this case, magnon excitations can be constructed \cite{SHSRS02,habil}
that are localized in some directions, but not all.
The frustrated square lattice can accommodate $L/2$ local magnon excitations
\cite{SHSRS02,habil} if the linear extent of the lattice is $L$,
leading to a finite-size jump $\delta m = L/(2\, N \, s)$.
A finite-size jump of height $\delta m = L/N$ is indeed
observed in exact diagonalization studies of the $s=1/2$ frustrated square
lattice at $J_2 = J_1/2$ \cite{frusH01,habil}.
However, due to the incomplete localization, the height of the jump
vanishes in the thermodynamic limit, i.e.\ the transition to
saturation remains continuous in such a case. Although
the magnetization curve should be exceptionally steep just below
saturation, the precise asymptotic
form has been discussed controversially \cite{YaMue,frusH01,SHSRS02}.
A recent diagrammatic analysis of the condensation problem into the
one-magnon dispersion
yields a square-root dependence with a logarithmic correction
for the frustrated square lattice at $J_2 = J_1/2$ \cite{JackZh03}.

\subsection{Shastry-Sutherland model versus SrCu$_2$(BO$_3$)$_2$}

\label{ExpShaSu}

\begin{figure}[tb]
\begin{center}
\includegraphics[angle=270,width=9.5 cm]{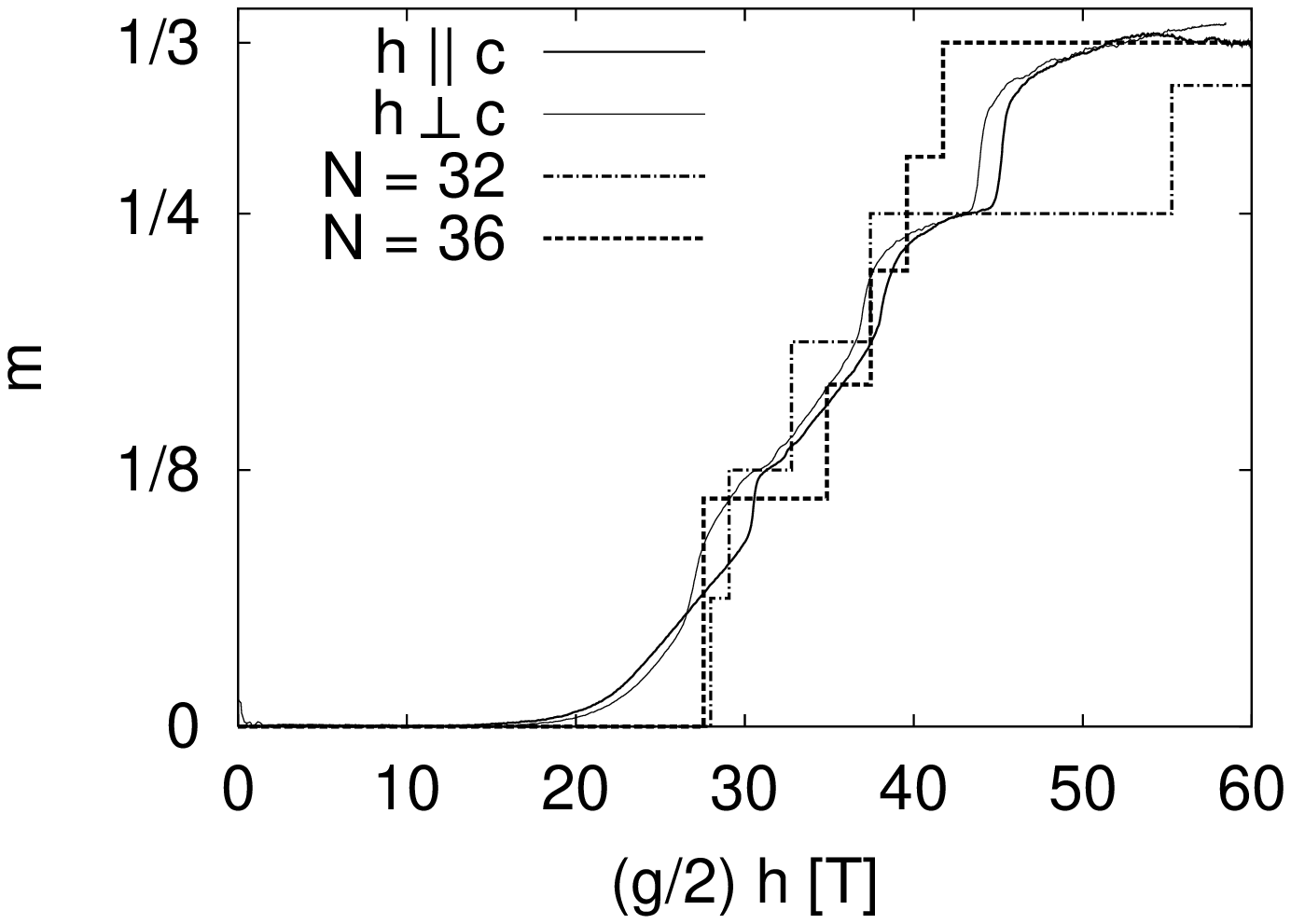}
\end{center}
\caption{
\label{figMscbo}
Magnetization curves of a SrCu$_2$(BO$_3$)$_2$ single crystal scaled by
$g=2.28$ for $h \parallel c$ and $g=2.05$ for $h \perp c$ \cite{OKNKUG}.
Also shown are magnetization curves of the $s=1/2$ Heisenberg antiferromagnet
on $N=32$ and $36$ Shastry-Sutherland lattices for $J_2 = 57$T,
$J_1 = 0.6\, J_2$.
}
\end{figure}

For the purpose of high-field magnetization experiments one does not only
need materials which realize a given lattice structure, but in addition $J$
must be small in order to be able to achieve full or at least a macroscopic
polarization of the sample in (pulsed) magnetization
experiments. SrCu$_2$(BO$_3$)$_2$ is an $s=1/2$ material whose
lattice structure corresponds to the tiling T6 and where the exchange
constants are sufficiently small to close the spin gap by an external
magnetic field and study the material at finite magnetizations in a
laboratory.
The magnetization process of SrCu$_2$(BO$_3$)$_2$ has attracted considerable
attention because plateaux are observed in the magnetization
curve\footnote{Magnetization experiments
are controlled by a material-dependent and anisotropic $g$-factor.
The $s=1/2$ spins in SrCu$_2$(BO$_3$)$_2$ are localized on Cu$^{2+}$-ions,
hence $g \approx 2$ -- see e.g.\ \cite{OKNKUG} for more details.}
at $m = 1/8$, $1/4$ and $1/3$
\cite{kageyama99dd,OKNKUG,KTHBKUMBM,JSJHBeSBDG03} (see Fig.\ \ref{figMscbo}).

By contrast, the tiling T6 at $J=J_1 = J_2$ has a smooth
magnetization curve (see Fig.\ \ref{mArch}),
hence we need to consider the Shastry-Sutherland model with $J_1 \ne J_2$.
The theoretical analysis of the magnetization process of the
two-dimensional Shastry-Sutherland model \cite{shastry81dd}
has been summarized in \cite{miyahara03dd}
-- here we discuss only some selected aspects.

For $J_2 \to 0$, the Shastry-Sutherland model reduces
to the square lattice antiferromagnet which is \Neel ordered in the transverse
components for all magnetic fields (see section \ref{secSquare}).
As discussed in section \ref{qpt}, this \Neel phase extends beyond
$J_2 = J_1$ for $m=0$. For $m \to 1$, \Neel order in the transverse
components is stable for
$J_2 \le J_1$ \cite{MoTo00}. We have performed a finite-size
analysis of the widths of the $m=1/8$, $1/4$, $1/3$ and $1/2$ steps
and found no indications for plateaux in the thermodynamic limit
for $J_2 = J_1$. These considerations indicate the absence of
quantum phase transitions between $J_2 = 0$ and $J_2 = J_1$
for any value of the field $h$ such that \Neel order
persists for the tiling T6 with $J_2 = J_1$ at all magnetic fields. 

This is one indication that SrCu$_2$(BO$_3$)$_2$ should be described
by $J_2 > J_1$ since several plateaux are observed in its magnetization curve
Fig.\ \ref{figMscbo}, namely at $m = 0$, $1/8$, $1/4$
and $1/3$ \cite{kageyama99dd,OKNKUG,KTHBKUMBM,JSJHBeSBDG03}.
In this regime, one can perform perturbation expansions around the limit
of decoupled dimers $J_1 = 0$ and indeed perturbation theory
plays a central role in the theoretical
approaches \cite{MiUe00,momoi00,MoTo00,FuOg00,Fukumoto01}.
Plateaux at $m=0$, $1/2$, $1/3$ and $1/4$ then arise in zeroth,
first, second and fourth order perturbation theory in $J_1$,
as has been clearly pointed out in \cite{Fukumoto01}.

For a direct comparison with the Shastry-Sutherland model,
we adopt the estimates $J_1 \approx 0.6\,J_2$ and
$J_2 \approx 70-75$K obtained by analyzing
inelastic neutron scattering data \cite{KBMHU00,KNAOYNK00,KnUh04} or
the specific heat in a magnetic field \cite{JSJHBeSBDG03}.
The magnetization curves for the
Shastry-Sutherland model shown in Fig.\ \ref{figMscbo}
were computed by choosing first $J_1 = 0.6\,J_2$ and
then setting the overall scale to $J_2 = 57$T ($\approx 77$K with $g=2$).
The $m=1/8$ and $1/4$ plateaux
(present only for $N=32$ in Fig.\ \ref{figMscbo}), the $m=1/3$ plateau
(present only for $N=36$ in Fig.\ \ref{figMscbo}) and the
$m=0$ plateau agree roughly with the experimental results \cite{OKNKUG}.
We have also performed computations for the value $J_1 = 0.68\,J_2$
proposed in \cite{OKNKUG} and have found less good agreement. 



\begin{figure}[tb]
\begin{center}
\includegraphics[width=9.5 cm]{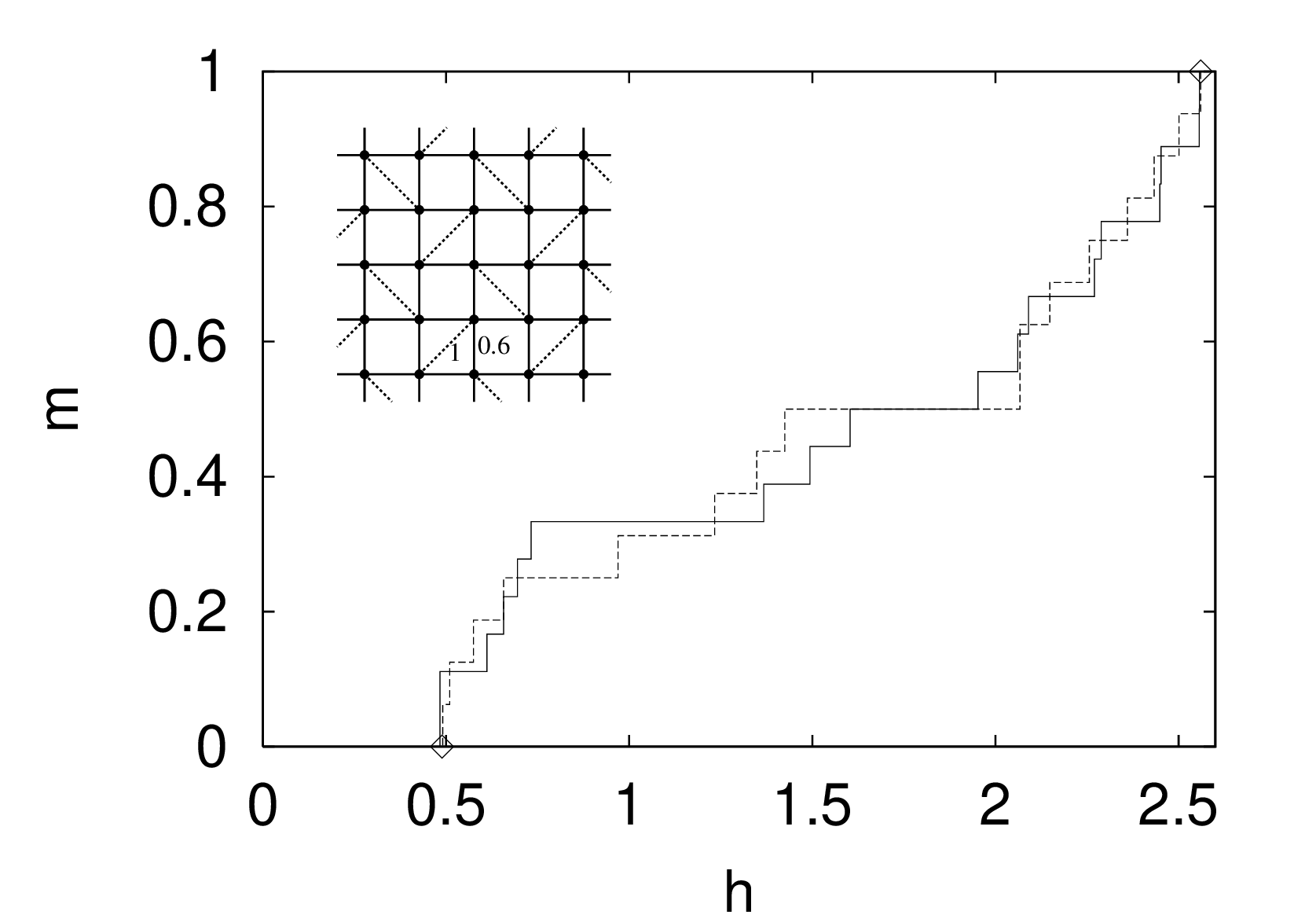}
\end{center}
\caption{
\label{figShaSu}
Magnetization curve of the $s=1/2$ Heisenberg antiferromagnet on the
Shastry-Sutherland lattice for $J_2 = 1$, $J_1 = 0.6$.
The dashed and solid lines are for $N=32$ and $N=36$ sites, respectively.
The diamonds denote the exact value of the saturation field at $m=1$
\cite{MoTo00} and a series expansion result for the gap to $S=1$ excitations
at $m=0$ \cite{KBMHU00}, respectively.
}
\end{figure}

Only the region with $m \le 1/3$ has so far been accessed
with magnetization experiments on SrCu$_2$(BO$_3$)$_2$.
Hence, the magnetization curves for the $N=32$ and $N=36$ Shastry-Sutherland
lattices are also restricted to $m \le 1/3$ in Fig.\ \ref{figMscbo}.
Fig.\ \ref{figShaSu} shows the corresponding complete magnetization curves
for  $J_2 = 0.6\, J_1$. Note that we have chosen an $N=32$ lattice
which is compatible with the structure of the $m=1/8$ plateau in
SrCu$_2$(BO$_3$)$_2$ as determined by NMR \cite{KTHBKUMBM}. 
For both finite lattices, the saturation field agrees well with the analytical
result \cite{MoTo00} shown by one of the diamonds in Fig.\ \ref{figShaSu}.
Also the boundary of the $m=0$ plateau is in good agreement with the
spin gap (i.e.\ the gap to $S=1$ excitations) computed by expansion
around the dimer limit $J_1 = 0$ \cite{KBMHU00} (compare the second diamond
in Fig.\ \ref{figShaSu}).

As on other lattices, it is more difficult to draw unambiguous
conclusions from finite-size data for intermediate values of $m$.
One complication which the Shastry-Sutherland model shares with the
trellis lattice are incommensurate ground states arising from the spiral
phase for $J_1 < J_2$ in the classical model (see section \ref{qpt}).
A more general aspect is that given magnetizations are realized
only for a limited number of sizes $N$. For example,
lattices with $N=32$ and $36$ share only $m=1/2$ in addition to $m=0$ and $1$.
Even for $m=1/2$ finite-size
effects are still important in Fig.\ \ref{figShaSu} although
the presence of a plateau at $m=1/2$ is well
established in the Shastry-Sutherland model (see above and \cite{miyahara03dd}).
$m=1/8$ is realized only for $N=16$ apart from $N=32$. From these two
lattice sizes one may estimate a plateau width 
$\lesssim J_2/10$ for the Shastry-Sutherland model
at $J_1 = 0.6\,J_2$, but the evidence in favor
of a plateau at $m=1/8$ is not very strong yet although its structure
has already been analyzed in detail \cite{KTHBKUMBM,MBM03}. 


\subsection{Summary of plateaux \& related topics}

Let us summarize the findings of this section. Firstly, in
section \ref{secSquare}
we have discussed the square lattice which we believe to be representative
for the non-frustrated bipartite tilings T2 (square), T6 (honeycomb), T10 (SHD)
and T11 (CaVO). In these cases, \Neel order probably persists in the
transverse components for all magnetic fields up to saturation,
leading to a smooth magnetization curve. The frustrated 
tilings T4 (maple leaf), T5 (trellis) and T6 (SrCuBO) may behave similarly.
At least their $N=36$ magnetization curves shown in Fig.\ \ref{mArch}
appear smooth and provide no evidence for any plateaux or jumps.

Also the triangular (T1) and bounce lattices (T7) are magnetically
ordered at $h=0$. However, in these two cases a plateau appears at $m=1/3$
in the magnetization curves (see section \ref{secTriag}).
In both cases, the appearance of a plateau
at $m=1/3$ may be attributed to the fact that these lattices are built
from triangles. Nevertheless, the structure of the $m=1/3$ state
on the bounce lattice may be different from the one of the triangular lattice
which corresponds to a long-range ordered collinear up-up-down spin
configuration.

The tilings T8 (\Kagome) and T9 (star) have the most interesting
magnetization curves. According to section \ref{seckagome}, at $h=0$ the
\kagome lattice is expected to have a small spin gap whereas the star
lattice has a large one.
This gives rise to a narrow and pronounced plateau at $m=0$, respectively.
Comparison of results for the $s=1/2$ Heisenberg 
antiferromagnet on the \kagome lattice
with different sizes $N$ (see section \ref{secKag}) shows that the magnetization
curve has a plateau at $m=1/3$. Evidence has been provided recently
\cite{CGHHPRS03} that the state of this $m=1/3$ plateau on the \kagome
lattice has a structure of the valence-bond-crystal type.
The $N=36$ magnetization curve of the star lattice shown in Fig.\ \ref{mArch}
indicates an $m=1/3$ plateau, too. Since the lattice T9 also consists of
triangles, it is plausible that this $m=1/3$ plateau persists in the limit
$N \to \infty$. Further plateaux are suspected on both lattices,
including one at $m=5/9$ on the \kagome lattice (see section \ref{secKag})
and a similar one at $m=7/9$ on the star lattice (compare Fig.\ \ref{figT9})
even if the currently available numerical data do not allow definite
conclusions. 

Close to saturation, exact eigenstates can be constructed for
the strongly frustrated tilings T8 (\Kagome) and T9 (star) -- see
section \ref{secImag}. For general $s$ they give rise to a jump
below saturation of height $\delta m = 1/(9\,s)$ (T8) and
$\delta m = 1/(18\,s)$ (T9). Furthermore, a plateau is expected
directly below this jump and such a plateau is indeed observed in
the $s=1/2$ Heisenberg 
antiferromagnet at $m=7/9$ for the \kagome lattice (T8)
and possibly at $m=8/9$ for the star lattice (T9). 

Although there are still open issues concerning the
magnetization process on the 11 Archimedean lattices, it is already
clear that even richer behavior is found if one allows different
exchange constants on non-equivalent bonds or adds further couplings.
Examples include the following:
\begin{itemize}
\item
The two-dimensional Shastry-Sutherland model has been discussed
in section \ref{ExpShaSu}. Here plateaux with $m=0$, $1/4$, $1/3$
and $1/2$ have been found and a further one is expected at $m=1/8$.
In contrast, its ancestor, the tiling T6
has a smooth magnetization curve with no particular features
(see Fig.~\ref{mArch}).
\item
A similar situation arises in the CaVO
lattice (T11) if one allows for two different exchange constants
$J$ and $J'$ as shown in Fig.\ \ref{cavo1}. It is clear at least
in the limit $J' \to 0$ that plateaux can then arise for $m=0$ and $1/2$
\cite{fukumoto99}.
Further plateaux with $m=1/4$, $3/4$ and $1/8$ arise in some parameter
regions if one adds a second-neighbor interaction $J_2$
\cite{fukumoto99,momoi00}.
\item
Not only the zero-field properties of the frustrated square lattice
have attracted considerable attention (see section \ref{qpt}), but
also its magnetization process has been studied intensively
\cite{LoNo93,gluzman94,YaMue,ZHP00,frusH01,HPZ01,FleMue01,ChaYa02,%
habil,JackZh03}.
In this model, a collinear up-up-up-down state arises at half the
saturation field \cite{ZHP00,HPZ01}. For $s=1/2$, this state is
found to be stabilized in the region
$0.5 \lesssim J_2/J_1 \lesssim 0.66$  where it gives rise to
a plateau with $m=1/2$ \cite{ZHP00,frusH01,HPZ01,habil}.
A further plateau at $m=1/3$ is predicted by a Chern-Simons theory
\cite{ChaYa02} and might also be observable in exact diagonalization
studies although the latter do not allow definite conclusions about
the presence or absence of an $m=1/3$ plateau yet \cite{habil}.
\item
Another variant of the square lattice is the checkerboard lattice,
a planar projection of the pyrochlore lattice. The $s=1/2$ checkerboard
lattice has a pronounced spin gap at $h=0$ \cite{PaCh01,FMSL03,BrHo02},
i.e.\ a plateau at $m=0$. In the limit of decoupled four-spin units
\cite{BrHo02} another plateau arises at $m=1/2$. Numerical
data for $N=32$, $40$ \cite{RSHSS03} and $36$ sites support the
presence of an $m=1/2$ plateau also in the checkerboard model where all
coupling constants are equal.
The construction of section \ref{secImag} predicts another plateau at
$m=3/4$ in the $s=1/2$ checkerboard model although here the numerical evidence
\cite{RSHSS03} is less clear.
\item
One can also add multi-spin interactions. On the triangular lattice,
inclusion of four-spin cyclic exchange terms in the $s=1/2$ model
gives rise to an additional plateau at $m=1/2$
\cite{MBLW98,MSK99,LiMMSL00,lhuillier01sep}. This $m=1/2$ plateau is already
present in the classical model where one also finds an $m=1/3$
plateau for a suitable choice of parameters \cite{KuMo97}. The
latter differs from the $m = 1/3$ plateau of section \ref{secTriag}
which arises only in the quantum Heisenberg 
antiferromagnet on the triangular lattice
and is absent in the classical limit.
\end{itemize}
All the aforementioned plateaux for $m\ne 0$ give rise to ordered
ground states (at least in those cases where the state of the plateau
is sufficiently well understood). The unit cell of the ground state then
has a volume $V$ such that the magnetization $m$ satisfies the quantization
condition (\ref{condM}). Hence, also in two dimensions this quantization
condition seems to hold generically.

The transitions to saturation in 9 of the 11 Archimedean lattices appear
to be continuous quantum phase transition (see Fig.\ \ref{mArch}). Generically,
the dispersion of the one-magnon excitations above the ferromagnetically
polarized state should be
quadratic close to their minima. An analysis of the associated
condensation problem then predicts the following universal asymptotic
behavior of the magnetization curve close to the saturation field
$h_{\rm sat}$ \cite{Popov72,gluzman93,SSS94}
\begin{equation}
1 - m \sim \left({h_{\rm sat} - h \over J}\right)
     \ln\left( {b \, J \over h_{\rm sat} - h} \right) 
\label{MagLog}
\end{equation}
where $b$ is a non-universal constant. The logarithmic correction
in (\ref{MagLog}) is characteristic for two dimensions and
arises because of a logarithmic singularity in the density of states
\cite{Popov72}.
The functional form (\ref{MagLog}) has been verified by a first-order
spin-wave approximation for the square lattice \cite{ZhNi} and
numerically for the $s=1/2$ Heisenberg antiferromagnet on the
square, honeycomb and triangular lattices \cite{hon99,habil}.
We note that a behavior of the form (\ref{MagLog}) is expected to be valid
at generic continuous transitions at plateau boundaries in two dimensions
\cite{SSS94} (at least in those cases where the fundamental excitations
are magnons).

Deviations from (\ref{MagLog}) are expected if the one-magnon dispersion
is not quadratic close to the minimum which in general requires fine-tuning
of parameters. Nevertheless, completely flat bands arise in
two Archimedean lattices, namley the \kagome and star lattices (T8 and T9).
In these cases, we find a macroscopic jump in the magnetization curve just
below saturation as we have discussed in section \ref{secImag}

\vspace{1cm}
\noindent {\bf Acknowledgments}

\noindent
The authors are indebted to R.~Ferchmin and P.~Tomczak for bringing our
attention to the 11 Archimedean tilings.
We thank A.~Harrison to inform us on  Ref.\ \cite{suding99}.
We would like to thank H.-U.~Everts and R.~Schumann for critical reading
the manuscript and  P.~Tomczak and H.-U.~Everts for useful discussions.
Thanks go to H.~Kageyama for sending us the magnetization data of
\cite{OKNKUG} which we reproduced in Fig.~\ref{figMscbo}.
The more complicated computations presented in this article have
been performed on the compute-servers {\tt wildfire}, {\tt marvel}
and {\tt cfgauss} at the computing centers of Magdeburg University
and the TU Braunschweig, respectively. We are particularly grateful
to J.\ Sch\"ule of the Rechenzentrum at the TU Braunschweig
for technical support.
This work was partly supported by the DFG (project Ri615/10-1).



\end{document}